\documentstyle[12pt,epsfig,axodraw,psfrag]{article}
\begin{document}

\newcommand{\ftsz}{\footnotesize}
\newcommand{\be}[1]{\begin{equation}\label{#1}}
\newcommand{\beq}{\begin{equation}}
\newcommand{\ee}{\end{equation}}
\newcommand{\beqn}[1]{\begin{eqnarray}\label{#1}}
\newcommand{\bea}{\begin{eqnarray}}
\newcommand{\eea}{\end{eqnarray}}
\newcommand{\bd}{\begin{displaymath}}
\newcommand{\ed}{\end{displaymath}}
\newcommand{\mat}[4]{\left(\begin{array}{cc}{#1}&{#2}\\{#3}&{#4}
\end{array}\right)}
\newcommand{\matr}[9]{\left(\begin{array}{ccc}{#1}&{#2}&{#3}\\
{#4}&{#5}&{#6}\\{#7}&{#8}&{#9}\end{array}\right)}
\newcommand{\matrr}[6]{\left(\begin{array}{cc}{#1}&{#2}\\
{#3}&{#4}\\{#5}&{#6}\end{array}\right)}
\def\lsim{\raise0.3ex\hbox{$\;<$\kern-0.75em\raise-1.1ex
\hbox{$\sim\;$}}}
\def\gsim{\raise0.3ex\hbox{$\;>$\kern-0.75em\raise-1.1ex
\hbox{$\sim\;$}}}
\def\abs#1{\left| #1\right|}
\def\simlt{\mathrel{\lower2.5pt\vbox{\lineskip=0pt\baselineskip=0pt
           \hbox{$<$}\hbox{$\sim$}}}}
\def\simgt{\mathrel{\lower2.5pt\vbox{\lineskip=0pt\baselineskip=0pt
           \hbox{$>$}\hbox{$\sim$}}}}
\def\unity{{\hbox{1\kern-.8mm l}}}
\def\epr{E^\prime}
\def\16p{16\pi^2}
\def\ga{\gamma}
\def\Ga{\Gamma}
\def\la{\lambda}
\def\La{\Lambda}
\def\al{\alpha}

\newcommand{\ov}{\overline}
\renewcommand{\to}{\rightarrow}
\renewcommand{\vec}[1]{\mbox{\boldmath$#1$}}
\def\mcirc{{\stackrel{o}{m}}}
\newcommand{\tanb}{\tan\beta}
\def\dfrac#1#2{{\displaystyle\frac{#1}{#2}}}

\def\bc{\begin{center}}
\def\ec{\end{center}}
\def\dd{\displaystyle}
\def\nn{\nonumber}
\def\mta{m_{\tau}}
\def\cl{{\cal L}}
\def\cm{{\cal M}}
\def\co{{\cal O}}
\def\gol{\tilde{G}}
\def\pho{\tilde{\gamma}}
\def\lab{\ov\lambda}
\def\gf{G_{\small F}} 
\def\fb{\ov{f}}
\def\fc{f^c}
\def\fcb{\ov\fc}
\def\ft{\tilde{f}}
\def\fct{\tilde{\fc}}
\def\fl{\tilde{f}_L}
\def\fr{\tilde{f}_R}
\def\lt{\tilde{\ell}}
\def\qt{\tilde{q}}
\def\mt{\tilde{m}}
\def\mmt{{\bf\mt}}
\def\staur{\tilde{m}_{\tilde{\tau}_R}}
\def\staul{\tilde{m}_{\tilde{\tau}_L}}
\def\smur{\tilde{m}_{\tilde{\mu}_R}}
\def\smul{\tilde{m}_{\tilde{\mu}_L}}
\def\mll{\tilde{m}^2_{LL}}
\def\mrr{\tilde{m}^2_{RR}}
\def\mlr{\tilde{m}^2_{LR}}
\def\mrl{\tilde{m}^2_{RL}}
\def\th{\theta}
\def\thb{\ov\theta}
\def\dmu{\partial^\mu}
\def\dmd{\partial_\mu}
\def\dnu{\partial^\nu}
\def\dnd{\partial_\nu}
\def\dslash{\not{\! \partial}}
\def\Dslash{\not{\! D}}
\def\lsq{{\Lambda^2}}
\def\smd{\sigma_{\mu}}
\def\smu{\sigma^{\mu}}
\def\smdb{{\bar\sigma}_{\mu}}
\def\smub{{\bar\sigma}^{\mu}}
\def\snd{\sigma_{\nu}}
\def\snu{\sigma^{\nu}}
\def\sndb{{\bar\sigma}_{\nu}}
\def\snub{{\bar\sigma}^{\nu}}
\def\smn{\sigma^{\mu\nu}}
\def\smnb{{\bar\sigma}^{\mu\nu}}

\begin{titlepage}

\begin{flushright}
{DFPD-04/TH/06} \\

\end{flushright}

\vspace{2.0cm}

\begin{center}

{\Large  \bf 
Anatomy and Phenomenology of \\ 
\vspace{0.5cm}
$\mu$-$\tau$  Lepton Flavour Violation in the MSSM
}

\vspace{0.7cm}

{\large Andrea Brignole\footnote{ E-mail address: andrea.brignole@pd.infn.it} 
and Anna Rossi\footnote{ E-mail address: anna.rossi@pd.infn.it}}

\vspace{10mm}

{\it  Dipartimento di Fisica `G.~Galilei',  
Universit\`a di Padova and \\
\vspace{0.1cm}
INFN, Sezione di Padova,
Via Marzolo 8, I-35131 Padua, Italy}

\end{center}

\vspace{8mm}

\begin{abstract}

\vspace{0.6cm}
\noindent
We perform a detailed analysis of several 
$\mu$-$\tau$ lepton flavour 
violating (LFV) processes, namely $\tau \to \mu X$ 
($X=\ga,  e^+ e^-, \mu^+ \mu^-, \rho, \pi, \eta, \eta'$), 
$Z\to \mu \tau$ and  Higgs boson decays into $\mu \tau$.
First we present a model independent operator analysis relevant 
to such decays, then we explicitly compute 
the LFV operator coefficients [and  $(g_\mu -2)$] 
in a general  {\it unconstrained} MSSM framework, 
allowing  slepton mass 
matrices to have large  $\mu$-$\tau$ entries. 
We systematically study the role and the interplay 
of dipole and non-dipole operators, showing 
how the rates and the mutual correlations of those LFV decays 
change in different regions of the MSSM parameter space.
Values of the LFV branching ratios  in the experimentally 
interesting range $10^{-9}-10^{-7}$ can be achieved.
For at least two  MSSM Higgs bosons, the
branching ratio of the LFV decay into $\mu \tau$ can 
reach values of order $10^{-4}$.

\end{abstract}

\end{titlepage}

\setcounter{footnote}{0}

\section{Introduction}
The search for lepton flavour violating processes 
is an important window into physics beyond the Standard Model.
In the Minimal Supersymmetric extension of the Standard Model (MSSM) 
several  such  processes can have non-negligible 
rates, provided the mass matrices of the leptons and of the sleptons are 
not aligned. 
The experimental bounds strongly constrain the amount 
of misalignment in the $e$-$\mu$ sector, whilst the constraints 
are weaker for the $\mu$-$\tau$ or $e$-$\tau$ sector 
(see {\it e.g.} \cite{fv,GGMS}). 
In this work we thoroughly explore the case  of large 
$\mu$-$\tau$ lepton flavour violation (LFV) in a general unconstrained 
MSSM framework, assuming negligible LFV in the other sectors.
The case of large $\mu$-$\tau$ LFV is partly motivated also by  
the observed atmospheric neutrino anomaly, which can be interpreted 
in terms of $\nu_\mu \to \nu_\tau$ oscillations with maximal 
mixing angle \cite{nus}.
The prototype and  most studied LFV process involving $\mu$ and $\tau$ 
is the radiative decay $\tau \to \mu\ga$, which proceeds 
via dipole operators. 
Besides this decay, we also consider other interesting LFV processes, 
such as $\tau \to \mu e e,~ \tau \to \mu \mu \mu, ~\tau \to \mu \rho, ~
\tau \to \mu \pi, ~\tau \to \mu \eta, ~\tau \to \mu \eta', ~ 
Z\to \mu \tau$, which may have different degrees of 
correlation with $\tau\to \mu \ga$.    
The present experimental bounds on all these decays are:
\bea
 BR(\tau^- \to \mu^-\ga) & < & 3.1\times 10^{-7}  ~~~~ \cite{exp-rad} ,
\label{e-rad}  \\ 
 BR(\tau^- \to \mu^- e^+ e^- ) & <& 1.9 \times 10^{-7} ~~~~ \label{e-mue} 
\cite{belle},\\ 
 BR(\tau^- \to \mu^- \mu^+ \mu^-)& <&  1.9 \times 10^{-7} ~~~~ \label{e-3mu} 
\cite{babar,belle}, \\ 
 BR(\tau^- \to \mu^- \rho^0)& < & 6.3 \times 10^{-6}  \label{e-muro}
~~~~ \cite{pdg},  \\ 
 BR(\tau^- \to \mu^- \pi^0) &< & 4.0 \times 10^{-6}  \label{e-mupi} 
~~~~ \cite{pdg}, \\
 BR(\tau^- \to \mu^- \eta) &< & 3.4 \times 10^{-7}  \label{e-mueta} 
~~~~ \cite{Enari}, \\
 BR(Z\to \mu^+\tau^-) & < & 1.2 \times 10^{-5}  \label{e-z} 
~~~~ \cite{pdg}. 
\eea
(We are not aware of experimental bounds on $\tau \to \mu \eta'$.)
In the MSSM with large $\mu$-$\tau$ flavour violation,  
a combined analysis of all such processes is  
very instructive, especially in view of the future sensitivity 
on those branching ratios, which may  reach 
$10^{-8} - 10^{-9}$ \cite{kekb,tesla}.  
In particular, it is interesting to study the interplay 
between dipole and non-dipole operators.
In addition, we  consider the LFV decays of MSSM  
Higgs bosons into $\mu \tau$ \cite{BR}, which 
are correlated to some of the above processes and 
can give testable signatures at future colliders. 

The outline of this paper is as follows. In Section~2 we introduce 
the relevant effective operators and compute the branching ratios
of the LFV decays.
In Section~3 we focus on the MSSM, exhibit the sources of 
LFV in the left and right slepton mass matrices   
and discuss the computation of 
the operator coefficients [including the Higgs-$\mu$-$\tau$ and 
($g_\mu-2$) ones], whose analytical expressions are displayed 
in Appendix. 
In Section~4   we give a numerical discussion of the 
branching ratios in  the case of 
large LFV in the left sector. In Section~5 we present an analogous  
discussion for large LFV in the right sector.   
Finally, in Section~6 we summarize our results.


\section{Effective operators and branching ratios}
This section is devoted to the `model-independent' calculation 
of the LFV branching ratios. Namely,  
first we parametrize the basic effective operators (Section 2.1), 
then  construct 
appropriate effective lagrangians   and 
compute  the  branching ratios in terms of the coefficients 
of the effective operators   (Section 2.2).
We also discuss correlations among different processes in cases of 
single-operator-dominance (Section~2.3).

\subsection{Parametrization of LFV effective operators \label{plfv}}
Our first step, in the derivation of the LFV branching ratios,
is the parametrization of the LFV operators 
that contain a muon, a tau and either a $Z$ boson, or a photon, 
or an additional $f$-fermion pair. 
The leading contributions to these operators arise from  $d=6$ 
$SU(2)_W\times U(1)_Y$-invariant operators \cite{BW}, possibly 
with Higgs fields set at their vacuum expectation values (VEVs). 
It is useful to keep this in mind, although we will not 
write explicitly the operators in the unbroken phase (except for 
a few examples).  
We also recall that the MSSM contains two Higgs doublets 
$H_1, H_2$, whose VEVs define the ratio 
$\tanb=\langle H^0_2\rangle/\langle H^0_1\rangle$. 
We postpone  the discussion of 
$d=4$ Higgs-muon-tau effective operators to Section~\ref{hmt}.
The parametrization given below assumes 
fermions to be on-shell, whereas gauge bosons could also 
be off-shell. We keep the tau mass $m_\tau$ at the leading order 
and neglect $m_\mu, m_e$ as well as the light-quark masses. 
We adopt  two-component spinor notation, so for example  
$\mu$ and $\tau$ ($\bar{\mu}^c$ and $\bar{\tau}^c$) are the left-handed 
(right-handed) components of the muon 
and tau fields, respectively\footnote
{For instance, in terms of the four-component spinors 
$\psi^T_\mu = (\mu ~\ov{\mu}^c)$, 
$\psi^T_\tau = (\tau ~\ov{\tau}^c)$,   
the bilinears $\ov{\mu} \snub \tau $ and 
$\mu^c \snu \ov{\tau}^c$ correspond  to 
$\ov{\psi}_{\mu} \ga^\nu P_L \psi_{\tau}$ and $\ov{\psi}_\mu \ga^\nu 
P_R \psi_\tau $ [$P_{L,R} = \frac12 (1\mp \ga^5)$] 
respectively.
We  take $\smu \equiv (1,\vec{\sigma})$, $\smub\equiv 
(1,-\vec{\sigma})$, $\smn  \equiv \frac14 (\smu \snub -\snu\smub)$, 
$\smnb  \equiv \frac14 (\smub \snu -\snub\smu)$ and 
$g_{\mu\nu}= {\rm diag}(+1, -1, -1, -1)$, where  
$\vec{\sigma} = (\sigma^1, \sigma^2, \sigma^3)$ are the 
$2\times 2$ Pauli matrices.}.
Sometimes ({\it e.g.} in Figs.~\ref{f1}, \ref{f2}, \ref{fh2}) 
symbols like $\mu, \tau, f$ 
will generically refer to the particles, not to 
specific chiralities. 
Finally, it is understood that the coefficients of all the LFV 
operators below should carry a flavour subscript $\mu\tau$, which 
we omit for brevity.
\vspace{0.3cm}

\noindent 
\underline{{\large \it $\tau \mu Z$ effective operators}}
\vspace{0.3cm}

We distinguish the following operators: 
\be{zm0}
g_Z m_Z^2 \left[ A_L^Z \, \ov{\mu} \smub \tau 
+ A_R^Z \, \mu^c \smu \ov{\tau}^c  + {\rm h.c.}\right]  Z_{\mu} , 
\ee
\be{zm2} 
g_Z \left[ C_L^Z \,  \ov{\mu} \smub \tau 
+ C_R^Z \, \mu^c \smu \ov{\tau}^c  + {\rm h.c.}\right] \Box Z_{\mu} , 
\ee
\be{zd}
g_Z \mta \left[ i D_L^Z \,  \ov{\mu} \smnb  \ov{\tau}^c 
+ i D_R^Z \, \mu^c \smn \tau + {\rm h.c.}\right] Z_{\mu\nu} , 
\ee 
where $g_Z= \sqrt{g^2 +g'^2}$ ($g, g'$ being the 
$SU(2)_W$ and $U(1)_Y$  coupling constants, respectively) 
and $Z_{\mu\nu}= \dmd Z_\nu - \dnd Z_\mu$. 
The   operators in (\ref{zm0})  are chirality conserving (monopole) and 
have no derivatives,  so they originate from 
$SU(2)_W\times U(1)_Y$-invariant operators with at least two Higgs fields, 
which  reflects in the factor $m^2_Z$ we have extracted out.
For instance, the leading contributions to (\ref{zm0})  
 come from operators such as:
\be{invs}
\left(\ov{L}_\mu \smub L_\tau\right)\left(H^\dagger_1 i D_\mu H_1\right) , 
~~~~~
\left(\ov{L}_\mu \smub \sigma^a L_\tau\right)\left(H^\dagger_1 \sigma^a 
i D_\mu 
 H_1\right) , ~~~~ 
\left(\mu^c \smu \ov{\tau}^c\right) \left(H^\dagger_1 i D_\mu H_1\right) ,
\ee
and analogous ones with $ H^\dagger_1 H_1$ replaced by 
$H^\dagger_2 H_2$ or $H_2 H_1 (+{\rm h.c.})$, where $L_\mu, L_\tau$ 
are the   $SU(2)_W$ lepton doublets.
The   operators in (\ref{zm2})  are also chirality conserving (monopole) and 
have two derivatives,  so they originate from 
$SU(2)_W\times U(1)_Y$-invariant operators with no Higgs fields 
at the leading order. 
The   operators in (\ref{zd}) are chirality flipping (dipole) and come 
from  $SU(2)_W\times U(1)_Y$-invariant operators with at least 
one Higgs field, hence we have  pulled out a factor $m_\tau$.
In the case of virtual $Z$, only the $A^Z$-operators need to be 
considered since the remaining ones (and other ones not listed above) 
give suppressed contributions. In the case of real  $Z$, the  $A^Z$- and 
$C^Z$-operators give  contributions of the same order\footnote{
Sometimes this feature has been overlooked in the literature, 
by  assuming the chirality 
conserving operators to be dominated by the zero-momentum 
component, {\it i.e.} $A^Z$ in our notation 
(see {\it e.g.} \cite{vis}).}
(notice that $\Box \to - m^2_Z$). 
The $D^Z$-operators  give comparable 
contributions if the $m_\tau$ suppression is compensated by a large 
$\tanb$ factor, 
induced by the Higgs field $H_2$ (notice that 
${50}\times m_\tau \sim m_Z$).
\vspace{0.3cm}

\noindent 
\underline{{\large \it  $\tau \mu \ga$ effective operators }}

\vspace{0.3cm}
We distinguish the following operators:
\be{phm} 
e \left[ C_L^{\ga} \,  \ov{\mu} \smub \tau 
+ C_R^{\ga} \, \mu^c  \smu \ov{\tau}^c + {\rm h.c.}\right] \Box A_{\mu} , 
\ee
\be{phd}
e \mta \left[i D_L^{\ga} \, \ov{\mu} \smnb  \ov{\tau}^c 
+ i D_R^{\ga} \,  \mu^c \smn \tau + {\rm h.c.}\right] F_{\mu\nu} .
\ee 
The operators in eqs.~(\ref{phm}),(\ref{phd}) are analogous to 
the $C^Z$ and $D^Z$ operators, respectively. 
We recall that the $Z$ and photon operators  have a common 
origin in terms of $SU(2)_W\times U(1)_Y$-invariant operators. 
However, the $A^Z$-operators (\ref{zm0}) have no photon counterpart 
because of the unbroken $U(1)_{em}$ symmetry.
In the case of virtual photon, both the monopole and dipole 
operators contribute, whereas in the case of real 
photon only the latter ones do.
\vspace{0.3cm}

\noindent 
\underline{{\large \it $\tau \mu f f$ effective operators }}

\vspace{0.3cm}
We parametrize  four-fermion operators  as follows\footnote{
Another discussion on the phenomenological implications
of $\mu$-$\tau$ LFV four-fermion operators appeared 
in \cite{Black}.}:
\be{box}
\sum_f \left[ 
(\ov{\mu} \smub \tau)  
\left( B_L^{f_L} \,\fb \smdb f + B_L^{f_R} \, f^c \smd  \fcb \right)
+ ( \mu^c \smu \ov{\tau}^c ) 
\left( B_R^{f_L} \,\fb \smdb f + B_R^{f_R} \,  f^c \smd  \fcb \right)
+ {\rm h.c.} \right] .  
\ee
Here we have  retained  chirality conserving  operators only and 
neglected chirality flipping ones, {\it e.g.} 
$(\mu^c \tau) (f f^c)$, because the latter are expected 
to suffer from a double chiral suppression and so  
would give subleading contributions to our LFV processes 
(in particular this holds in the MSSM, 
even for large $\tanb$). 
We emphasize that eq.~(\ref{box}) does not yet include
either the contributions induced by $Z$ and photon exchange,
to be added in the next section, or 
the chirality flipping operators induced by Higgs exchange,
which are relevant for large $\tanb$ and will
be discussed in Section~\ref{hmt}.

\subsection{Effective lagrangians and branching ratios}
All the coefficients of the operators listed above ($A^Z, C^Z, D^Z, C^\ga, 
D^\ga, B^f$) have mass dimension $-2$, and encode information on the 
underlying physics. 
Such    coefficients are the building blocks for the amplitudes 
of our LFV processes, as schematically shown  in Fig.~\ref{f1}.
Now we present for each  LFV decay the relevant lagrangian and 
the corresponding branching ratio.

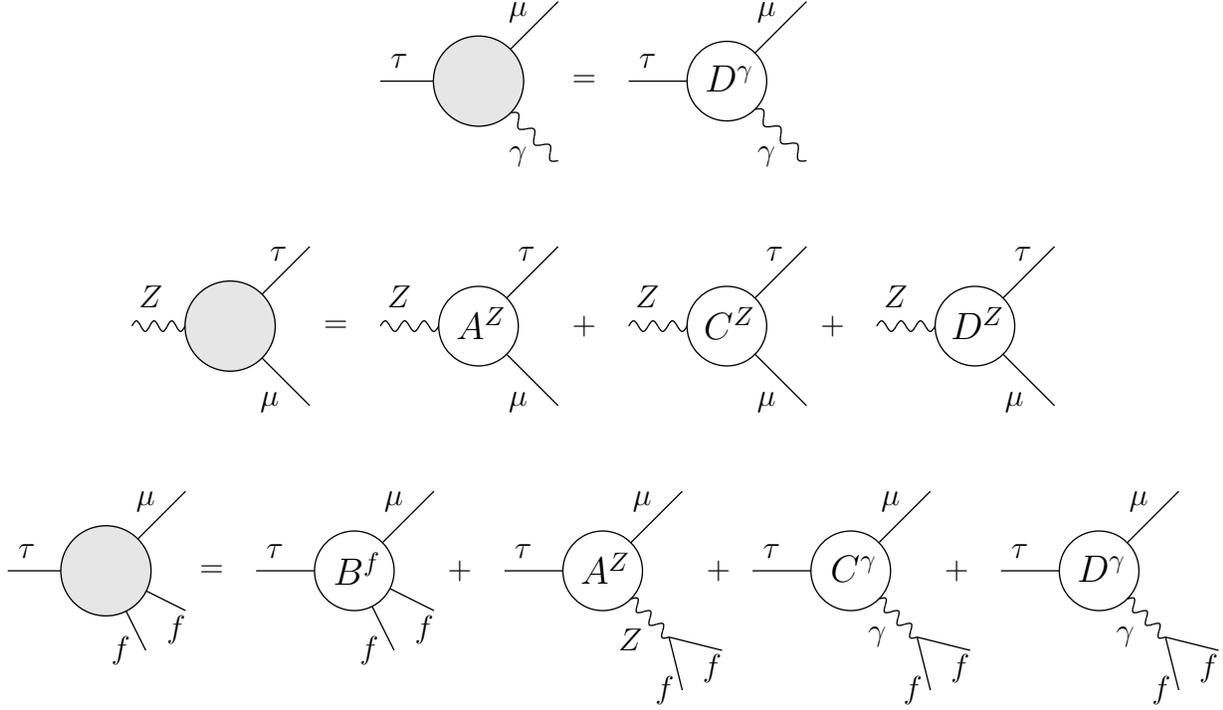
\begin{figure}[ht]
\begin{center}
\begin{picture}(90,60)(-45,-30)
\GCirc(0,0){17}{0.9}
\Line(-37,0)(-17,0)
\Line(12,12)(30,30)
\Photon(12,-12)(30,-30){2}{3}
\Text(-30,5)[b]{$\tau$}
\Text(15,30)[t]{$\mu$}
\Text(15,-30)[b]{$\gamma$}
\Text(40,0)[]{$=$}
\end{picture}
\begin{picture}(90,60)(-45,-30)
\CArc(0,0)(15,0,360)
\Line(-37,0)(-15,0)
\Line(10.6,10.6)(30,30)
\Photon(10.6,-10.6)(30,-30){2}{3}
\Text(-30,5)[b]{$\tau$}
\Text(15,30)[t]{$\mu$}
\Text(15,-30)[b]{$\gamma$}
\Text(1,1)[]{{\large $D^{\gamma}$}}
\end{picture}
\end{center}

\vspace{0.2 cm}
\begin{center}
\begin{picture}(90,60)(-45,-30)
\GCirc(0,0){17}{0.9}
\Photon(-37,0)(-17,0){2}{3}
\Line(12,12)(30,30)
\Line(12,-12)(30,-30)
\Text(-30,7)[b]{$Z$}
\Text(15,30)[tl]{$\tau$}
\Text(15,-30)[b]{$\mu$}
\Text(40,0)[]{$=$}
\end{picture}
\begin{picture}(90,60)(-45,-30)
\CArc(0,0)(15,0,360)
\Photon(-37,0)(-15,0){2}{3}
\Line(10.6,10.6)(30,30)
\Line(10.6,-10.6)(30,-30)
\Text(-30,7)[b]{$Z$}
\Text(15,30)[tl]{$\tau$}
\Text(15,-30)[b]{$\mu$}
\Text(1,1)[]{{\large $A^Z$}}
\Text(40,0)[]{$+$}
\end{picture}
\begin{picture}(90,60)(-45,-30)
\CArc(0,0)(15,0,360)
\Photon(-37,0)(-15,0){2}{3}
\Line(10.6,10.6)(30,30)
\Line(10.6,-10.6)(30,-30)
\Text(-30,7)[b]{$Z$}
\Text(15,30)[tl]{$\tau$}
\Text(15,-30)[b]{$\mu$}
\Text(1,1)[]{{\large $C^Z$}}
\Text(40,0)[]{$+$}
\end{picture}
\begin{picture}(90,60)(-45,-30)
\CArc(0,0)(15,0,360)
\Photon(-37,0)(-15,0){2}{3}
\Line(10.6,10.6)(30,30)
\Line(10.6,-10.6)(30,-30)
\Text(-30,7)[b]{$Z$}
\Text(15,30)[tl]{$\tau$}
\Text(15,-30)[b]{$\mu$}
\Text(1,1)[]{{\large $D^Z$}}
\end{picture}
\end{center}

\vspace{0.2 cm}
\begin{center}
\begin{picture}(90,75)(-45,-45)
\GCirc(0,0){17}{0.9}
\Line(-37,0)(-17,0)
\Line(12,12)(30,30)
\Line(15.2,-7.6)(30,-15)
\Line(7.6,-15.2)(15,-30)
\Text(-30,5)[b]{$\tau$}
\Text(15,30)[t]{$\mu$}
\Text(30,-25)[br]{$f$}
\Text(2,-30)[l]{$f$}
\Text(40,0)[]{$=$}
\end{picture}
\begin{picture}(90,75)(-45,-45)
\CArc(0,0)(15,0,360)
\Line(-37,0)(-15,0)
\Line(10.6,10.6)(30,30)
\Line(13.4,-6.7)(30,-15)
\Line(6.7,-13.4)(15,-30)
\Text(-30,5)[b]{$\tau$}
\Text(15,30)[t]{$\mu$}
\Text(30,-25)[br]{$f$}
\Text(2,-30)[l]{$f$}
\Text(1,1)[]{{\large $B^f$}}
\Text(40,0)[]{$+$}
\end{picture}
\begin{picture}(90,75)(-45,-45)
\CArc(0,0)(15,0,360)
\Line(-37,0)(-15,0)
\Line(10.6,10.6)(30,30)
\Photon(10.6,-10.6)(25,-25){2}{3}
\Line(25,-25)(45,-30)
\Line(25,-25)(30,-45)
\Text(-30,5)[b]{$\tau$}
\Text(15,30)[t]{$\mu$}
\Text(15,-30)[br]{$Z$}
\Text(45,-40)[br]{$f$}
\Text(20,-45)[l]{$f$}
\Text(1,1)[]{{\large $A^Z$}}
\end{picture}
\begin{picture}(90,75)(-45,-45)
\CArc(0,0)(15,0,360)
\Line(-37,0)(-15,0)
\Line(10.6,10.6)(30,30)
\Photon(10.6,-10.6)(25,-25){2}{3}
\Line(25,-25)(45,-30)
\Line(25,-25)(30,-45)
\Text(-30,5)[b]{$\tau$}
\Text(15,30)[t]{$\mu$}
\Text(10,-25)[]{$\gamma$}
\Text(45,-40)[br]{$f$}
\Text(20,-45)[l]{$f$}
\Text(1,1)[]{{\large $C^{\gamma}$}}
\Text(-50,0)[]{$+$}
\Text(40,0)[]{$+$}
\end{picture}
\begin{picture}(90,75)(-45,-45)
\CArc(0,0)(15,0,360)
\Line(-37,0)(-15,0)
\Line(10.6,10.6)(30,30)
\Photon(10.6,-10.6)(25,-25){2}{3}
\Line(25,-25)(45,-30)
\Line(25,-25)(30,-45)
\Text(-30,5)[b]{$\tau$}
\Text(15,30)[t]{$\mu$}
\Text(10,-25)[]{$\gamma$}
\Text(45,-40)[br]{$f$}
\Text(20,-45)[l]{$f$}
\Text(1,1)[]{{\large $D^{\gamma}$}}
\end{picture}
\end{center}
\caption{\ftsz The different  contributions to 
$\tau \to \mu \ga$, $ Z\to \mu\tau$ and $\tau\to \mu f f$ decays.
}
\label{f1}
\end{figure}

\vspace{0.3cm}

\noindent 
\underline{{\large \bf $Z \to  \mu \tau $}}
\vspace{0.3cm}
 
The effective lagrangian for the $Z \to  \mu \tau $ decay is easily 
obtained from eqs.~(\ref{zm0}-\ref{zd}):
\be{lzmt}
{\cal L}^{\rm eff}_{Z\mu\tau}  =  
g_Z \left[m_Z^2 \left( F_L^Z \, \ov{\mu} \smub \tau 
+ F_R^Z \, \mu^c \smu \ov{\tau}^c  \right)Z_\mu - 
2m_\tau\left( D_L^Z \,  \ov{\mu} \smnb  \ov{\tau}^c + 
D_R^Z \, \mu^c \smn \tau\right) i \partial_\nu Z_\mu\right]
+ {\rm h.c.} , 
\ee
where 
\be{fz}
F^Z_{L (R)} = A^Z_{L (R)} - C^Z_{L (R)} .
\ee
The branching ratio reads as: 
\be{bz}
BR(Z \to \mu^+ \tau^-)  =   c\, m_Z^4
\left[ |F_L^Z|^2 + |F_R^Z|^2 + \frac12
\left|{ \mta \over  m_Z}D_L^Z\right|^2 + \frac12\left| 
{\mta \over  m_Z} D_R^Z \right|^2 \right] \,
BR(Z \to \ell^+ \ell^-)
\ee
where $c \equiv (1/4-s_W^2+2 s_W^4)^{-1} \simeq 7.9$,
$BR(Z \to \ell^+ \ell^-) \simeq 3.4 \% $ and lepton masses have been 
neglected in the kinematics\footnote{Notice that the factors $m_\tau$ 
in eq.~(\ref{bz}) come from the $D^Z$-operators, which may contain 
a (compensating) large $\tanb$ factor, as mentioned above. 
}.

\vspace{0.3cm}

\noindent 
\underline{{\large \bf $\tau \to  \mu \ga $}}
\vspace{0.3cm}

In this case the expression of the effective  lagrangian
is already in eq.~(\ref{phd}). The related branching ratio is: 
\be{rate1}
BR(\tau^- \to \mu^- \ga) = \frac{48 \pi^3 \alpha}{G_F^2}
\left[ |D_L^{\ga}|^2 +|D_R^{\ga}|^2 \right] \,
BR( \tau^- \to \mu^- \bar{\nu}_{\mu} \nu_{\tau})
\ee
where $\alpha=e^2/(4\pi)$ is the fine-structure constant, $G_F$ 
is the Fermi constant and 
$BR( \tau^- \to \mu^- \bar{\nu}_{\mu} \nu_{\tau}) \simeq 17 \% $.

\vspace{0.3cm}

\noindent 
\underline{{\large \bf $\tau \to \mu \ell^+\ell^-, ~\tau\to\mu \rho, ~
\tau\to\mu P ~(P=\pi, \eta, \eta')$ }}
\vspace{0.3cm}

The four-fermion effective lagrangian relevant for these decays 
is obtained by combining the contributions 
in eq.~(\ref{box}) with those induced 
by $Z$ and photon exchange (see also  Fig.~\ref{f1}). 
Other contributions, induced by Higgs boson exchange, will be 
discussed in Section~\ref{hmt}.
We have:
\beqn{leff}
{\cal L}^{\rm eff}_{\tau\mu ff} & = & 
\sum_f \left[ 
(\ov{\mu} \smub \tau)  
\left( F_L^{f_L} \,\fb \smdb f + F_L^{f_R} \,  f^c \smd  \fcb \right)
+ ( \mu^c \smu \ov{\tau}^c ) 
\left( F_R^{f_L} \,\fb \smdb f + F_R^{f_R} \,  f^c \smd  \fcb \right)
\right] 
\nonumber
\\
& - & 2 e^2 \left( D_L^{\ga} \, \ov{\mu} \smnb  \ov{\tau}^c 
+ D_R^{\ga} \, \mu^c \smn \tau \right) 
 {\mta i \dnd \over \Box}  
\sum_f Q_f (\ov{f} \smdb f +  f^c \smd  \fcb) 
+ {\rm h.c.}
\eea
where 
\beqn{fff}
F_{L(R)}^{f_L} & = & B_{L(R)}^{f_L} 
+ (T^3_{f_L}-Q_f s^2_W) g_Z^2 A^Z_{L(R)} + Q_f e^2 C^{\ga}_{L(R)}
\\
F_{L(R)}^{f_R} & = & B_{L(R)}^{f_R} 
- Q_f s^2_W g_Z^2 A^Z_{L(R)} + Q_f e^2 C^{\ga}_{L(R)} .
\eea
For the sake of brevity, we will often refer to the operators 
in the first line of eq.~(\ref{leff}) as 
`monopole' ones, to be contrasted with the `dipole' ones 
in the second line.  

The effective lagrangian (\ref{leff}) can be applied
either to leptonic transitions, such as the decays 
$\tau^- \to \mu^- e^+ e^-$ and $\tau^- \to \mu^- \mu^+ \mu^- $, 
or to semileptonic transitions involving hadrons,
after taking the appropriate matrix elements of
quark bilinears. 
In the processes involving a neutral 
pseudoscalar meson $P$ ($P=\pi, \eta, \eta'$)
the relevant matrix elements are 
$\langle 0 |J^{\mu a}_5(0)|P(p)\rangle = i f_P^a p^\mu$ ($a = 0,3,8$), 
where $f^a_P$ are the decay constants and 
$J^{\mu a}_5$ are the axial-vector currents\footnote{
In terms of four-component spinors 
$\psi^T_q = (q ~\ov{q}^c)~  (q=u,d,s)$, 
collected in a triplet $\Psi^T= (\psi^T_u ~ \psi^T_d ~ \psi^T_s)$, 
the vector and axial-vector currents read as 
$J^{\mu a} = \ov{\Psi} \ga^\mu  \frac{\la^a}{2}\Psi$ and 
$J^{\mu a}_5 = \ov{\Psi} \ga^\mu \ga^5 \frac{\la^a}{2}\Psi$ 
($a=0,1, \ldots,8$), respectively. 
The neutral currents contain the matrices
$\la^0 = \sqrt{\frac23}{\rm diag}(1,1,1), ~
\la^3 = {\rm diag}(1,-1,0), ~
\la^8 = \frac{1}{\sqrt3}{\rm diag}(1,1,-2)$.}.
Among the neutral vector mesons, we have selected the $\rho$ meson 
and so the relevant matrix element is 
$\langle 0 |J^{\mu 3} (0)|\rho \rangle = 
\kappa_\rho m^2_\rho \epsilon^\mu$, where 
$J^{\mu 3}$ is the third component of the vector current, 
$ \kappa_\rho\simeq 0.2$ and 
$m_\rho,  \epsilon^\mu$ are the $\rho$ mass and polarization vector, 
respectively. 
Our results can be easily extended to the other vector mesons, 
$\phi$ and $\omega$.

\begin{itemize}
\item 
\underline{{ \bf $\tau \to  \mu  e e, ~\tau \to  \mu  \mu \mu $}}
\vspace{0.3cm}

The branching ratios are directly computed from eq.~(\ref{leff}):
\beqn{btmuee}
BR(\tau^- \to \mu^- e^+ e^-) \!\!\!\! & = & \!\! \!\! \frac{1}{8 G_F^2 }
\left[ |F^{e_L}_L|^2 + |F^{e_R}_L|^2 
+ |F^{e_L}_R|^2 + |F^{e_R}_R|^2 
\right.
\nonumber \\
& + & \!\!\!\!\!\!
4 e^2 {\rm Re}\left( D_L^{\ga}(\bar{F}^{e_L}_L + \bar{F}^{e_R}_L)
+  D_R^{\ga}(\bar{F}^{e_L}_R + \bar{F}^{e_R}_R) \right)
\nonumber \\
& + & \left.\!\!\!\!\!\! 
8 e^4 ( |D_L^{\ga}|^2 +|D_R^{\ga}|^2 )
\left(\log{\mta^2 \over m_e^2} - 3 \right) \right]
BR( \tau^- \to \mu^- \bar{\nu}_{\mu} \nu_{\tau}) , 
\eea
\beqn{bt3mu}
BR(\tau^- \to \mu^- \mu^+ \mu^-)\!\! \!\! & = & \!\!\!\! \frac{1}{8 G_F^2 }
\left[ 2 |F^{\mu_L}_L|^2 + |F^{\mu_R}_L|^2 
+  |F^{\mu_L}_R|^2 + 2 |F^{\mu_R}_R|^2 
\right.
\nonumber \\
& + &  \left.\!\!\!\!\!\!
4 e^2 {\rm Re}\left(D_L^{\ga}(2 \bar{F}^{\mu_L}_L + \bar{F}^{\mu_R}_L)
+  D_R^{\ga}(\bar{F}^{\mu_L}_R + 2 \bar{F}^{\mu_R}_R) \right) \right.
\nonumber \\
& + & \left.  \!\!\!\!\!\!
8 e^4 ( |D_L^{\ga}|^2 +|D_R^{\ga}|^2 )
\left(\log{\mta^2 \over m^2_{\mu}} - {11 \over 4} \right) \right]
BR( \tau^- \to \mu^- \bar{\nu}_{\mu} \nu_{\tau}) .
\eea
The electron mass $m_e$ and the muon mass $m_\mu$ 
regularize an infrared singularity 
in phase space in $\tau\to\mu ee$ and $\tau\to\mu \mu \mu$, respectively.

\item 
\underline{{ \bf $\tau \to  \mu  \rho $}}
\vspace{0.3cm}

The effective lagrangian for $\tau \to \mu \rho$ is also obtained 
from eq.~(\ref{leff}): 
\be{lmuro}
{\cal L}^{\rm eff}_{\tau\mu\rho} =  
\kappa_{\rho}\left[ m_{\rho}^2
\left( F_L^{\rho}\ov{\mu} \smub \tau 
+ F_R^{\rho} \mu^c \smu \ov{\tau}^c \right)  
\rho_{\mu}^0
+  2 e^2 \mta 
\left( D_L^{\ga}  \ov{\mu} \smnb  \ov{\tau}^c 
+ D_R^{\ga} \mu^c  \smn \tau  \right) i \dnd \rho_{\mu}^0
\right] + {\rm h.c.}
\ee
where 
\beqn{fro}
F_{L(R)}^{\rho} & = & 
{ 1\over 2}(F_{L(R)}^{u_L}-F_{L(R)}^{d_L}) 
+ { 1\over 2}(F_{L(R)}^{u_R}-F_{L(R)}^{d_R})
\nonumber\\
& = & 
{ 1\over 2}\left[(B_{L(R)}^{u_L}-B_{L(R)}^{d_L}) 
+ (B_{L(R)}^{u_R}-B_{L(R)}^{d_R})
+ (1 - 2 s_W^2) g_Z^2 A_{L(R)}^Z
+ 2 e^2 C_{L(R)}^{\ga}\right] . \nonumber \\
& & 
\eea
The related branching ratio is:
\beqn{btmuro}
BR(\tau^- \to \mu^- \rho^0) &  = & \frac{1}{4 G_F^2 \ c^2_c}
\left[ |F_L^{\rho}|^2 + |F_R^{\rho}|^2 -
\frac{6 e^2}{1+2 x_{\rho}}
{\rm Re}\left( D_L^{\ga} \bar{F}_L^{\rho} 
+ D_R^{\ga} \bar{F}_R^{\rho} \right)
\right.
\nonumber
\\
& + & \left. 
\frac{e^4 (2 + x_{\rho})}{x_{\rho}(1+2 x_{\rho})}
( |D_L^{\ga}|^2 +|D_R^{\ga}|^2 ) \right] \,
BR(\tau^- \to \nu_{\tau} \rho^-)
\eea
where $c^2_c\equiv \cos^2 \theta_c \simeq 0.95$, 
$ x_{\rho}\equiv m_{\rho}^2/\mta^2 \simeq 0.19$ and
$BR(\tau^- \to \nu_{\tau} \rho^-) \simeq 25 \% $.

\item 
\underline{{ \bf $\tau \to  \mu  \pi $}}
\vspace{0.3cm}

From eq.~(\ref{leff}) we also derive the effective lagrangian 
for $\tau \to \mu \pi$. It does not receive 
contributions from photon exchange, and reads as:
\be{lmupi}
{\cal L}^{\rm eff}_{\tau\mu\pi} =  
f_{\pi} 
\left( F_L^{\pi} \ov{\mu} \smub \tau 
+ F_R^{\pi} \mu^c  \smu \ov{\tau}^c \right)  
\dmd \pi^0 + {\rm h.c.} 
\ee
where $f_{\pi}\equiv f^3_\pi \simeq 92~{\rm MeV}$ 
(we neglect $f^0_\pi,  f^8_\pi$) and  
\beqn{fpi}
F_{L(R)}^{\pi} & = & 
{ 1\over 2}(F_{L(R)}^{u_L}-F_{L(R)}^{d_L}) 
- { 1\over 2}(F_{L(R)}^{u_R}-F_{L(R)}^{d_R})
\nonumber\\
& = & 
{ 1\over 2}\left[(B_{L(R)}^{u_L}-B_{L(R)}^{d_L}) 
- (B_{L(R)}^{u_R}-B_{L(R)}^{d_R})
+ g_Z^2 A_{L(R)}^Z \right].
\eea
The branching ratio is:
\be{rate2}
BR(\tau^- \to \mu^- \pi^0) = \frac{1}{4 G_F^2  c^2_c}
\left[|F_L^{\pi}|^2 + |F_R^{\pi}|^2 \right] \,
BR(\tau^- \to \nu_{\tau} \pi^-)
\ee
where $BR(\tau^- \to \nu_{\tau} \pi^-) \simeq 11 \%$.

\item 
\underline{{ \bf $\tau \to  \mu  \eta, ~\tau \to  \mu  \eta'$}}
\vspace{0.3cm}

Each of the two  mesons $\eta$ and $\eta'$ has both octet and singlet 
$SU(3)_{\rm flav}$ components. 
The relevant decay constants are 
$f_P^a$, where 
$P=\eta, \eta'$ and $a=8,0$ (we neglect $f^3_P$)
\cite{chiral, TF}. For the ratios 
$f_P^a/f_\pi$  
we take the following numerical values \cite{TF}:
$f^8_\eta \sim 1.2 f_\pi, ~
f^0_\eta \sim 0.2 f_\pi, ~
f^8_{\eta'} \sim - 0.45 f_\pi, ~
f^0_{\eta'} \sim  1.15 f_\pi$.

The effective lagrangian 
for $\tau \to \mu \eta$, which does not receive 
contributions from photon exchange, is derived from eq.~(\ref{leff}).
It  reads as:
\be{lmueta}
{\cal L}^{\rm eff}_{\tau\mu \eta} =
\left[  
\left(f^8_{\eta} F^{\eta,8}_L +f^0_{\eta} F^{\eta,0}_L \right)  
 \ov{\mu} \smub \tau 
+\left(f^8_{\eta} F^{\eta,8}_R +f^0_{\eta} F^{\eta,0}_R \right) 
 \mu^c  \smu \ov{\tau}^c \right]  
\dmd \eta + {\rm h.c.}
\ee
where 
\beqn{feta}
F_{L(R)}^{\eta, 8} & = & \!\! 
{ 1\over 2 \sqrt{3}}(F_{L(R)}^{u_L}+F_{L(R)}^{d_L} -2F_{L(R)}^{s_L}) 
- { 1\over 2\sqrt{3}}(F_{L(R)}^{u_R} + F_{L(R)}^{d_R} -2 F_{L(R)}^{s_R})
\nonumber\\
& = & 
\!\!{ 1\over 2\sqrt{3}} 
\left[ (B_{L(R)}^{u_L}+B_{L(R)}^{d_L} -2B_{L(R)}^{s_L}) 
- (B_{L(R)}^{u_R}+ B_{L(R)}^{d_R} -2B_{L(R)}^{s_R})
+ g_Z^2 A_{L(R)}^Z \right] , \\ 
F_{L(R)}^{\eta, 0} & = & \!\! 
{ 1\over  \sqrt{6}}(F_{L(R)}^{u_L}+F_{L(R)}^{d_L} + F_{L(R)}^{s_L}) 
- { 1\over \sqrt{6}}(F_{L(R)}^{u_R} + F_{L(R)}^{d_R} + F_{L(R)}^{s_R})
\nonumber\\
& = & 
\!\!{ 1\over \sqrt{6}} 
\left[ (B_{L(R)}^{u_L}+B_{L(R)}^{d_L}  + B_{L(R)}^{s_L}) 
- (B_{L(R)}^{u_R}+ B_{L(R)}^{d_R} + B_{L(R)}^{s_R})
-\frac12 g_Z^2 A_{L(R)}^Z \right] . \label{feta0}
\eea
The branching ratio can be expressed as:
\be{rate3}
BR(\tau^- \to \mu^- \eta) = \frac{1}{4 G_F^2  c^2_c}
\left[ \left| \frac{f^{8}_\eta}{f_\pi} F_L^{\eta,8} +
\frac{f^{0}_\eta}{f_\pi} 
F_L^{\eta,0} \right|^2 + (L\to R) \right]
(1 - x_\eta)^2
\,
BR(\tau^- \to \nu_{\tau} \pi^-)
\ee
where $x_\eta \equiv m^2_\eta/m^2_\tau \simeq 9.5\times 10^{-2}$ 
and the ratio $m^2_\pi/m^2_\tau$ has been neglected.
For the decay $\tau \to \mu \eta'$, one has to make the 
replacement $\eta \to \eta'$ in eqs.~(\ref{lmueta}), 
(\ref{feta}), (\ref{feta0}) and (\ref{rate3}).
We anticipate that the equalities $F_{L(R)}^{\eta,8}
=F_{L(R)}^{\eta',8}$, $F_{L(R)}^{\eta,0}=F_{L(R)}^{\eta',0}$,
which follow from eq.~(\ref{leff}), will not hold in the
presence of sizeable Higgs-mediated contributions 
[see Section~\ref{hmt}]. 
 
\end{itemize}

\subsection{Correlations}
In the previous subsection we have seen that, on the one hand, 
each operator can  contribute  
to different processes, hence correlations exist.
On the other hand, the correlation pattern is not trivial  
because  each process can get contributions from different 
operators,  so making general predictions  is not straightforward.
Nevertheless, it is interesting to 
see what happens when several processes are dominated by the same 
operator. 
Such a situation may or may not  be realized  in a specific underlying 
model or in a specific portion of the model parameter space.

\begin{description}
\item {\it $D^\ga$-dominance.} 
If photon dipole contributions are the dominant ones in
the decays $\tau \to \mu e e, ~ 
\tau \to \mu \mu \mu$ and $\tau \to \mu \rho$, 
then such processes can
directly be compared to $\tau \to \mu \ga$:
\bea \label{dg1}
\frac{ BR(\tau^- \to \mu^- e^+ e^-)}{BR(\tau^- \to \mu^- \ga)}
\!& \simeq &\! 
\frac{\alpha}{3 \pi} \,  
\left(\log{\mta^2 \over m_e^2} - 3 \right)  
\simeq   10^{-2}   
\\
& & \nonumber \\
\frac{ BR(\tau^- \to \mu^- \mu^+ \mu^-)}{BR(\tau^- \to \mu^- \ga)}
\!& \simeq & \!
\frac{\alpha}{3 \pi} \,  
\left(\log{\mta^2 \over m_{\mu}^2} - {11 \over 4} \right)  
\simeq   2.2 \times 10^{-3} \label{dg2}  
\\
& & \nonumber \\
\frac{ BR(\tau^- \to \mu^- \rho^0)}{BR(\tau^- \to \mu^- \ga)}
\!& \simeq &\! 
\frac{\alpha}{12 \pi c^2_c} \,
\frac{(2 + x_{\rho})}{x_{\rho}(1+2 x_{\rho})} \,
\frac{ BR(\tau^- \to \nu_{\tau} \rho^-)}
{BR( \tau^- \to \mu^- \bar{\nu}_{\mu} \nu_{\tau})}
\simeq  2.5 \times 10^{-3} . \label{dg3} 
\eea 
In particular, the present bound (\ref{e-rad}) on $\tau\to\mu\ga$ would 
imply the following bounds:
\bea
\label{dom-dge}
BR(\tau^- \to \mu^- e^+ e^-)& \lsim & 3\times 10^{-9} , \\
\label{dom-dgm}
 BR(\tau^- \to \mu^- \mu^+ \mu^-)&\lsim& 7 \times 10^{-10} , \\
\label{dom-dgr}
 BR(\tau^- \to \mu^- \rho^0) & \lsim & 8 \times 10^{-10} .
\eea

\item {\it $C^\ga$-dominance.}
Suppose now that 
$D^\ga$-contributions are suppressed and that monopole contributions to 
 $\tau \to \mu e e, ~\tau \to \mu \mu \mu,~  
\tau \to \mu \rho$ are dominated by $C^\ga$-operators. Then:
\be{dom-c}
{ BR(\tau^- \to \mu^- \rho^0)} \simeq 
 { BR(\tau^- \to \mu^- \mu^+ \mu^-)}
\simeq 1.5\times BR(\tau^- \to \mu^-  e^+ e^-) .
\ee

\item {\it $A^Z$-dominance.}
Suppose again that dipole 
contributions are suppressed and that monopole contributions 
to $\tau \to \mu e e, ~ \tau \to \mu \mu \mu, ~ 
\tau \to \mu \rho, ~\tau \to \mu \pi, ~ 
\tau \to \mu \eta, ~\tau \to \mu \eta'$ 
and $Z\to  \mu\tau$ are dominated by $A^Z$-operators. Then:
\beqn{dom-a}
 BR(Z\to \mu^+\tau^-)& \simeq & 3 \times BR(\tau^- \to \mu^- e^+ e^- ) , \\
BR(\tau^- \to \mu^- \rho^0 ) & \simeq & 1.8 \times
 BR(\tau^- \to \mu^- e^+ e^- ) , \\
 BR(\tau^- \to \mu^- \pi^0)& \simeq & 2.7 \times
 BR(\tau^- \to \mu^- e^+ e^- ) ,\\
 BR(\tau^- \to \mu^- \eta)& \simeq & 0.8 \times
 BR(\tau^- \to \mu^- e^+ e^- ) ,\\
BR(\tau^- \to \mu^- \eta')& \simeq & 0.7 \times
 BR(\tau^- \to \mu^- e^+ e^- ) ,\\
  BR(\tau^- \to \mu^- \mu^+ \mu^-)& \simeq&
\pmatrix{1.6 \cr 1.4 } \times BR(\tau^- \to \mu^- e^+ e^- ) .
\eea
The result for   $\tau \to \mu \mu \mu$  depends on the relative 
amount of  $A^Z_L$ and $A^Z_R$ contributions: 
the upper (lower) estimate   refers to 
the case of pure $A^Z_L$ ($A^Z_R$)  dominance,  with $A^Z_R=0$ ($A^Z_L=0$).  

\item {\it $B^f$-dominance.}
Finally,  consider the case in which all contributions are suppressed 
except for those  $B^f$-induced. 
Then correlations are in general weakened or lost, since 
different $B^f $ coefficients  appear in different processes. 
In some specific case, however, simple correlations may emerge. 
For example, consider the  limit of flavour universality 
in the down-quark sector, {\it i.e.}   
$B_{L(R)}^{d_L}= B_{L(R)}^{s_L}$ and $B_{L(R)}^{d_R}= B_{L(R)}^{s_R}$ 
(which holds in the MSSM with down-squark mass universality).  
Then we obtain $BR(\tau \to \mu\pi) \sim 3 \times
BR(\tau \to \mu \eta)$ by  
neglecting $f^0_\eta$ in eq.~(\ref{rate3}).
Another example of correlation is that  between 
$\tau \to \mu \rho$ and $\tau \to \mu \pi$ in the absence of  
LFV in the left sector ($B^{u_{L(R)}}_L= B^{d_{L(R)}}_L =0$). 
In this case,   
in the $SU(2)_W$-symmetric limit ($B^{u_L}_R = B^{d_L}_R$), 
both  decays only depend on the 
combination $B^{u_R}_R - B^{d_R}_R$,  hence 
$BR(\tau \to \mu \rho) \sim 2.3 \times BR(\tau \to \mu \pi)$.

\end{description}

\section{The MSSM framework \label{susy}}
In the above sections we have given a model-independent 
description of effective operators that contribute to 
our selected set of LFV processes. Here our purpose is to compute the 
coefficients of such operators in the framework of the MSSM. 
The source of LFV in the MSSM is the potential misalignment 
between the lepton and slepton mass matrices.
More precisely, in the superfield basis in which the 
charged lepton mass matrix is diagonal, the sources of LFV are the 
off-diagonal entries of the soft-breaking matrices 
$\tilde{{\cal M}}^2_L, \tilde{{\cal M}}^2_R$ and $\tilde{{\cal A}}$. 
If we restrict our attention to the second/third generation, 
the latter matrices have the form (see also Appendix~\ref{a1}):
\be{lfvm}
\tilde{{\cal M}}^2_L =
\pmatrix{
\tilde{m}^2_{L \mu \mu} & \tilde{m}^2_{L \mu \tau} \cr
 & \cr
\tilde{m}^{2*}_{L \mu \tau} & \tilde{m}^2_{L \tau \tau} }, \, \,~~~~
\tilde{{\cal M}}^2_R = \pmatrix{
\tilde{m}^2_{R \mu \mu} & \tilde{m}^2_{R \mu \tau} \cr
 & \cr
\tilde{m}^{2*}_{R \mu \tau} & \tilde{m}^2_{R \tau \tau} } ,  \, \,~~~~
\tilde{{\cal A}} = \pmatrix{
h_\mu A_\mu  & h_\tau A^R_{\mu \tau}\cr
 & \cr
 h_\tau A^L_{\mu \tau} & h_\tau A_\tau } .
\ee
Notice that we have parametrized  the entries of 
$\tilde{{\cal A}}$ by extracting 
a suitable  Yukawa coupling (either  $h_\mu$ or $ h_\tau$).
We will use the symbols $({\rm LFV})_L$ and  
$({\rm LFV})_R$ to characterize LFV in the left and right sectors, 
respectively:
\bea
&& ({\rm LFV})_L: ~~ \tilde{m}^2_{L \mu \tau} \neq 0 ~~~{\rm and/or}~~~ 
A^L_{\mu \tau} \neq 0 \label{fvl} ,\\
&& ({\rm LFV})_R: ~~ \tilde{m}^2_{R \mu \tau} \neq 0 ~~~ {\rm and/or} ~~~
A^R_{\mu \tau} \neq 0 \label{fvr} . 
\eea
In the case of pure (LFV)$_L$ [(LFV)$_R$] the parameter 
$A^R_{\mu \tau}$ ($A^L_{\mu \tau}$) is expected to be suppressed 
by $h_\mu/h_\tau$ with respect to  
$A^L_{\mu \tau}$ ($A^R_{\mu \tau}$).

In Section~\ref{gao} we  describe some general features of 
the MSSM contributions to 
the operator coefficients listed in Section~\ref{plfv}.
We also present the Higgs-$\mu$-$\tau$ effective operators and discuss 
their impact on $\tau \to 3 \mu,~\tau \to \mu \pi,~ 
\tau \to \mu \eta, ~\tau \to \mu \eta'$  (Section~\ref{hmt}). 
Finally, we discuss the effect of LFV on 
the muon anomalous magnetic moment (Section~\ref{mamm}).
The explicit results of the calculations for the operator coefficients 
are presented in Appendices~A.2-A.10.

\subsection{General aspects of operator contributions \label{gao}}
The leading contributions to 
the coefficients $A^Z, C^Z, C^\ga, D^Z, D^\ga, 
B^f$ arise from one-loop diagrams that 
involve the exchange of gauginos, 
Higgsinos and sleptons (and also $\tilde{f}$ sfermions in the 
case of $B^f$).
These diagrams are shown schematically in Fig.~2 and 
in more detail in Appendices~A.3-A.8, 
where the explicit analytical results 
for the coefficients are also presented.
The relevant parameters involved are those in eq.~(\ref{lfvm}) (or equivalent 
ones, see below), the $SU(2)_W\times U(1)_Y$ gaugino masses 
($M_2, M_1$), the $\mu$ parameter\footnote
{We take a superpotential term of the form $W= \mu H_1 H_2= \mu (
H^0_1 H^0_2 - H^-_1 H^+_2)$, so our sign convention 
for the $\mu$ parameter is opposite to the one commonly used (for this and 
other conventions see also 
Appendix~A.1).},                     
other sfermion masses and $\tanb$. We neglect contributions 
to the LFV operator coefficients coming from other sources\footnote{
Additional contributions arise, for instance, from the interactions 
of the MSSM fields with the goldstino supermultiplet \cite{BR1}. 
Such effects are suppressed, unless the supersymmetry 
breaking scale $\sqrt{F}$ is close to the electroweak scale.}.
 
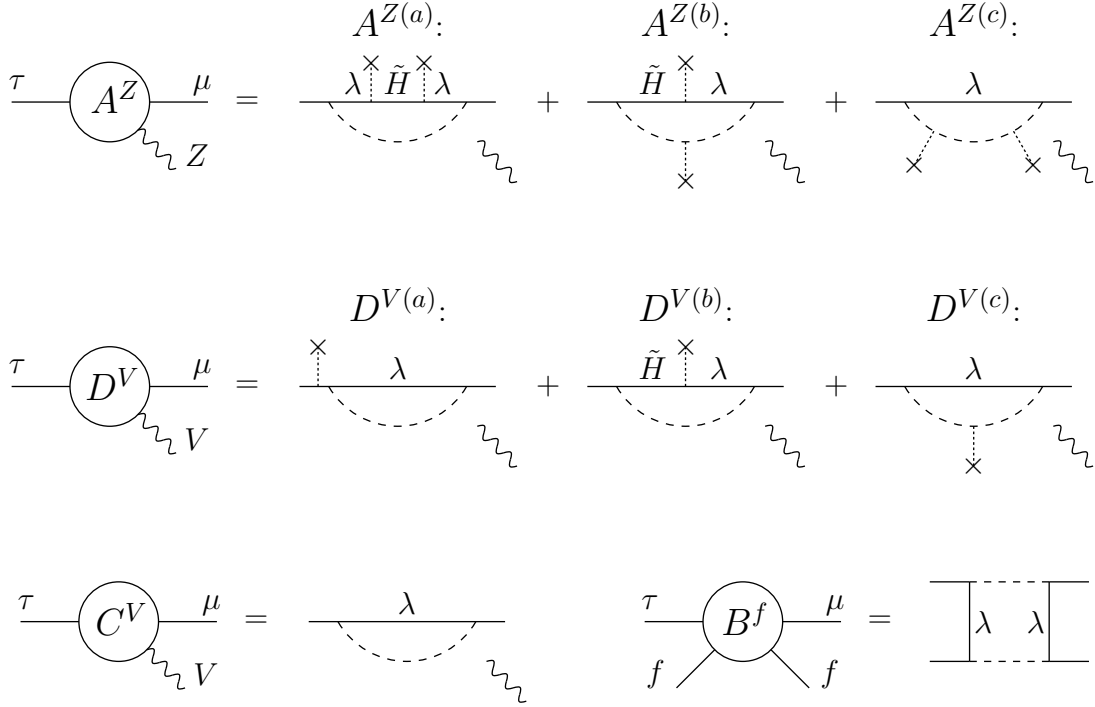
\begin{figure}[htb]
\begin{center}
\begin{picture}(105,75)(-48,-35)
\Line(-37,0)(-15,0)
\Line(15,0)(37,0)
\CArc(0,0)(15,0,360)
\Photon(10.6,-10.6)(25,-25){2}{3}
\Text(2,2)[]{{\large $A^Z$}}
\Text(-35,5)[b]{$\tau$}
\Text(35,5)[b]{$\mu$}
\Text(33,-20)[]{$Z$}
\Text(52,0)[]{$=$}
\end{picture}
\begin{picture}(105,75)(-48,-35)
\Line(-37,0)(37,0)
\DashCArc(0,15)(30,210,330){3}
\DashLine(-10,0)(-10,15){1}
\DashLine(10,0)(10,15){1}
\Photon(30,-15)(45,-30){2}{3}
\Text(-10,15)[]{$\times$}
\Text(10,15)[]{$\times$}
\Text(-17,3)[b]{$\lambda$}
\Text(17,3)[b]{$\lambda$}
\Text(0,3)[b]{$\tilde{H}$}
\Text(0,25)[b]{{\large $A^{Z(a)}$}:}
\end{picture}
\begin{picture}(105,75)(-48,-35)
\Line(-37,0)(37,0)
\DashCArc(0,15)(30,210,330){3}
\DashLine(0,0)(0,15){1}
\DashLine(0,-15)(0,-30){1}
\Photon(30,-15)(45,-30){2}{3}
\Text(0,15)[]{$\times$}
\Text(0,-30)[]{$\times$}
\Text(-12.5,3)[b]{$\tilde{H}$}
\Text(12.5,3)[b]{$\lambda$}
\Text(0,25)[b]{{\large $A^{Z(b)}$}:}
\Text(-52,0)[]{$+$}
\end{picture}
\begin{picture}(105,75)(-48,-35)
\Line(-37,0)(37,0)
\DashCArc(0,15)(30,210,330){3}
\DashLine(-15,-11)(-22.5,-24){1}
\DashLine(15,-11)(22.5,-24){1}
\Photon(30,-15)(45,-30){2}{3}
\Text(-22.5,-24)[]{$\times$}
\Text(22.5,-24)[]{$\times$}
\Text(0,3)[b]{$\lambda$}
\Text(0,25)[b]{{\large $A^{Z(c)}$}:}
\Text(-52,0)[]{$+$}
\end{picture}
\end{center}
\vspace{0.2 cm}
\begin{center}
\begin{picture}(105,75)(-48,-35)
\Line(-37,0)(-15,0)
\Line(15,0)(37,0)
\CArc(0,0)(15,0,360)
\Photon(10.6,-10.6)(25,-25){2}{3}
\Text(1,1)[]{{\large $D^V$}}
\Text(-35,5)[b]{$\tau$}
\Text(35,5)[b]{$\mu$}
\Text(33,-20)[]{$V$}
\Text(52,0)[]{$=$}
\end{picture}
\begin{picture}(105,75)(-48,-35)
\Line(-37,0)(37,0)
\DashCArc(0,15)(30,210,330){3}
\DashLine(-30,0)(-30,15){1}
\Photon(30,-15)(45,-30){2}{3}
\Text(-30,15)[]{$\times$}
\Text(0,3)[b]{$\lambda$}
\Text(0,25)[b]{{\large $D^{V(a)}$}:}
\end{picture}
\begin{picture}(105,75)(-48,-35)
\Line(-37,0)(37,0)
\DashCArc(0,15)(30,210,330){3}
\DashLine(0,0)(0,15){1}
\Photon(30,-15)(45,-30){2}{3}
\Text(0,15)[]{$\times$}
\Text(-12.5,3)[b]{$\tilde{H}$}
\Text(12.5,3)[b]{$\lambda$}
\Text(0,25)[b]{{\large $D^{V(b)}$}:}
\Text(-52,0)[]{$+$}
\end{picture}
\begin{picture}(105,75)(-48,-35)
\Line(-37,0)(37,0)
\DashCArc(0,15)(30,210,330){3}
\DashLine(0,-15)(0,-30){1}
\Photon(30,-15)(45,-30){2}{3}
\Text(0,-30)[]{$\times$}
\Text(0,3)[b]{$\lambda$}
\Text(0,25)[b]{{\large $D^{V(c)}$}:}
\Text(-52,0)[]{$+$}
\end{picture}
\end{center}

\begin{center}
\begin{picture}(105,75)(-48,-35)
\Line(-37,0)(-15,0)
\Line(15,0)(37,0)
\CArc(0,0)(15,0,360)
\Photon(10.6,-10.6)(25,-25){2}{3}
\Text(1,1)[]{{\large $C^V$}}
\Text(-35,5)[b]{$\tau$}
\Text(35,5)[b]{$\mu$}
\Text(33,-20)[]{$V$}
\Text(52,0)[]{$=$}
\end{picture}
\begin{picture}(105,75)(-48,-35)
\Line(-37,0)(37,0)
\DashCArc(0,15)(30,210,330){3}
\Photon(30,-15)(45,-30){2}{3}
\Text(0,3)[b]{$\lambda$}
\end{picture}
\hspace{0.5 cm}
\begin{picture}(105,75)(-48,-35)
\Line(-37,0)(-15,0)
\Line(15,0)(37,0)
\CArc(0,0)(15,0,360)
\Line(-10.6,-10.6)(-25,-25)
\Line(10.6,-10.6)(25,-25)
\Text(1,1)[]{{\large $B^f$}}
\Text(-35,5)[b]{$\tau$}
\Text(35,5)[b]{$\mu$}
\Text(-33,-20)[]{$f$}
\Text(33,-20)[]{$f$}
\Text(52,0)[]{$=$}
\end{picture}
\begin{picture}(80,75)(-40,-35)
\Line(-30,15)(-15,15)
\Line(15,15)(30,15)
\Line(-30,-15)(-15,-15)
\Line(15,-15)(30,-15)
\DashLine(-15,15)(15,15){3}
\DashLine(-15,-15)(15,-15){3}
\Line(-15,15)(-15,-15)
\Line(15,15)(15,-15)
\Text(-10,0)[]{$\lambda$}
\Text(10,0)[]{$\lambda$}
\end{picture}
\end{center}
\caption{\ftsz Topologies of the MSSM diagrams contributing 
to the coefficients $A^Z$, $D^V$,  $C^V$ ($V = \ga, Z$) 
and  $B^f$. 
Internal solid lines denote gauginos ($\lambda$) 
or higgsinos ($\tilde{H}$).
Dashed lines denote sleptons (or other sfermions
in the box diagrams).  Dotted lines with a cross denote   
Higgs insertions. 
}
\label{f2}
\end{figure}

In the diagrams in Fig.~\ref{f2}, the flavour transition occurs 
along the slepton line. 
The generic behaviour of the related operator 
coefficients  at the linear 
order in the LFV parameters $\mt^2_{\mu\tau}, A_{\mu\tau}$ 
(the so-called mass-insertion approximation \cite{HKR})    
goes as follows:
\beqn{mia}
A^Z, C^\ga, C^Z, {B^f}/{{g}^2_w}& \sim &\frac{g^2_w}{16\pi^2} \cdot
\frac{1}{M^2_{S}}\cdot
\left(\frac{\mt^2_{\mu\tau}}{M^2_{S}}\right) , \\
 D^\ga, D^Z & \sim & \frac{g^2_w}{16\pi^2} \cdot \frac{1}{M^2_{S}}\cdot
\left( \frac{\mt^2_{\mu\tau}}{M^2_{S}}~ {\it or} ~
\frac{\mt^2_{\mu\tau}}{M^2_{S}} \tanb ~{\it or}~ 
\frac{A_{\mu\tau}}{M_{S}} \right) , 
\eea
where the constant $g_w$ can be either $g$ or $g'$ and $M_S$ is 
some effective sparticle mass. 
These relations show that generically all the coefficients have the 
same order of magnitude (except for a possible $\tanb$-enhancement 
in the dipole coefficients),  are suppressed by $M^2_S$ and are 
proportional to the relative amount of LFV.
Nevertheless, detailed computations are, of course, necessary 
to single out  the specific dependence of each coefficient 
on the MSSM parameters.
Upon addressing   such a computation one is 
concerned, in particular, with the treatment of LFV.
Since LFV in the second/third generation may  not be a small effect, 
we choose to go beyond the mass-insertion approximation. 
Specifically, we   diagonalize  
the mass matrices $\tilde{{\cal M}}^2_L, \tilde{{\cal M}}^2_R$ 
and work with their mass eigenstates, {\it i.e.} our results 
are expressed in terms of  the  corresponding 
eigenvalues $\mt^2_{L_\al}, \mt^2_{R_\al}$ ($\al =2,3$)  and 
mixing angles $\theta_L, \theta_R$. 
In particular, large $({\rm LFV})_L$ in $\tilde{{\cal M}}^2_L$ means 
$\tilde{m}^2_{L \mu \tau}\sim \tilde{m}^2_{L \mu \mu}\sim  
\tilde{m}^2_{L \tau\tau}$, which in turn implies 
$\theta_L \sim {\cal O}(1)$ and 
$(\mt^2_{L_2}-\mt^2_{L_3})/(\mt^2_{L_2}+\mt^2_{L_3})
 \sim {\cal O}(1)$ (with $\mt^2_{L_2}, \mt^2_{L_3}$ of the same order or
hierarchical). 
Analogous relations hold for the case of large  $({\rm LFV})_R$.
Another aspect we have to deal with is 
the inclusion of electroweak breaking effects.
We choose to treat such effects  at lowest order,
{\it i.e.} the only Higgs insertions we consider are those  explicitly 
depicted in the diagrams in Fig.~2 (and also in Appendix), 
which also corresponds to take only the $d=6$ operators in the 
$SU(2)_W\times U(1)_Y$-unbroken phase.
We believe that working in this approximation 
is more transparent  and appealing from the theoretical point of 
view, although this may imply some loss of numerical accuracy 
when $M^2_S$ is very close to $m^2_Z$. 

A few additional comments are in order about  some operator coefficients. 
In particular, we recall that  $A^Z$ 
is associated to operators like those in 
eq.~(\ref{invs}), where the $Z_\mu$ field is contained in a  
covariant derivative acting on a Higgs field ($D_\mu H$). 
Thus the operator coefficient can also be extracted 
from the part of the operator containing an ordinary derivative 
($\dmd H$), that is from diagrams with momentum dependent Higgs lines and 
no $Z$ lines.
This method, which we have used, has the advantage that one  
need not compute LFV diagrams 
corresponding to wave-function renormalization  
 of  lepton fields, which are  associated to  operators like 
$(\ov{L}_{\mu} \ov{\sigma}^\nu D_\nu L_\tau)$ 
with or without Higgs fields attached.
Also notice that $A^Z$ receives contributions of three different types:
\be{azsum}
A^Z_{L,R} = A^{Z(a)}_{L,R}  +A^{Z(b)}_{L,R}  +A^{Z(c)}_{L,R} .
\ee
The terms $A^{Z(b)}_{L,R}, A^{Z(c)}_{L,R}$ are parametrically suppressed 
by $h^2_\tau/g^2_w$ with respect to $A^{Z(a)}_{L,R}$, hence those 
contributions can only be relevant  for large $\tanb$.  
Also the dipole coefficients $D^V$ ($V=\ga, Z$)  receive contributions 
of three different types:
\be{dsum}
D^V_{L,R} = D^{V(a)}_{L,R}  +D^{V(b)}_{L,R}  +D^{V(c)}_{L,R} .
\ee 
Notice that the diagrams contributing to $D^{V(a)}_{L,R}$ 
can only contain the Higgs field $H_1$, which  comes from the  $\tau$ 
equation of motion, whereas either  $H_1$ or $H_2$ can appear 
in the diagrams contributing to $D^{V(b)}_{L,R}, D^{V(c)}_{L,R}$.
As a consequence only the latter coefficients  can receive 
a $\tanb$-enhancement. 
In the case of $D^Z$ it is sufficient to compute 
these $\tanb$-enhanced terms, as they are 
the only ones which can give contributions comparable to 
the monopole ones in the $Z\to \mu \tau$ decay [see eq.~(\ref{bz})]. 
Notice that $\tanb$-enhanced terms have a more dramatic effect in 
the case of $D^\ga$ \cite{HMTY}, since they generically make the dipole 
operators dominate over the monopole ones in decays 
such as $\tau \to \mu ee, \tau \to \mu \mu \mu$ and 
$\tau \to \mu \rho$.

\subsection{Higgs-muon-tau effective interactions \label{hmt}}
At large $\tanb$, another class of LFV interactions 
is relevant, namely those between 
a $\mu$-$\tau$ pair and Higgs bosons. 
For our purposes  it is sufficient to focus on the leading
effects, which arise from these dimension-four operators:
\be{leff-h}
-h_\tau \Delta^*_{L} {H}^{0*}_2 \tau^c \mu
-h_\tau \Delta_{R} {H}^{0*}_2 \mu^c \tau 
 +{\rm h.c.}  . 
\ee
The corresponding diagrams are sketched in Fig.~\ref{fh1} and
\begin{figure}[htb]
\begin{center}
\begin{picture}(105,75)(-48,-35)
\Line(-37,0)(37,0)
\DashCArc(0,15)(30,210,330){3}
\DashLine(0,0)(0,15){1}
\Text(0,18)[b]{$H_2^0$}
\Text(-12.5,3)[b]{$\tilde{H}$}
\Text(12.5,3)[b]{$\lambda$}
\Text(-60,5)[]{{\large $\Delta^{(b)}$}:}
\end{picture}
\hspace{2 cm}
\begin{picture}(105,75)(-48,-35)
\Line(-37,0)(37,0)
\DashCArc(0,15)(30,210,330){3}
\DashLine(0,-15)(0,-30){1}
\Text(3,-30)[bl]{$H_2^0$}
\Text(0,3)[b]{$\lambda$}
\Text(-60,5)[]{{\large $\Delta^{(c)}$}:}
\end{picture}
\end{center}
\caption{\ftsz Topologies of the MSSM diagrams contributing to 
the  $\Delta$ coefficients. Dashed lines denote sleptons.}
\label{fh1}
\end{figure}
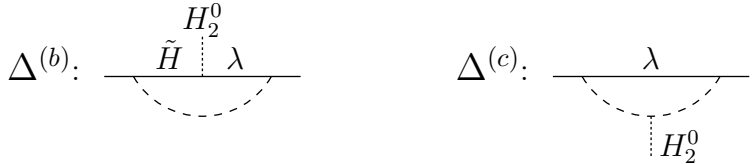
shown in more detail in Appendix~A.9, where the explicit expressions 
of the dimensionless coefficients 
$\Delta_{L,R}=\Delta_{L,R}^{(b)}+\Delta_{L,R}^{(c)}$ are given\footnote{
We should add that most of the properties discussed in this section 
do not rely on the specific MSSM origin of $\Delta_{L,R}$.
The discussion can also apply, more generally, 
to two-Higgs-doublet models in which lepton masses are
mainly generated through Yukawa couplings to $H_1$,
with smaller contributions from $H_2$.}.  
In the mass-eigenstate basis for both leptons and Higgs bosons, 
the LFV couplings read as:
\be{LFV-h}
{\cal L}^{\rm eff}_{{\rm Higgs}~\mu \tau} = - \frac{h_\tau}{\sqrt2 
{\rm c}_\beta} 
( \Delta^*_L ~\tau^c \mu +\Delta_R~ \mu^c \tau )~ 
[  {\rm c}_{\beta -\alpha} h
- {\rm s}_{\beta -\alpha} H  - {\rm i} A] 
+ {\rm h.c.}  , 
\ee
where 
$A$ is the physical CP-odd Higgs field, 
$\alpha$ is the  mixing angle in the CP-even Higgs sector 
[$\sqrt2 {\rm Re}(H^0_1 -  \langle H^0_1\rangle) = 
  {\rm c}_\al H -   {\rm s}_\al h$, 
$ ~ \sqrt2 {\rm Re}(H^0_2 -  \langle H^0_2\rangle) 
=   {\rm s}_\al H +   {\rm c}_\al h$],  and a short-hand notation is 
used for ${\rm c}_\xi = \cos \xi, {\rm s}_\xi = \sin \xi$ 
($\xi= \al, \beta, \beta-\al$).
The effective couplings (\ref{LFV-h}) directly 
contribute to LFV 
decays of the neutral Higgs bosons, 
$A,H, h\to \mu\tau $ \cite{BR}, and also induce Higgs mediated 
contributions to several LFV $\tau$ decays, 
such as  $\tau\to 3 \mu$ \cite{BK,BR}, $\tau\to \mu \eta$ 
\cite{sher1}, $\tau\to \mu \eta'$ and  $\tau\to \mu \pi$. 
The $\Delta$-contributions to all such processes 
are displayed in Fig.~\ref{fh2}. 
Here we discuss each of them.
 
\begin{figure}[htb]
\begin{center}
\begin{picture}(90,60)(-45,-30)
\CArc(0,0)(15,0,360)
\DashLine(-45,0)(-15,0){3}
\Line(10.6,10.6)(30,30)
\Line(10.6,-10.6)(30,-30)
\Text(-40,5)[b]{$A,H,h$}
\Text(15,30)[tl]{$\tau$}
\Text(15,-30)[b]{$\mu$}
\Text(1,1)[]{{\large $\Delta$}}
\end{picture}
\begin{picture}(100,75)(-50,-45)
\CArc(0,0)(15,0,360)
\Line(-37,0)(-15,0)
\Line(10.6,10.6)(30,30)
\DashLine(10.6,-10.6)(25,-25){3}
\Line(25,-25)(45,-30)
\Line(25,-25)(30,-45)
\Text(-30,5)[b]{$\tau$}
\Text(15,30)[t]{$\mu$}
\Text(35,-11)[]{$A,H,h$}
\Text(45,-39)[br]{$\mu$}
\Text(18,-45)[l]{$\mu$}
\Text(1,1)[]{{\large $\Delta$}}
\end{picture}
\begin{picture}(120,85)(-70,-55)
\CArc(0,0)(15,0,360)
\Line(-37,0)(-15,0)
\Line(10.6,10.6)(30,30)
\DashLine(10.6,-10.6)(25,-25){3}
\Line(25,-25)(45,-30)
\Line(25,-25)(30,-45)
\Text(-30,5)[b]{$\tau$}
\Text(15,30)[t]{$\mu$}
\Text(23,-10)[]{$A$}
\Text(15,-36)[]{$d,s$}
\Text(45,-21)[]{$d,s$}
\Text(40,-47)[]{$\underbrace{\phantom{pppppp}}$}
\Text(40,-60)[]{$\pi, \eta, \eta'$}
\Text(1,1)[]{{\large $\Delta$}}
\end{picture}
\begin{picture}(120,85)(-50,-55)
\CArc(0,0)(15,0,360)
\Line(-37,0)(-15,0)
\Line(10.6,10.6)(30,30)
\DashLine(10.6,-10.6)(25,-25){3}
\Line(25,-25)(37,-32)
\Line(25,-25)(28,-40)
\Line(28,-40)(37,-32)
\Gluon(37,-32)(52,-40){2}{2}
\Gluon(28,-40)(32,-55){2}{2}
\Text(-30,5)[b]{$\tau$}
\Text(15,30)[t]{$\mu$}
\Text(23,-10)[]{$A$}
\Text(37,-40)[]{$b$}
\Text(33,-22)[]{$b$}
\Text(21,-33)[]{$b$}
\Text(43,-57)[]{$\underbrace{\phantom{pppppp}}$}
\Text(43,-70)[]{$\pi, \eta, \eta'$}
\Text(1,1)[]{{\large $\Delta$}}
\end{picture}
\end{center}
\caption{\ftsz $\Delta$-contributions to the Higgs boson decays 
$ A,H,h\to \mu \tau$ and to the decays $\tau \to  3\mu$, 
$\tau \to \mu \pi$, $\tau \to \mu \eta$, $\tau \to \mu \eta'$. 
In the last diagram, curly lines denote gluons.
}
\label{fh2}
\end{figure}
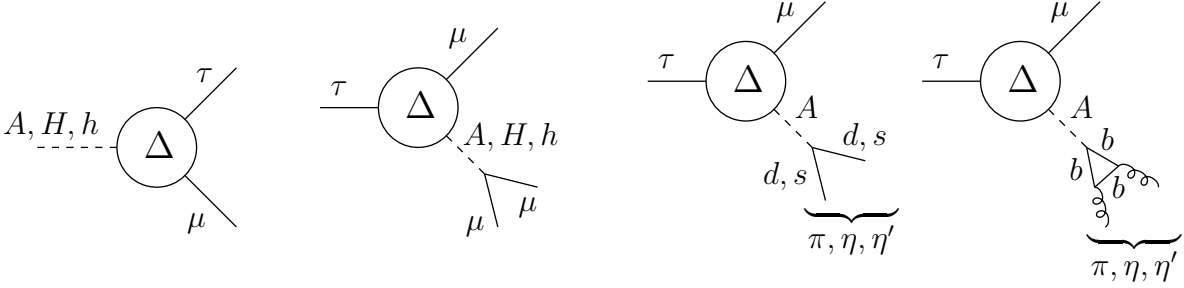

\begin{itemize}
\item 
\underline{ Higgs{\bf $\to \mu \tau $}}
\vspace{0.3cm}

Concerning the Higgs boson decays, we have 
\be{rphi}
{ BR(A\to \mu^+\tau^-)}
=  \tan^2\beta~ (|\Delta_L|^2 + |\Delta_R|^2) 
 { BR(A\to \tau^+\tau^-)}  \, ,
\ee
where we have approximated $1/{\rm c}^2_\beta \simeq \tan^2\beta$
since non-negligible effects can only arise in the 
large $\tan\beta$ limit. 
If $A$ is replaced with $H$ [or $h$] in eq.~(\ref{rphi}),
the r.h.s. should also be multiplied by a factor 
$({\rm c}_{\beta-\al}/{\rm s}_\al)^2$
[or $({\rm s}_{\beta-\al}/{\rm c}_\al)^2$]. 
These LFV decays and the related phenomenology have been extensively 
investigated in \cite{BR} (for other studies in two-Higgs doublet 
models see {\it e.g.} \cite{diaz}). 
We recall that $BR(A\to \mu \tau)$ can reach values of order 
$10^{-4}$. The same holds for the `non-standard' 
CP-even Higgs boson (either $H$ or $h$, depending on $m_A$).

\item 
\underline{{ \bf $\tau \to  \mu  \mu \mu $}}
\vspace{0.3cm}

Consider now the implications of virtual Higgs exchange for      
$\tau \to \mu \mu \mu$.   
The effective lagrangian ${\cal L}^{\rm eff}_{\tau \mu \mu\mu}$ 
[see eq.~(\ref{leff})]  receives 
the following additional contribution from Higgs exchange:
\be{leff-ff}
\delta {\cal L}^{\rm eff}_{\tau \mu \mu\mu} = -
\sqrt{2}G_F~ m_\mu m_\tau 
\left( {\cal C}_+ \ov{\mu} ~\ov{\mu}^c  + 
{\cal C}_- \mu^c \mu \right) \left( \Delta_R \mu^c \tau + 
  \Delta^*_L \mu \tau^c\right) + {\rm h.c.} , 
\ee
where the ${\cal C}_\pm$ coefficients are:
\be{ccf}
{\cal C}_\pm = \frac{1}{{\rm c}^3_\beta}\left( 
{ {\rm s}_\al  {\rm c}_{\beta-\al} \over m^2_h} +  
{{\rm c}_\al  {\rm s}_{\beta-\al} \over m^2_H} \pm 
{{\rm s}_\beta \over m^2_A} \right).
\ee
The  operators proportional to ${\cal C}_+$  
can be Fierz-rearranged and  
cast in a form already exhibited in eq.~(\ref{leff}).
In other words, these contributions can  equivalently 
be regarded as a shift in 
$ F^{\mu_R}_L$ and $ F^{\mu_L}_R$, that is 
\be{dflr}
\delta F^{\mu_R}_L= {G_F \over \sqrt{2}}~ m_\mu m_\tau 
 {\cal C}_+ \Delta_R , ~~~~~~~
\delta F^{\mu_L}_R= {G_F \over \sqrt{2}}~ m_\mu m_\tau 
 {\cal C}_+ \Delta_L ,
\ee
and the branching ratio (\ref{bt3mu}) is accordingly affected.
The operators proportional to  ${\cal C}_-$ have a different chiral 
structure and give an extra contribution 
$(m_\mu m_\tau {\cal C}_-)^2 (|\Delta_L|^2 + |\Delta_R|^2)/32 $ 
to $BR(\tau \to 3 \mu)/ BR(\tau \to \mu  \nu \nu )$. 
Notice, however, that the Higgs-mediated  contributions
are potentially relevant only for large $\tanb$, so that 
the lepton mass suppression is overcome.
More precisely, the truly enhanced operators are those
proportional to ${\cal C}_+$, which becomes 
${\cal C}_+\simeq 2 \tan^3\beta / m^2_A$.
In the limit in which the Higgs-mediated contribution dominates 
over the other ones (in particular, the dipole one), 
the decay $\tau \to 3 \mu$ is directly correlated 
to the decay $A \to \mu \tau$ \cite{BR}:
\be{crl}
BR(\tau^- \to \mu^- \mu^+ \mu^-) \sim 4\times 10^{-5} 
\left(\frac{\tanb}{40}\right)^4 ~
 \left(\frac{100~{\rm GeV}}{m_A}\right)^4~
\frac{BR(A\to \mu^+ \tau^-)}{BR(A\to \tau^+ \tau^-)} .
\ee

\item 
\underline{{ \bf $\tau \to  \mu  P$ $(P=\pi, \eta, \eta')$}}
\vspace{0.3cm}

Now consider the implications of virtual Higgs exchange for 
the decays $\tau \to \mu P$, where $P$ is a neutral pseudoscalar 
meson ($P=\pi, \eta, \eta')$.
Since we assume CP conservation in the Higgs sector, only 
the exchange of the $A$ Higgs boson is relevant.
Moreover, in the large $\tanb$ limit, only the $A$ couplings to 
down-type quarks are important. These  can be written as:
\be{lq}
-i (\sqrt{2} G_F)^{1/2}\tanb~ A(\xi_d m_d d^c d +
\xi_s m_s s^c s +\xi_b m_b b^c b) + {\rm h.c.} .
\ee
The  parameters $\xi_d, \xi_s, \xi_b$ are equal 
to one at tree level, 
but can significantly deviate from this value 
because of higher order corrections 
proportional to $\tanb$ \cite{bottom,all1}, generated by
integrating out superpartners\footnote{
In the limit of quark flavour conservation, 
each $\xi_q$ $(q=d,s,b)$ has the form 
$\xi_q = (1 +\Delta_q \tanb)^{-1}$, where $\Delta_q$ appears in
the loop-generated term $-h_q \Delta_q H^{0 *}_2 q^c q +{\rm h.c.}$.
We recall that the leading contribution 
to $\Delta_q$ arises from a gluino-squark loop and reads 
$\Delta_q \simeq - \frac{2 \alpha_s}{3 \pi}\mu M_3 
I_3(M^2_3, \mt^2_{\qt_L},\mt^2_{\qt_R})$. A Higgsino-stop loop 
also contributes to $\Delta_b$ through a term 
$- \frac{\alpha_t}{4 \pi} \mu A_t 
I_3(\mu^2, \mt^2_{\tilde{t}_L},\mt^2_{\tilde{t}_R})$.}.
At energies below the bottom mass, the $b$-quark can be integrated out, 
so in eq.~(\ref{lq}) the bilinear $-i m_b b^c b + {\rm h.c.}$ 
is effectively replaced by the gluon operator $\Omega = 
\frac{g^2_s}{64 \pi^2} \epsilon^{\mu \nu \rho \sigma} G^a_{\mu \nu}
G^a_{\rho \sigma}$, where $g_s$ and  $G^a_{\mu \nu}$ are the $SU(3)_C$ 
coupling constant and  field 
strength, respectively (see Fig.~\ref{fh2}).
The  effective lagrangian due to $A$ boson exchange, 
relevant for $\tau \to \mu P$ ($P=\pi, \eta, \eta'$), reads:
\be{leff-P}
\delta {\cal L}^{\rm eff} = i 
\sqrt{2}G_F~ m_\tau \frac{\tan^3\beta}{m^2_A} \left[
\xi_d m_d j^5_d  +\xi_s m_s j^5_s
 + \xi_b \Omega \right] \left( \Delta_R \mu^c \tau + 
  \Delta^*_L \mu \tau^c\right) + {\rm h.c.} ,  
\ee
where we have defined the quark pseudoscalar densities\footnote{
In four-component notation  the pseudoscalar densities read 
as $j^5_q= i \ov{\psi}_q \gamma^5 \psi_q$ with 
$\psi^T_q = (q~~\ov{q}^c)$.} 
$j^5_q = i(\ov{q}~ \ov{q}^c - q^c q)$, with $q=d,s$, and 
approximated $1/{\rm c}^2_\beta \simeq \tan^2\beta$.
The  matrix elements $\langle 0|j^5_q| P\rangle$ and 
$\langle 0|\Omega| P\rangle$ 
can be determined  
along the lines of \cite{ME}, {\it i.e.} by taking 
the  matrix elements of 
$\dmd J^{\mu 3}_5, \dmd J^{\mu 8}_5, \dmd J^{\mu 0}_5$ 
(the latter divergence is anomalous and contains the term 
$-\sqrt6 \Omega$). We find:
\beqn{matele}
\langle 0|m_d j^5_d| \pi\rangle & =& 
- \frac{f_\pi m^2_\pi}{1 +z}~ , ~~~~~~~~~
 \langle 0|m_s j^5_s| \pi\rangle = 
\langle 0|\Omega| \pi\rangle=
- \frac12 \left(\frac{1-z} {1 +z}\right)f_\pi m^2_\pi ~,  \\
\langle 0|m_s j^5_s| \eta\rangle & = & 
-\frac{\sqrt3}{2} f^8_\eta m^2_\eta ~, ~~~~~ 
\langle 0|\Omega| \eta\rangle= - \frac{1}{\sqrt{6}}
\left( \frac{1}{\sqrt2} f^8_\eta + f^0_\eta \right) m^2_\eta , \\
\langle 0|m_s j^5_s| \eta'\rangle & = & 
-\frac{\sqrt3}{2} f^8_{\eta'} m^2_{\eta'}~ , ~~~~
\langle 0|\Omega| \eta'\rangle= - \frac{1}{\sqrt{6}}
\left(f^0_{\eta'} + \frac{1}{\sqrt2} f^8_{\eta'}\right) 
m^2_{\eta'}~ , 
\eea
where $z = m_u/m_d$. In deriving  the matrix elements with 
$\eta, \eta'$ we have neglected $m_u, m_d$.
All these results translate into additional contributions to 
the effective meson lagrangians 
$ {\cal L}^{\rm eff}_{\tau \mu \pi}, 
{\cal L}^{\rm eff}_{\tau \mu \eta},  
{\cal L}^{\rm eff}_{\tau \mu \eta'}$ in Section~2.2. 
Such contributions   
can be interpreted as shifts in 
$F^\pi_{L(R)}$, $F^{\eta,8}_{L(R)}$, $F^{\eta,0}_{L(R)}$, 
$F^{\eta',8}_{L(R)}$, $F^{\eta',0}_{L(R)}$, that is
\beqn{dflpi}
\delta F^{\pi}_{L(R)}&=& -\sqrt{2}~ G_F ~m^2_\pi 
\left(\xi_d \frac{1}{1+z} +\frac{\xi_s +\xi_b}{2} \frac{1-z}{1+z}\right)
 {\tan^3\beta \over m^2_A}~
\Delta_{L(R)} ,  \\
\delta F^{\eta,8}_{L(R)}&=& -\sqrt{\frac32}~ G_F ~m^2_\eta 
\left(\xi_s+\frac13 \xi_b\right) {\tan^3\beta 
\over m^2_A}~
\Delta_{L(R)}~, \label{dfleta8}  \\
 \delta F^{\eta,0}_{L(R)}&=&-\frac{1}{\sqrt3}~ G_F ~m^2_\eta 
\xi_b {\tan^3\beta 
\over m^2_A}~
\Delta_{L(R)}~ , \label{dfleta0} 
\eea
and $\delta F^{\eta',8}_{L(R)}, ~\delta F^{\eta',0}_{L(R)}$ 
can be obtained from  
$\delta F^{\eta,8}_{L(R)}, ~\delta F^{\eta,0}_{L(R)}$ 
by the substitution $\eta \to \eta'$.
We have performed a cross-check 
by using the approach of  chiral 
perturbation theory and obtained results that are  
consistent with those above.

In the limit in which the processes $\tau \to 3\mu$ and  
$\tau \to \mu \eta$ are both dominated by   
Higgs-exchange, these decays are related as: 
\be{eta3mu}
\frac{BR(\tau^- \to \mu^- \eta)}{BR(\tau^- \to \mu^- \mu^+\mu^- )} 
\simeq	  
36 \pi^2 \left(\frac{f^8_\eta m^2_\eta}{m_\mu m^2_\tau}\right)^2  
(1 - x_\eta)^2 
\left[\xi_s +\frac{\xi_b}{3}\left(1 +\sqrt2 \frac{f^0_\eta}{f^8_\eta}
\right)\right]^2 .  
\ee
For $\xi_s \sim \xi_b \sim 1$ this ratio is about 5,
but it could also be a few times larger or smaller than that,
depending on the actual values of $\xi_s,\xi_b$.
Our result confirms the relevance of Higgs-exchange in 
$\tau \to \mu \eta$, previously emphasized in  \cite{sher1}. 
In the latter paper, however, neither the contribution of 
the (bottom-loop induced) gluon operator $\Omega$ nor the 
factors $\xi_q$ were included. 
Moreover, if these effects were disregarded, the ratio (\ref{eta3mu}) 
would be 3 times smaller than the one found in  \cite{sher1}
(notice that $F^8_\eta$ in \cite{sher1} corresponds to 
$\sqrt{2}f^8_\eta$).

Finally, let us compare $\tau \to \mu \eta'$ and  $\tau \to \mu \pi$ with  
$\tau \to \mu \eta$ in the limit of  Higgs-exchange domination:
\beqn{etaetap}
\frac{BR(\tau^- \to \mu^- \eta')}{BR(\tau^- \to \mu^- \eta)} 
& \simeq &	  
\frac29  \left( \frac{f^0_{\eta'}}{f^8_\eta}\right)^2 
\frac{m^4_{\eta'}}{m^4_\eta} \left(\frac{1 - x_{\eta'}}{1 - x_\eta} \right)^2
\left[\frac{
1 + \frac{3}{\sqrt2} \frac{f^8_{\eta'}}{f^0_{\eta'}} 
\left( \frac{\xi_s}{\xi_b}  +\frac13\right)}
{\frac{\xi_s}{\xi_b} + \frac13 + 
\frac{\sqrt2}{3} \frac{f^0_\eta}{f^8_\eta}}
\right]^2 , \\
\label{pieta}
\frac{BR(\tau^- \to \mu^- \pi)}{BR(\tau^- \to \mu^- \eta)} 
& \simeq & \frac43  \left(\frac{f_\pi}{f^8_\eta}\right)^2 
 {m^4_\pi \over m^4_\eta} ~   
(1 - x_\eta)^{-2}
\left[\frac{\frac{\xi_d}{\xi_b} \frac{1}{1+z} + \frac{1}{2}
(1 + \frac{\xi_s}{\xi_b}) 
\frac{1-z}{1+z}}
{\frac{\xi_s}{\xi_b} + \frac13 + 
\frac{\sqrt2}{3} \frac{f^0_\eta}{f^8_\eta}}
\right]^2 . 
\eea
Both ratios are suppressed, although for different reasons.
The ratio $BR(\tau \to \mu \eta')/BR(\tau \to \mu \eta)$, which  
 seems to be ${\cal O}(1)$,  is much smaller 
because the singlet and octet contributions to $\tau \to \mu \eta'$ 
tend to cancel against each other (we recall that 
${f^8_{\eta'}}/{f^0_{\eta'}} \sim -0.4$).
For $\xi_s/\xi_b \sim 1$, the ratio (\ref{etaetap}) is $6\times 10^{-3}$.
The ratio $BR(\tau \to \mu \pi)/BR(\tau \to \mu \eta)$ 
is small because it is parametrically  
suppressed by  $m^4_\pi/m^4_\eta \sim 10^{-2}$. 
The actual numerical value is sensitive to the parameter $z=m_u/m_d$, 
and to the ratios $\xi_d/\xi_b,~ \xi_s/\xi_b$. 
In the case  $\xi_d/\xi_b \sim 
\xi_s/\xi_b\sim 1$, if we let $z$ vary between $0.2$ and $0.7$ \cite{pdg} 
the ratio (\ref{pieta}) ranges from $4\times 10^{-3}$ to $10^{-3}$.
These results, combined with the present bound (\ref{e-mueta}) on 
$\tau \to\mu\eta$, imply that the Higgs mediated contribution to 
$BR(\tau \to\mu\eta')$ and $BR(\tau \to\mu\pi)$ can reach 
${\cal O}(10^{-9})$.

\end{itemize} 
 
\subsection{Muon anomalous magnetic moment \label{mamm}}
The flavour changing dipole operators in eq.~(\ref{phd}) have an 
obvious flavour conserving counterpart, which for the muon is 
\be{gm2}
e m_\mu i D_\mu^{\ga} \, {\mu}^c \sigma^{\rho \sigma} {\mu}~ 
F_{\rho\sigma}+ {\rm h.c.} 
\ee
The coefficient $D_\mu^{\ga}$ is directly related to 
the anomalous magnetic moment of the 
muon:  $a_\mu\equiv(g_\mu-2)/2=2m^2_\mu {\rm Re} D^\ga_\mu$.
In the MSSM the coefficient $D^\ga_\mu$ receives three types of 
one-loop contributions from superparticle exchange,
analogous to those shown in Fig.~\ref{f2}. In turn the (a)-type and 
(b)-type contributions can be further distinguished 
according to the chirality (either $L$ or $R$) of the sleptons circulating 
in the loop. In the  (c)-type contributions both $L$ and $R$ sleptons 
are simultaneously present. 
Therefore we can express the superparticle contribution to  $a_\mu$, 
denoted as $a^{\rm MSSM}_\mu$, as the sum of 
five terms: 
\be{amu}
a^{\rm MSSM}_\mu = 
a^{(a)}_{\mu L} +a^{(a)}_{\mu R} +a^{(b)}_{\mu L}+a^{(b)}_{\mu R}
+a^{(c)}_{\mu L R}.
\ee
Non-vanishing  contributions to  $a^{\rm MSSM}_\mu $ arise 
even in the absence of  LFV, of course \cite{fayet,moroi}.  
In this case the only sleptons involved are 
$\tilde{L}_{\mu}=(\tilde{\nu}_\mu , \tilde{\mu}_L)$ and $\tilde{\mu}_R$,  
and the contributions to 
$a_\mu$ are generically proportional to  $ ({g^2_w}/{16\pi^2})~ 
(m^2_\mu/M^2_{S})$, possibly with an extra $\tanb$ factor.
In the case of mixing between second and third slepton generations 
the no-mixing results need to be  generalized (see Appendix~A.10).  
One of the effects of LFV is that (c)-type  contributions 
have  extra terms proportional to $m_\tau/m_\mu$,  
as already observed in \cite{moroi,g2}. 
These contributions are potentially the largest ones if 
 (LFV)$_L$ and (LFV)$_R$ are both large. 
However, the latter situation does not seem very natural 
if the smallness of $m_\mu/m_\tau$ is to be explained by an 
underlying flavour symmetry. In fact, we rather expect 
LFV to be large in at most one sector ($L$ or $R$).   
In such a case the apparent $m_\tau/m_\mu$-enhancement is 
compensated by suppressing factors, so that the size 
of such extra contributions to $a^{\rm MSSM}_\mu$ does not exceed that
of the other ones.

We also recall that, at present, it is not clear whether a discrepancy 
exists between the experimental determination of $a_\mu$ and the 
Standard Model prediction (see  {\it e.g.} \cite{muon}).
Hence some caution is needed in deriving constraints 
on  the MSSM parameter space \cite{MW}. 

\newpage

\section{Large (LFV)$_L$: numerical analysis \label{lfvl}}
In this section we perform a detailed numerical 
analysis in the case of large  (LFV)$_L$, assuming  
vanishing  (LFV)$_R$, {\it i.e.} 
$\mt^2_{R \mu \tau} = A^R_{\mu \tau}= 0$.  
We remark that all operator coefficients depend on 
$\mt_{L_\al}$,  $\theta_L$ and  gaugino masses.
Some coefficients also depend on additional parameters. 
In particular, $A^{Z(a)}_L$ and $D^{\ga (b)}_L$ depend on $\mu$ and 
$\beta$; $D^{\ga (c)}_L$ depends on $\mu$, $\beta$, $\mt_{\tilde{\tau}_R}$, 
$A_\tau, A^L_{\mu\tau}$;  $B^{f}_L$ depends on $\mt_{\tilde{f}}$.

The lightest  eigenvalue of
$\tilde{{\cal M}}^2_L$ is  conventionally chosen to be $\mt^2_{L_3}$ 
(although our formulae in Appendix do not depend on such a choice).
To enhance (LFV)$_L$,  in all our numerical 
examples we will take 
 maximal mixing, $\theta_L=\pi/4$ ({\it i.e.} 
$\mt^2_{L \mu \mu}= \mt^2_{L \tau \tau} $ in $\tilde{{\cal M}}^2_L$), 
and widely split eigenvalues.
The mass parameters will be varied in such a way that 
charged sparticle   masses  be $\gsim 100~{\rm GeV}$,   
to respect the LEP constraints \cite{pdg,lep}.
We will also check that  neutralino 
production at LEP be either kinematically forbidden or 
adequately suppressed \cite{delphi}. Incidentally, 
we should add that we have performed such checks 
using the full mass eigenvalues (not just those at zero-th order 
in electroweak breaking).
For simplicity, we take a common (soft-breaking) 
mass $\mt_{\tilde{e}}$ for 
selectrons of both chiralities, as well as a common mass 
$\mt_{\qt}$ for first and second generation 
squarks of both chiralities.

\subsection{(LFV)$_L$ with small $\tanb$} 

The decay $\tau \to \mu \ga$ poses  
significant constraints on the MSSM parameter space, especially 
for  large (LFV)$_L$.   
Moreover, the decays $\tau \to \mu ee,~ \tau \to \mu \mu \mu, ~
\tau \to \mu \rho$ are often dominated by the dipole contribution.
One of our purposes is to study to which extent the 
latter property holds
in the MSSM parameter space or, more generally, 
to study the interplay between dipole 
and non-dipole contributions.
We start our numerical analysis by considering 
in more detail the contributions to the dipole operator, 
taking into account that the present bound (\ref{e-rad}) on  
$\tau \to \mu \ga$ translates into the bound  
$|D^\ga_L| \lsim 5\times 10^{-9}~ {\rm GeV}^{-2}$.
\begin{figure}[hb]
\begin{center}
\psfrag{m2}{\footnotesize $M_2$~[GeV]}
\psfrag{ml3}{\footnotesize $\begin{array}{c}
\tilde{m}_{L_3} \\
{\rm { [GeV]}}\end{array}$}
\psfrag{c1}{\footnotesize 50}
\psfrag{c2}{\footnotesize 30}
\psfrag{c3}{\footnotesize 20}
\psfrag{s1}{\footnotesize 7}
\psfrag{s2}{\footnotesize 5}
\psfrag{s3}{\footnotesize 3}
\psfrag{D}{\small  $D^{\gamma (b)}_L~{[10^{-9}\cdot {\rm GeV}^{-2}]}$}
\vskip -1.7cm
\hglue -0.8cm
\scalebox{0.77}{
{\mbox{\epsfig{file=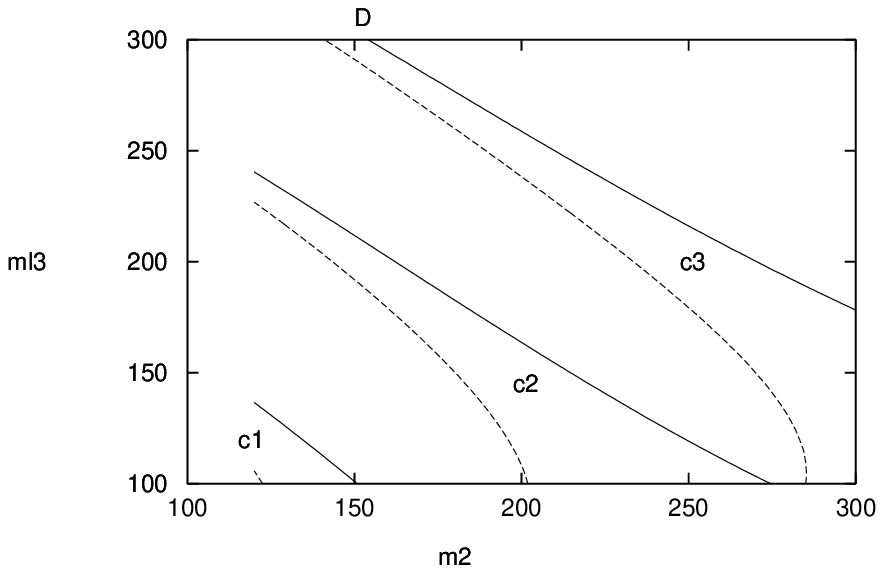}}}
\hglue -3.5cm
{\mbox{
\epsfig{file=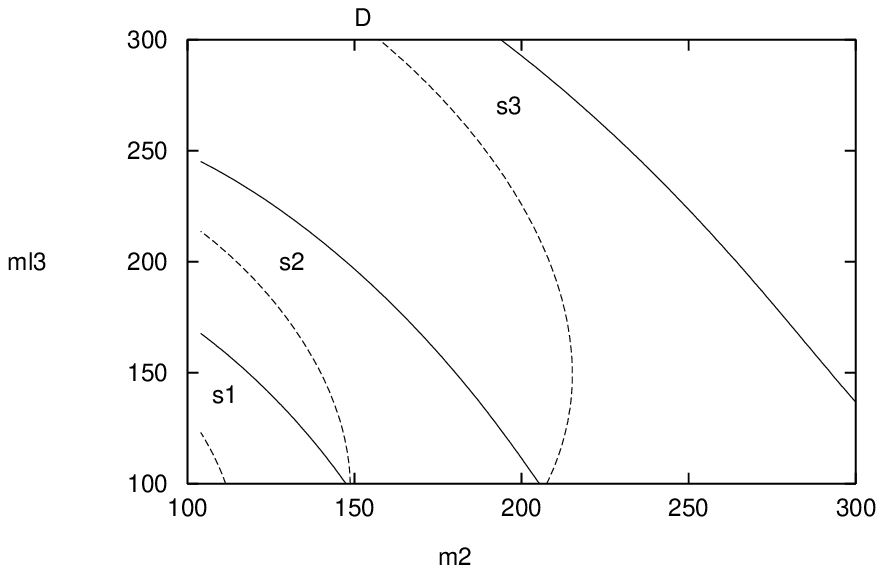}}}}
\vglue -1.2cm
\caption{{\footnotesize $D^{\gamma(b)}_L$ 
 isocontours with $\tanb =3$, $\mt_{L_2} = 1~{\rm TeV}$,  
$\theta_L=\pi/4$ 
and $\mu =140~{\rm GeV}$ (1~TeV) in the left (right) panel. 
The solid and dashed lines refer to  $M_1=100~{\rm GeV}$ and 
$M_1= -100~{\rm GeV}$, respectively.}}
\label{f3}
\end{center}
\end{figure}

The coefficient $|D^{\ga(a)}_L|$ is at most  
$10^{-9}~ {\rm GeV}^{-2}$, so it fulfils the bound. 
In contrast, 
$D^{\ga(b)}_L$ and  $D^{\ga(c)}_L$ may separately exceed the bound 
in some regions of the parameter space, especially  for large $\tanb$. 
In this section we focus on small $\tanb$. 
Incidentally, notice that monopole coefficients do not have a 
strong $\tanb$ dependence.
In Fig.~\ref{f3} we  show the contours of the coefficient 
$D^{\ga (b)}_L$ in the plane $(M_2, \mt_{L_3})$ with $\tanb =3$, 
$\theta_L =\pi/4$,  $\mt_{L_2} = 1~{\rm TeV}$, $|M_1|=100~{\rm GeV}$ 
and $\mu =140~{\rm GeV}$ (1~TeV) in the left (right) panel.
For the  range of $M_2$ and $\mt_{L_3}$ 
shown in the figure, $D^{\ga (b)}_L$ is well above the bound for 
small $\mu$,   whilst it is of the same order of the bound
for large $\mu$. 
Also $D^{\ga (c)}_L$  may or may not exceed the bound, depending on 
the range of the extra parameters $\mt_{\tilde{\tau}_R}$, 
$A_\tau$, $A^L_{\mu\tau}$. 
Even if $D^{\ga(b)}_L$ and $D^{\ga(c)}_L$ should 
separately exceed the bound, however, 
mutual cancellations could bring the total dipole contribution 
$D^{\ga }_L$ below the bound.
%
\begin{figure}[hb]
\begin{center}
\psfrag{atau}{\footnotesize $A_\tau$~[TeV]}
\psfrag{mr3}{\footnotesize $\begin{array}{c}
\tilde{m}_{\tilde{\tau}_R} \\
 {\rm { [GeV]}}
\end{array}
$}
\vglue -2.cm
\scalebox{1.10}{
{\mbox{\epsfig{file=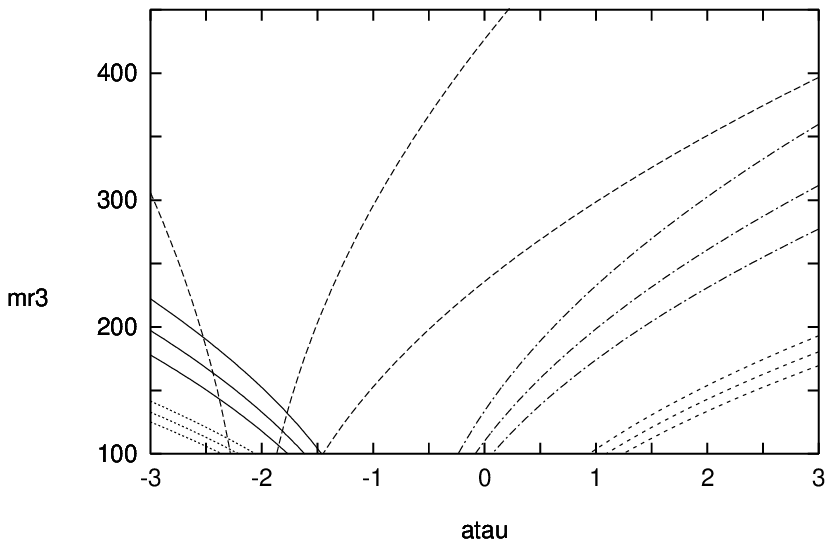}}}}
\vglue -1.8cm
\caption{{\footnotesize $D^\gamma_L$ contours for $\tanb=3$, 
$\mt_{L_2} =1~{\rm TeV}$, $\theta_L=\pi/4$, 
$A^L_{\mu\tau}=0$ and  five choices of $(\mt_{L_3}, \mu, M_2, M_1)$ in GeV: 
$(100,140,120,100)$ (dotted),
$(100,140,120,-100)$ (short-dashed),
$(100,250,150,100)$ (solid), 
$(100,250,150,-100)$ (dot-dashed),
$(200,700,300,-100)$ (dashed).  
For each example the two external lines correspond to  
 $|D^{\ga }_L| = 5 \times 10^{-9}~
{\rm GeV}^{-2}$ and  the middle one  to $D^{\ga }_L=0$.
}}
\label{f4}
\end{center}
\end{figure}
To illustrate this, in Fig.~\ref{f4} we depict 
the  $D^{\ga }_L$ contours in the plane $(A_\tau, 
\mt_{\tilde{\tau}_R})$ for $\tanb=3, \mt_{L_2} =1~{\rm TeV}, 
A^L_{\mu\tau}=0$ and  several choices of $(\mt_{L_3}, \mu, M_1, M_2)$.
For each case we show three  lines: the two external ones 
 delimit the allowed region ($|D^{\ga }_L| \leq 5 \times 10^{-9}~
{\rm GeV}^{-2}$), the middle one corresponds to full
cancellation ($D^{\ga }_L=0$).
The coefficient  $D^{\ga(b)}_L$  
is  above the bound  in the four cases with small $\mu$,  
while it is below 
in the example with large $\mu$, hence in the latter case the 
allowed region is wider.
In the examples with $M_1>0$,  
large and negative $A_\tau$ values\footnote{A situation with 
$|A_\tau| \gg \mt_{L_3},~ \mt_{\tilde{\tau}_R}$ could destabilize 
the scalar potential and induce VEVs for slepton fields. 
To avoid this one can take, for instance, a sufficiently 
large $m_A$. 
The inequality $ m_\tau |A_\tau + \mu \tanb| < \sqrt{2}\mt_{L_3} \staur$
is another requirement to avoid tachyonic sleptons 
(here $\theta_L = \pi/4$). 
This has been 
verified throughout all our numerical examples.
Similar considerations apply to the 
(LFV)$_R$ case in Section 5, with  $L\leftrightarrow R$. 
} 
are required to make the sign of $D^{\ga(c)}_L$ 
opposite to that of $D^{\ga(b)}_L$. 
We also note that, in general, the role of $A_\tau$ would be
shared by  $A_\tau$ and $A^L_{\mu \tau}$ [see eq.~(\ref{a6dc})].
\begin{figure}[p]
\begin{center}
\psfrag{atau}{\footnotesize $A_\tau$~[TeV]}
\psfrag{br}{\footnotesize $BR$}
\psfrag{mr}{\footnotesize $\tilde{m}_{\tilde{\tau}_R}$~[GeV]
}
\psfrag{t1}{\scriptsize $10^{-9}$}
\psfrag{t2}{\scriptsize $10^{-8}$}
\psfrag{t3}{\scriptsize $10^{-7}$}
\psfrag{t4}{\scriptsize $10^{-6}$}
\psfrag{br1}{\footnotesize $\tau\rightarrow \mu\gamma$}
\psfrag{br2}{\footnotesize $Z\rightarrow \mu\tau$}
\psfrag{br3}{\footnotesize $\tau\rightarrow \mu e e$}
\psfrag{br4}{\footnotesize $\tau\rightarrow 3 \mu$}
\psfrag{br5}{\footnotesize $\tau\rightarrow \mu\rho$}
\psfrag{br6}{\footnotesize $\tau\rightarrow \mu\pi$}
\psfrag{exp}{\scriptsize $3\times 10^{-7}$}
\vglue -1.0cm
\scalebox{1.0}{
{\mbox{\epsfig{file=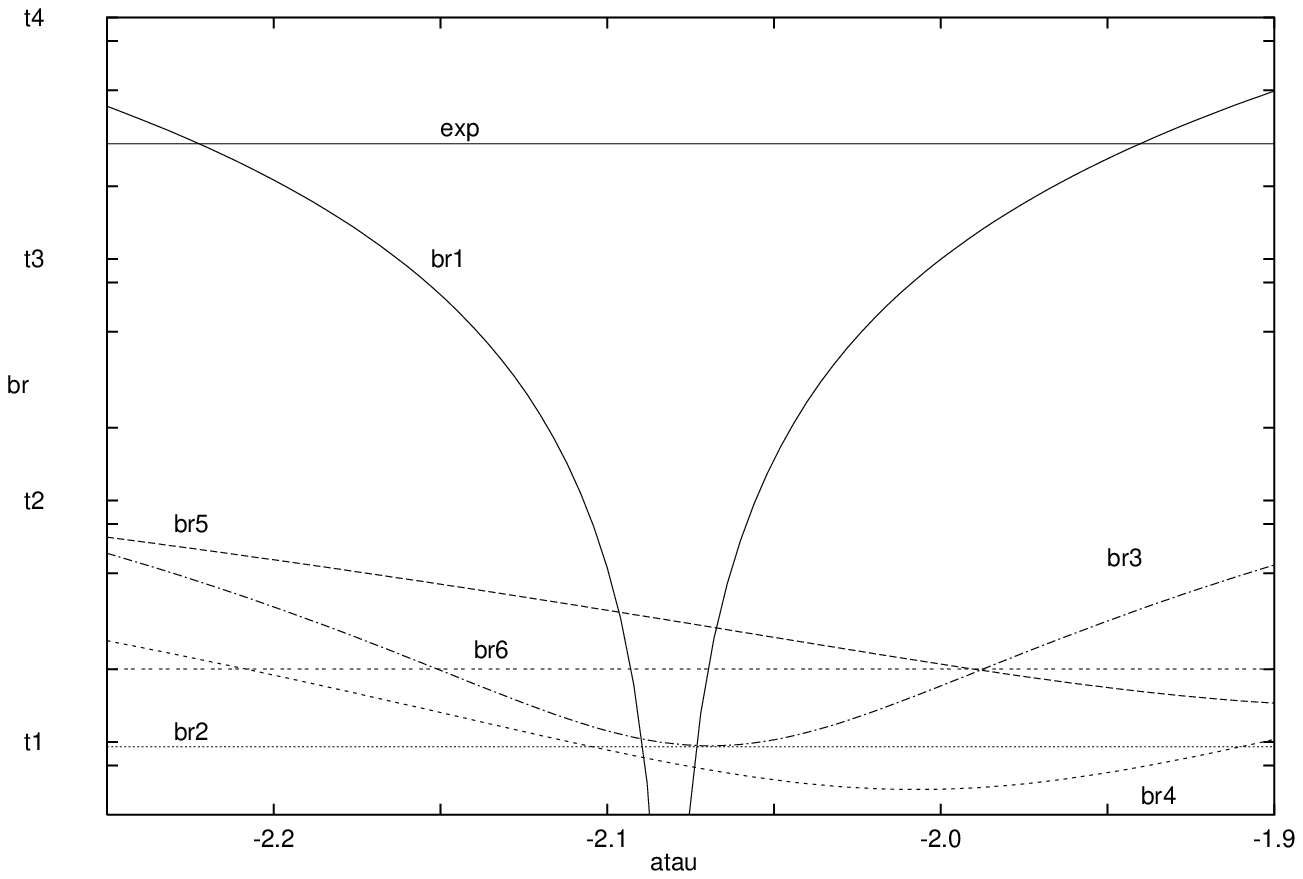}}}}

\vglue 0.8cm
\scalebox{1.0}{
{\mbox{\epsfig{file=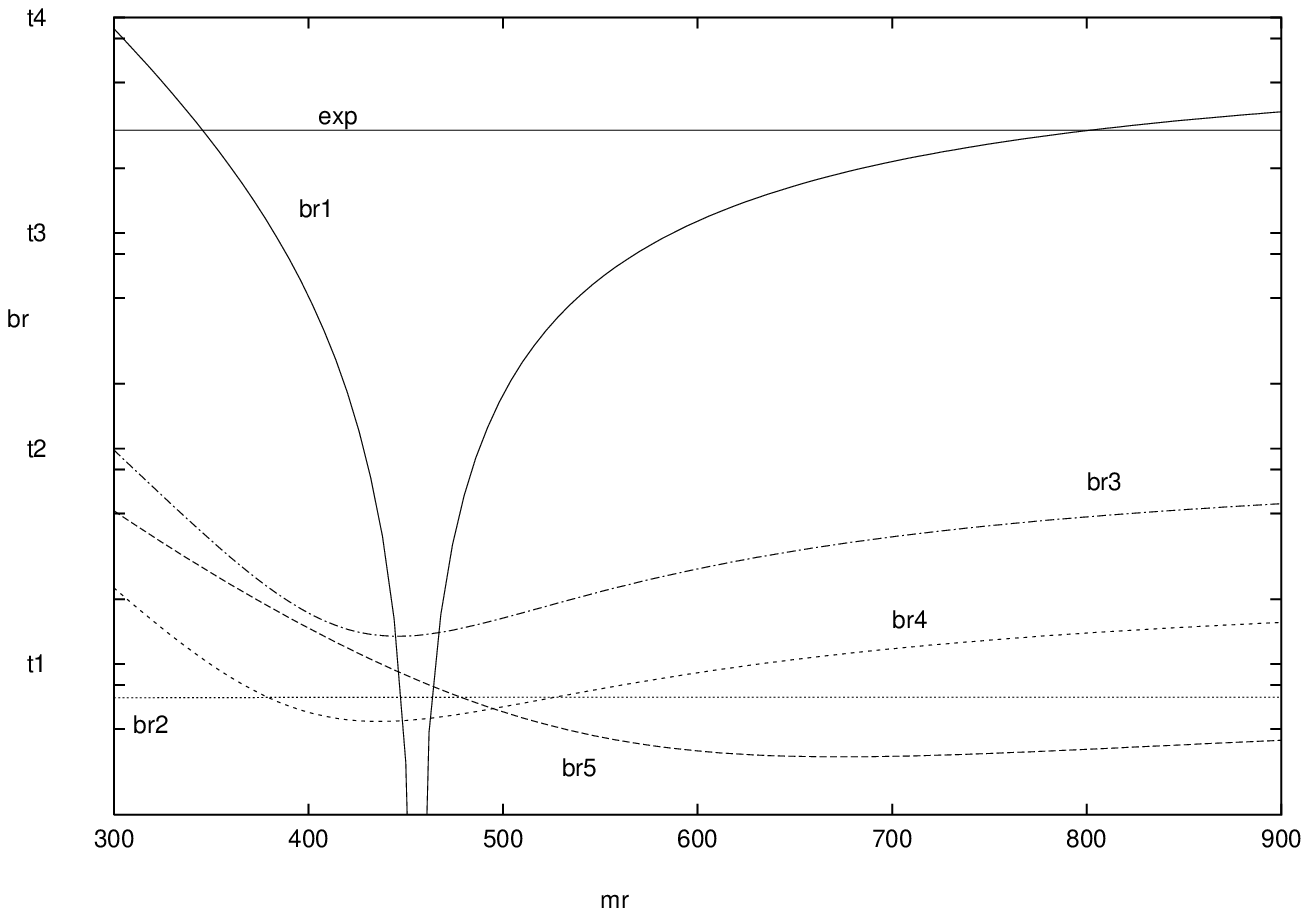}}}}
\vglue 0.5cm
\caption{{\footnotesize Branching ratios of  
LFV decays for $\tanb=3$, $\mt_{L_2}= 1~{\rm TeV}$, 
 $\mt_{L_3}= 100~{\rm GeV}$, $\theta_L= \pi/4$, $A^L_{\mu \tau}=0$. 
In the upper panel, the remaining parameters are: 
$\mu =120~{\rm GeV}$, $M_2=150~{\rm GeV}$, $M_1=100~{\rm GeV}$, 
$\mt_{\tilde{\tau}_R}= \mt_{\tilde{\mu}_R} = 100~{\rm GeV}$, 
$\mt_{\tilde{e}}= \mt_{\tilde{q}}= 1~{\rm TeV}$. 
In the lower panel: 
$\mu =800~{\rm GeV}$, $M_2=120~{\rm GeV}$, $M_1=-100~{\rm GeV}$, 
$A_\tau=0$, 
$\mt_{\tilde{e}}= 120~{\rm GeV}$, $\mt_{\tilde{q}}= 1~{\rm TeV}$ and 
$\mt_{\tilde{\mu}_R}=\mt_{\tilde{\tau}_R}$. 
The solid horizontal line indicates the present bound on 
$BR(\tau\to \mu\ga)$. In the upper example:  
$BR(\tau \to \mu \eta) \sim 6\times 10^{-10},~ 
BR(\tau \to \mu \eta') \sim 5\times 10^{-10} $.
In the lower example:  
$BR(\tau \to \mu \pi) \sim 2\times 10^{-12},~
BR(\tau \to \mu \eta) \sim BR(\tau \to \mu \eta') \sim 10^{-12} $.
}}
\label{f5}
\end{center}
\end{figure}

A remark is in order about the muon anomalous magnetic moment. 
There are obvious analogies between the contributions to  $D^\ga_\mu$ 
({\it i.e.} to $a^{\rm MSSM}_\mu$) and those to $D^\ga$, 
such as a similar diagrammatic origin and 
the dependence on common parameters 
(see Appendices~A.6 and A.10).
In the case of large (LFV)$_L$, for instance, we have 
typically that 
$a^{(a, b)}_{\mu L}~\sim~-0.02\times (D^{\ga (a, b)}_L/{\rm GeV}^{-2})$. 
So if $D^{\ga (b)}_L$ saturates the present bound on $D^\ga$, then 
$|a^{(b)}_{\mu L}| \lsim 10^{-10}$, while if 
 $D^{\ga (b)}_L$ is ten times larger than the bound 
(which should be fulfilled through cancellations), then 
$|a^{(b)}_{\mu L}| \sim 10^{-9}$. This is  of the order of the 
experimental error on $a_\mu$ and could either be acceptable on
its own or play a role in explaining 
a possible discrepancy between the experimental 
determination of $a_\mu$ and the Standard Model prediction. 
The connection between  other contributions to  $a^{\rm MSSM}_\mu$ and 
$D^\ga_L$ is less direct, since some 
parameters differently affect  each of them.
For instance, in the (LFV)$_L$ case that we are considering here, 
$a^{\rm MSSM}_\mu$ depends also on ${\smur}$ 
and $A_\mu$, which do not enter the $D^\ga_L$-coefficients 
(at least at leading order).
In particular, cancellations
among different contributions to $a^{\rm MSSM}_\mu$ 
(analogous to those in $D^\ga_L$) may or may not occur, 
depending on the choice of such parameters.
All such features loosen the correlation between  
$a^{\rm MSSM}_\mu$ and $D^\ga$.
Since we work in an unconstrained MSSM framework,
we content ourselves 
with verifying that the numerical examples of this Section are
consistent with values of 
$|a^{\rm MSSM}_\mu|$ of order $10^{-9}$ or smaller. 
Similar considerations  apply to the (LFV)$_R$ case discussed 
in Section 5 (where we shall not come back to this point).

After this digression on $a_\mu$, let us continue the discussion
on the  LFV operators. 
It is interesting to see how the branching ratios of the 
different  LFV processes behave when  the $D^{\ga}_L$ dipole 
contribution varies and the monopole ones are basically fixed.
In Fig.~\ref{f5}  we plot two such 
examples, where $\mu$ is either small (upper panel) or  large 
(lower panel). The specific values of the parameters involved 
are given in  the caption.
The parameter that varies on  the horizontal axis (either  $A_\tau$ 
or  $\mt_{\tilde{\tau}_R}$) has the only role  to induce 
a variation of the $D^{\ga}_L$  dipole  contribution.
The $BR$s of $Z\to \mu \tau$,  $\tau\to \mu \pi, ~
\tau\to \mu \eta, ~ \tau\to \mu \eta'$ are constant\footnote{
In fact, $BR(Z\to \mu \tau)$ depends on $\mt_{\tilde{\tau}_R}$ 
through the coefficients $A^{Z(b,c)}_L, D^{Z(b,c)}_L$. 
Similarly, $BR(\tau\to \mu \pi), ~BR(\tau\to \mu \eta)$ 
and $BR(\tau\to \mu \eta')$  
depend on $\mt_{\tilde{\tau}_R}$ through the coefficients $A^{Z(b,c)}_L,
\Delta^{(c)}_L$. 
However, for small $\tanb$ these contributions are suppressed.}. 
In the other decays we can observe an interplay between the 
dipole and monopole contributions, which are comparable in 
magnitude in these examples. For instance, 
in both examples the pure monopole contribution to 
$BR(\tau\to \mu ee)$ amounts to about $10^{-9}$, as 
we can read in correspondence of the dipole cancellation 
($\tau \to \mu\ga$ `dip'). At the points
where $\tau \to \mu\ga$ saturates the bound, 
the pure dipole contribution 
is $3\times 10^{-9}$ [see eq.~(\ref{dom-dge})] and the combined 
contribution can reach  $5\times 10^{-9}$. 
In the case of $\tau \to \mu \rho$ we can also notice a 
strong interference effect between the dipole and the 
monopole contributions. 
Consider, for instance, the first example.
In correspondence of the  $\tau \to \mu\ga$-saturation points, 
$BR(\tau \to \mu \rho)$ 
is about $6\times 10^{-9}$ (left point) and  $1.5\times 10^{-9}$ 
(right point). These numbers should be compared with  the 
pure dipole contribution at those points, $0.8\times 10^{-9}$ 
[eq.~(\ref{dom-dgr})], and the pure monopole one,  $3\times 10^{-9}$.

The previous examples have shown that monopole operators
can play an important role. 
We recall that the monopole coefficients ($F^Z, F^\mu, F^e, F^\rho, 
F^\pi, F^\eta, F^{\eta'}$) 
receive different contributions (from $A^Z, C^Z, C^\ga, B^f$),  
which have specific parameter dependences. Notice that
all these contributions depend on $\mt_{L_3}$ and $M_2$,
so  latter masses  should 
not be too large, if we want to avoid a strong suppression of 
monopole coefficients. 
The relevance of each contribution, as well as the possibility
of mutual cancellations, is also influenced by other
parameters, such as $\mu$ (in $A^Z$), the selectron masses 
 (in $B^e$) or the squark masses (in $B^u, B^d, B^s$).
In the first example of Fig.~\ref{f5}, for instance, 
the large values of $\mt_{\tilde{e}}$ and  $\mt_{\tilde{q}}$ 
imply suppressed box contributions, while small $\mu$ implies 
unsuppressed $A^Z$-contribution (which combines with those of 
$C^Z$ or $C^\ga$). In the second example, $A^Z$ is suppressed 
by the large $\mu$ value, whereas $B^e$ is unsuppressed 
because $\mt_{\tilde{e}}$ is small.
In the following we shall analyse in more detail the parameter
dependence of monopole contributions, considering one process
at a time. 

\begin{figure}[ht]
\begin{center}
\psfrag{mu} {\Large $\mu$~[GeV]}
\psfrag{t1}{\large $10^{-11}$}
\psfrag{t2}{\large $10^{-10}$}
\psfrag{t3}{ \large $10^{-9}$}
\psfrag{t4}{\large $10^{-8}$}
\psfrag{m12}{ \large $M_2=120$~GeV}
\psfrag{m2}{\footnotesize $M_2$~[GeV]}
\psfrag{ml3}{\footnotesize $\begin{array}{c}
\tilde{m}_{L_3} \\
{\rm { [GeV]}}\end{array}$}
\psfrag{m16}{\large $M_2=160$~GeV}
\psfrag{br1}{\Large $BR(Z\rightarrow \mu\tau)$}
\psfrag{zmu}{\small $BR(Z\rightarrow \mu\tau)~[10^{-9}]$}
\vglue -0.5cm
\scalebox{0.47}{
{\mbox{\epsfig{file=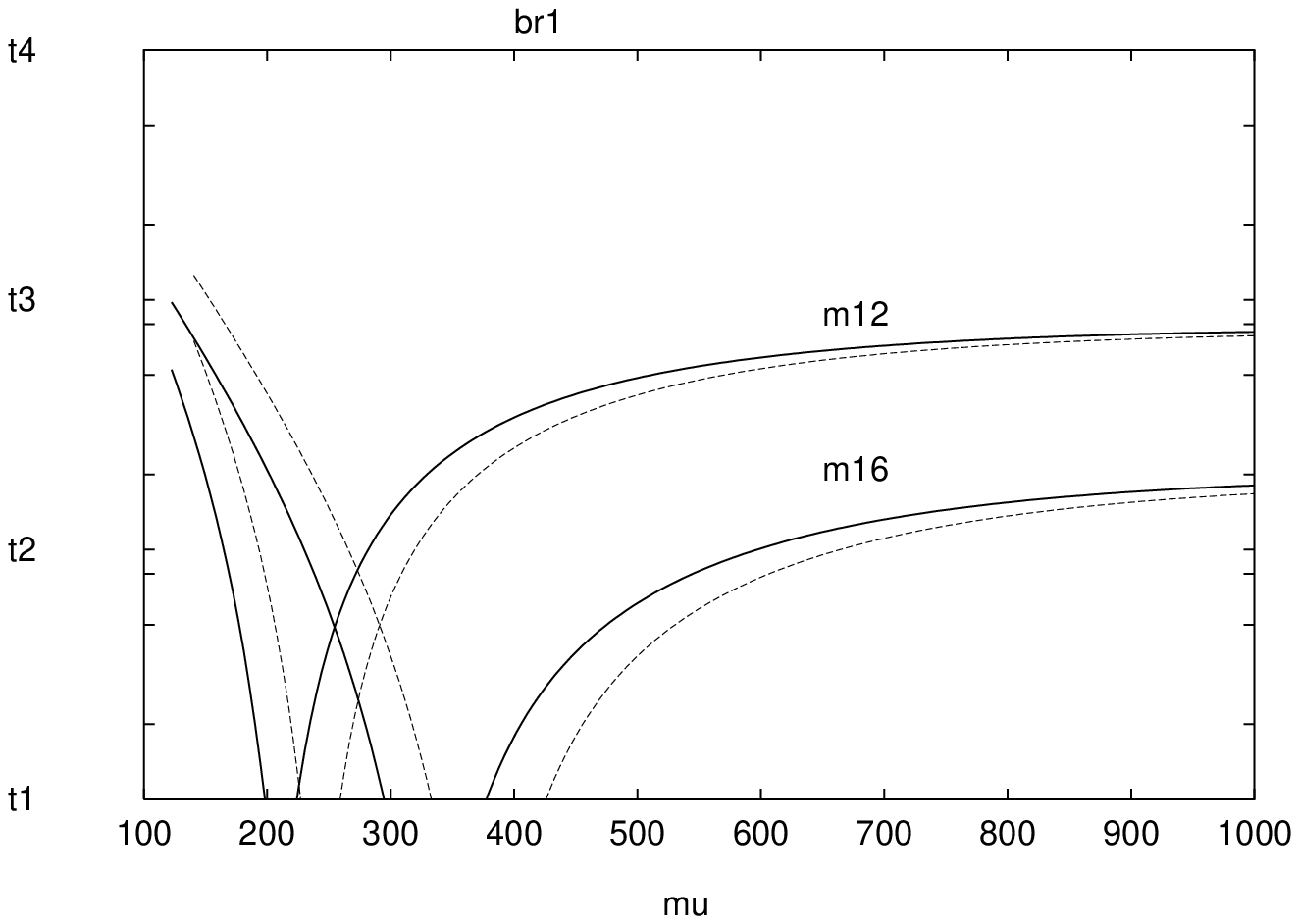}}}}
\vskip -0.8cm
\hglue -0.8cm
\scalebox{0.77}{
{\mbox{\epsfig{file=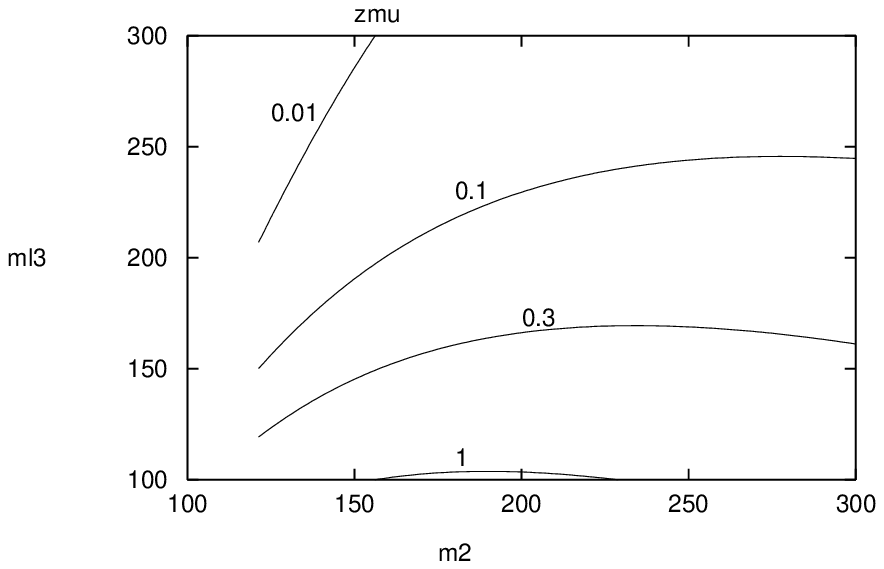}}}
\hglue -3.5cm
{\mbox{\epsfig{file=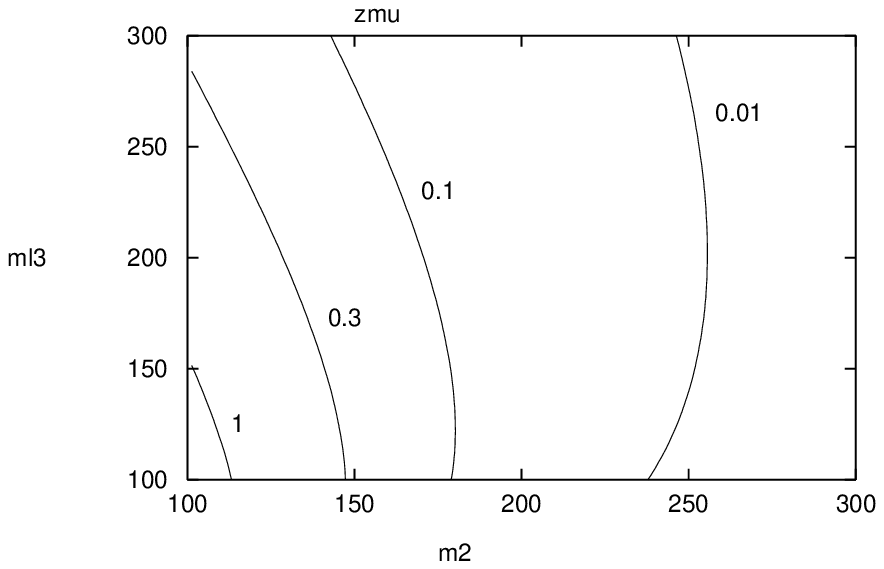}}}}
\vglue -1.2cm
\caption{{\footnotesize $BR(Z\to \mu \tau)$ for $\tanb =3$, $\mt_{L_2}= 
1~{\rm TeV}$, $\theta_L= \pi/4$. 
In the upper panel:   $\mt_{L_3}= 100~{\rm GeV}, \staur=  100~{\rm GeV}$, 
$M_2$ as shown, $M_1= 100~(-100)~{\rm GeV}$ for the solid (dashed) lines.  
In the left (right) lower panel: $\mu =  120~(1000)~{\rm GeV}, 
M_1=  100~(-100)~{\rm GeV}$ and $\staur = 100~(500)~{\rm GeV}$.
}}
\label{f6}
\end{center}
\end{figure}

In Fig.~\ref{f6} $BR(Z\to \mu \tau)$ is studied\footnote{
For earlier studies on $Z\to \mu \tau$ in the MSSM, 
see {\it e.g.} \cite{zeta}. 
A more recent computation, with related discussion,  
appeared in \cite{IM}.}
as a function of 
$\mu$ (upper panel) and in the $(M_2, \mt_{L_3})$ plane
for small $\mu$ ( left  lower panel) or large $\mu$ (right lower panel).
The behaviour in the upper panel shows the destructive interference 
of the $A^Z$ and $C^Z$ contributions. 
\begin{figure}[p]
\begin{center}
\psfrag{mu} {\Large $\mu$~[GeV]}
\psfrag{t1}{\large $10^{-11}$}
\psfrag{t2}{\large $10^{-10}$}
\psfrag{t3}{ \large $10^{-9}$}
\psfrag{t4}{\large $10^{-8}$}
\psfrag{m12}{ \large $M_2=120$~GeV}
\psfrag{m2}{\footnotesize $M_2$~[GeV]}
\psfrag{ml3}{\footnotesize $\begin{array}{c}
\tilde{m}_{L_3} \\
{\rm { [GeV]}}\end{array}$}
\psfrag{m16}{\large $M_2=160$~GeV}
\psfrag{br1}{\Large $BR(\tau\rightarrow \mu\mu\mu)_{D^\gamma=0}$}
\psfrag{t3mu}{\small $BR(\tau\rightarrow \mu\mu\mu)_{D^\gamma =0}~[10^{-9}]$}
\vglue -2.cm
\scalebox{0.47}{
{\mbox{\epsfig{file=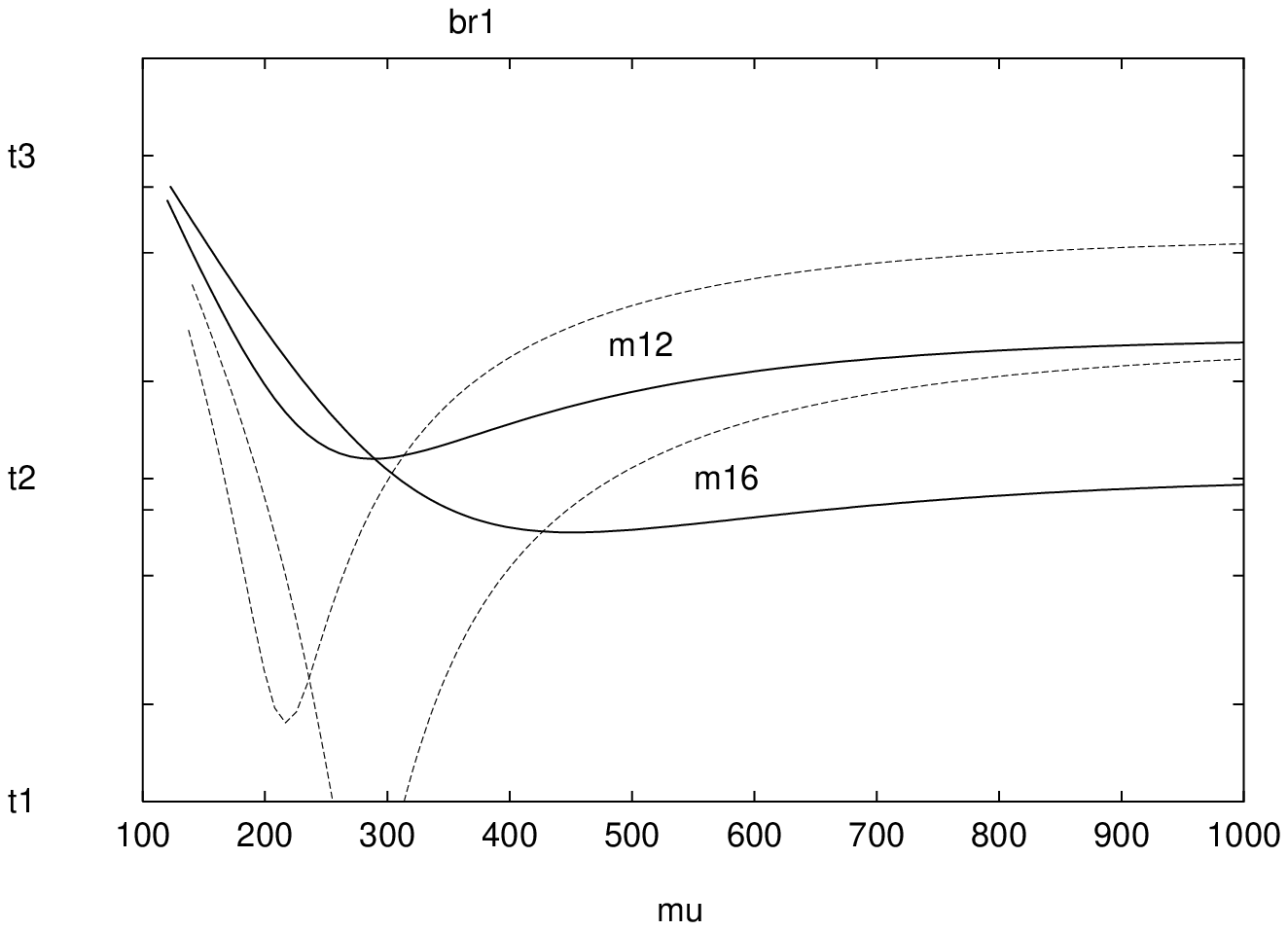}}}}
\vskip -0.8cm
\hglue -0.8cm
\scalebox{0.77}{
{\mbox{\epsfig{file=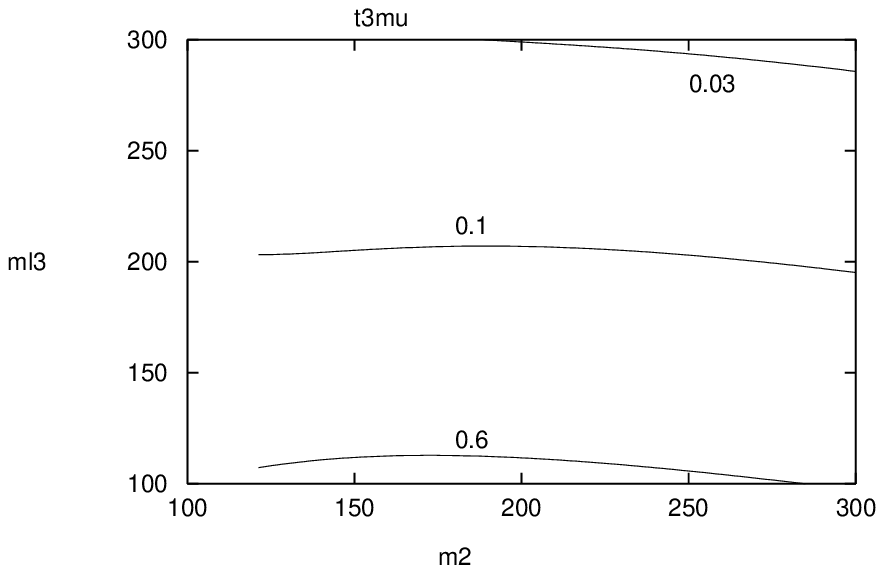}}}
\hglue -3.5cm
{\mbox{\epsfig{file=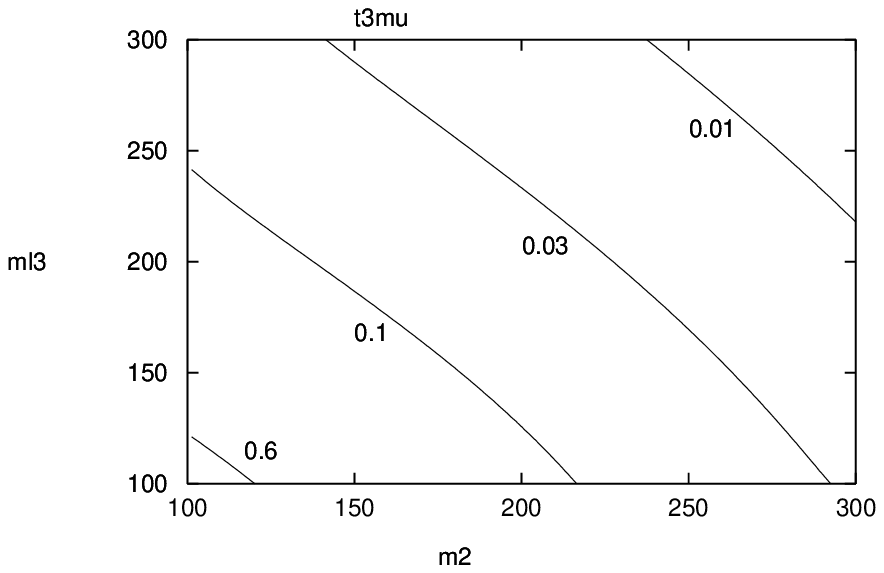}}}}
\vglue -1.4cm
\caption{{
\footnotesize Monopole contribution to $BR(\tau \to \mu \mu\mu)$ 
for $\tanb =3$, $\mt_{L_2}= 
1~{\rm TeV}$, $\theta_L= \pi/4$. 
In the upper panel:   $\mt_{L_3}= 100~{\rm GeV}, \staur=\smur=   
100~{\rm GeV}$, 
$M_2$ as shown, $M_1= 100~(-100)~{\rm GeV}$ for the solid (dashed) lines.  
In the left (right) lower panel: $\mu =  120~(1000)~{\rm GeV}, 
M_1=  100~(-100)~{\rm GeV}$ and $\staur = 100~(500)~{\rm GeV}$.
}}
\label{f7}
\psfrag{mu} {\footnotesize  $\mu$~[GeV]}
\psfrag{m2}{\footnotesize $M_2$~[GeV]}
\psfrag{ml3}{\footnotesize $\begin{array}{c}
\tilde{m}_{L_3} \\
{\rm { [GeV]}}\end{array}$}
\psfrag{sel}{\footnotesize $\begin{array}{c}
\tilde{m}_{\tilde{e}} \\
{\rm { [GeV]}}\end{array}$}
\psfrag{tmue}{\small $BR(\tau\rightarrow \mu e e)_{D^\gamma =0}~[10^{-9}]$}
\vglue -0.5cm 
\scalebox{0.77}{
{\mbox{\epsfig{file=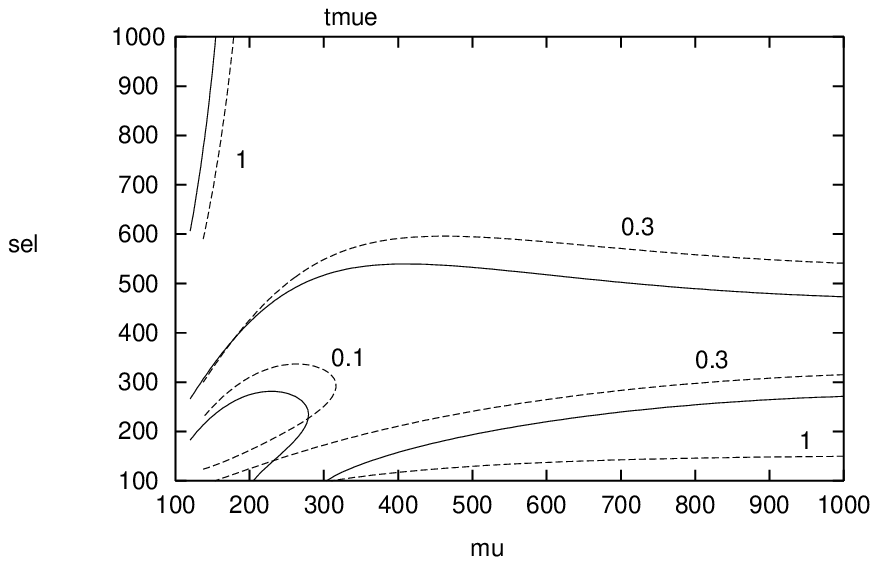}}}}
\vskip -2.cm
\hglue -0.8cm
\scalebox{0.77}{
{\mbox{\epsfig{file=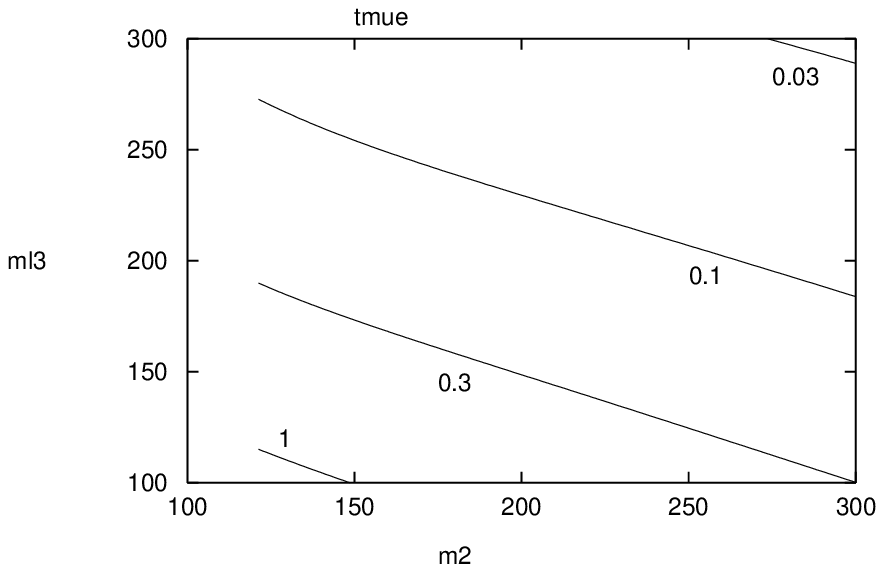}}}
\hglue -3.5cm
{\mbox{\epsfig{file=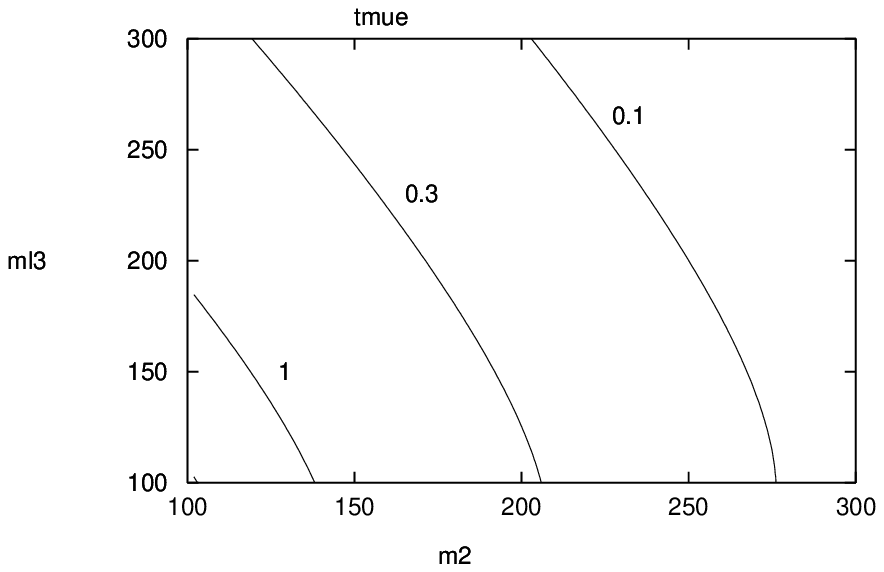}}}}
\vglue -1.4cm
\caption{{\footnotesize 
Monopole contribution to $BR(\tau \to \mu e e)$ 
for $\tanb =3$, $\mt_{L_2}= 
1~{\rm TeV}$, $\theta_L= \pi/4$. 
In the upper panel:   $\mt_{L_3}= 100~{\rm GeV}, \staur=  100~{\rm GeV}$, 
$M_2 = 120~{\rm GeV}$, $M_1= 100~(-100)~{\rm GeV}$ 
for the solid (dashed) lines.  
In the left (right) lower panel: $\mu =  120~(1000)~{\rm GeV}, 
M_1=  100~(-100)~{\rm GeV}$, 
$\mt_{\tilde{e}} =1000~(120)~{\rm GeV}$ and 
$\staur = 100~(500)~{\rm GeV}$.
}}
\label{f8}
\end{center}
\end{figure}
For small $\mu$ we have 
$|A^Z_L| > |C^Z_L|$. For large $\mu$,  $A^Z_L$ is suppressed 
and  $BR(Z\to \mu \tau)$ approaches an asymptotic value 
essentially determined by $C^Z_L$. 
The lower panels also show 
that  $BR(Z\to \mu \tau)$ depends on  $(M_2, \mt_{L_3})$ in a different 
way for small and large  $\mu$. In either case, 
this $BR$ can reach $\sim 10^{-9}$.

In Fig.~\ref{f7} the monopole contribution to 
$BR(\tau \to \mu \mu\mu)$ is drawn as a function of 
$\mu$ (upper panel) and in the $(M_2, \mt_{L_3})$ plane
for small
$\mu$ (left  lower panel) or large $\mu$ (right lower panel).
The monopole coefficients $F^{\mu_{L(R)}}_L$ receive contributions 
of different signs from  $A^Z_L, C^\ga_L$ and $B^{\mu_{L (R)}}_L$. 
The upper panel shows the changes in the  interference pattern, since 
$A^Z_L$ is substantial for small  $\mu$ and  suppressed for  large $\mu$, 
while  $C^\ga_L$ and $B^{\mu_{L(R)}}_L$ are constant.
The lower panels show how the monopole-induced  
$BR(\tau\to \mu\mu\mu)$ depends on  $(M_2, \mt_{L_3})$ 
for small and large  $\mu$. 
We can notice that the maximal monopole  contribution 
to $BR(\tau\to \mu\mu\mu)$ in these examples is comparable to 
the maximal allowed dipole contribution [eq.~(\ref{dom-dgm})], 
so the combination can be around $10^{-9}$ (see also Fig.~\ref{f5}).

In Fig.~\ref{f8} the monopole contribution to 
$BR(\tau \to \mu e e)$ is plotted 
in the $(\mu, \mt_{\tilde{e}})$ plane
(upper panel) and in the $(M_2, \mt_{L_3})$ plane
for small $\mu$ and large $\mt_{\tilde{e}}$ 
(left  lower panel) or large $\mu$ 
and small  $\mt_{\tilde{e}}$ 
(right lower panel).
The contours in the upper panel reflect 
the interference effects among 
the different contributions to $F^{e_{L(R)}}_L$,
 {\it i.e.} $A^Z_L, C^\ga_L$ and $B^{e_{L(R)}}_L$.
The monopole-induced $BR(\tau \to \mu e e)$ can be ${\cal O}(10^{-9})$  
either for small $\mu$ and large $\mt_{\tilde{e}}$ 
 or for large $\mu$ and small  $\mt_{\tilde{e}}$.
In the former case $A^Z_L$ is substantial and  
 box contributions are suppressed, in the latter case the opposite 
situation occurs. 
In both cases there is the constant $C^\ga$ contribution. 
In other regions the $BR$ is smaller than  $10^{-9}$ 
because of mutual cancellations.
The maximal monopole  contribution 
in these examples is comparable to 
the maximal allowed dipole contribution [eq.~(\ref{dom-dge})], 
and  the combination can be a few times  $10^{-9}$ (see also Fig.~\ref{f5}).

\begin{figure}[ht]
\begin{center}
\psfrag{mu} {\footnotesize $\mu$~[GeV]}
\psfrag{t1}{\large $10^{-11}$}
\psfrag{t2}{\large $10^{-10}$}
\psfrag{t3}{ \large $10^{-9}$}
\psfrag{t4}{\large $10^{-8}$}
\psfrag{m2}{\footnotesize $M_2$~[GeV]}
\psfrag{ml3}{\footnotesize $\begin{array}{c}
\tilde{m}_{L_3} \\
{\rm { [GeV]}}\end{array}$}
\psfrag{msq}{\footnotesize $\begin{array}{c}
\tilde{m}_{\tilde{q}} \\
{\rm { [GeV]}}\end{array}$}
\psfrag{tmur}{\small $BR(\tau\rightarrow \mu\rho)_{D^\gamma =0}
~[10^{-9}]$}
\vglue -1.3cm
\scalebox{0.77}{
{\mbox{\epsfig{file=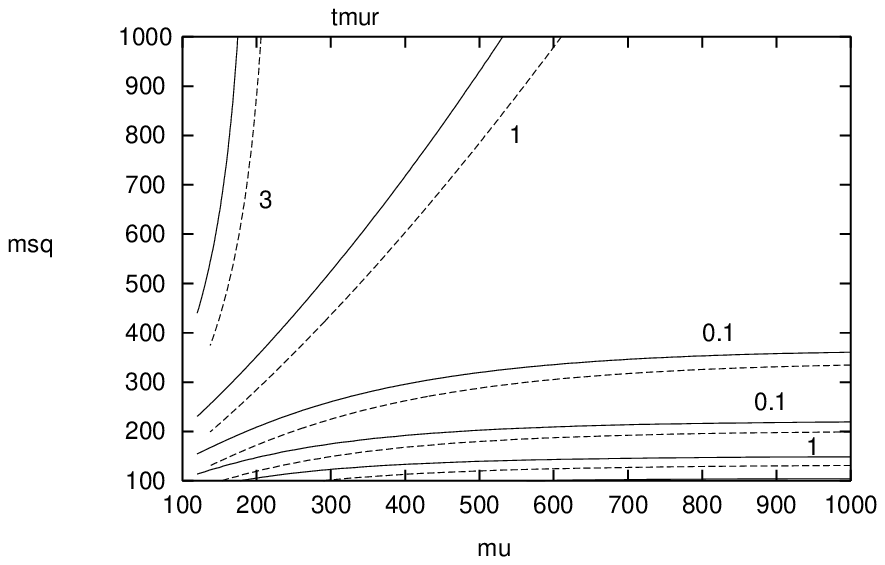}}}
\hglue -3.5cm
{\mbox{\epsfig{file=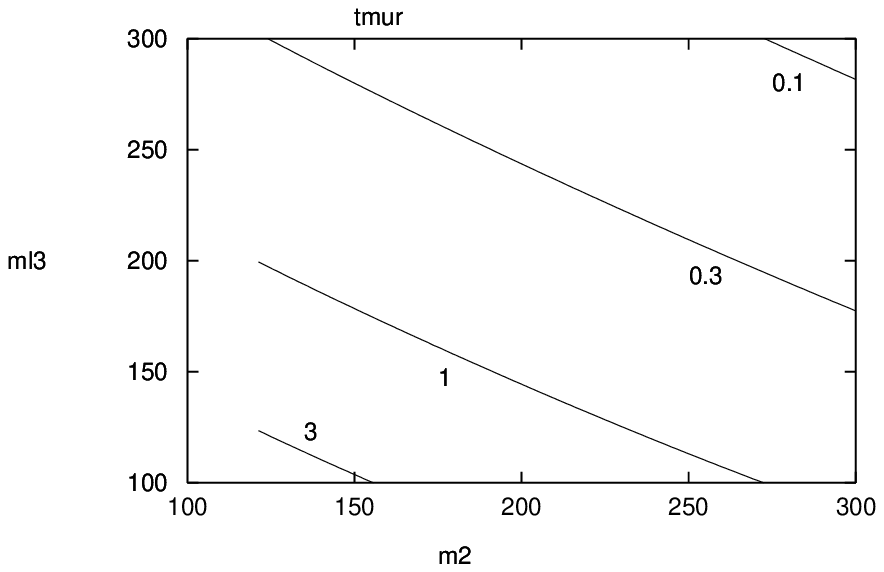}}}}
\vglue -1.2cm
\psfrag{mu} {\footnotesize $\mu$~[GeV]}
\psfrag{t1}{\large $10^{-11}$}
\psfrag{t2}{\large $10^{-10}$}
\psfrag{t3}{ \large $10^{-9}$}
\psfrag{t4}{\large $10^{-8}$}
\psfrag{br}{\Large $BR$~~~~~~~~~~}
\psfrag{m2}{\footnotesize $M_2$~[GeV]}
\psfrag{ml3}{\footnotesize $\begin{array}{c}
\tilde{m}_{L_3} \\
{\rm { [GeV]}}\end{array}$}
\psfrag{msq}{\footnotesize $\begin{array}{c}
\tilde{m}_{\tilde{q}} \\
{\rm { [GeV]}}\end{array}$}
\psfrag{tmupi}{\small $BR(\tau\rightarrow \mu\pi)~[10^{-9}]$}
\vglue -0.7cm
\scalebox{0.77}{
{\mbox{\epsfig{file=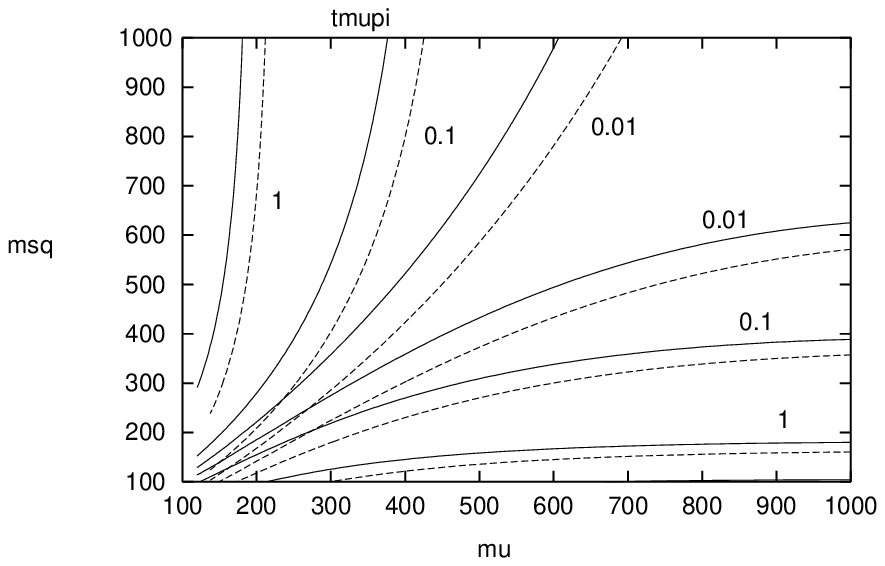}}}
\hglue -3.5cm
{\mbox{\epsfig{file=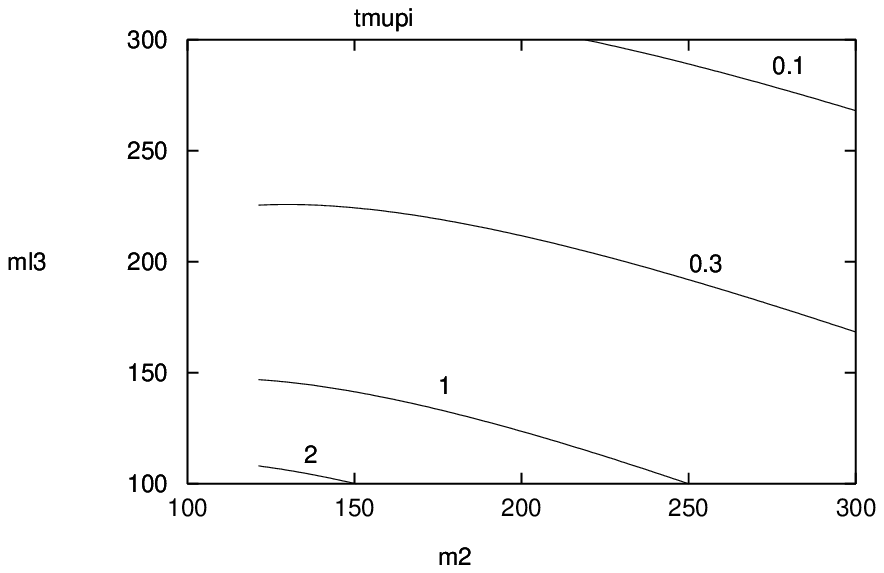}}}}
\vglue -1.2cm
\caption{{\footnotesize 
Monopole contribution to $BR(\tau \to \mu \rho)$ (upper panels)
and $BR(\tau \to \mu \pi)$ (lower panels), 
for $\tanb =3$, $\mt_{L_2}= 
1~{\rm TeV}$, $\theta_L= \pi/4$. 
In the left panels:   $\mt_{L_3}= 100~{\rm GeV}, \staur=  100~{\rm GeV}$, 
$M_2 = 120~{\rm GeV}$, $M_1= 100~(-100)~{\rm GeV}$ 
for the solid (dashed) lines.  
In the right  panels: $\mu =  120~{\rm GeV}, 
M_1=  100~{\rm GeV}$, $\mt_{\tilde{q}} =1~{\rm TeV}$ and 
$\staur = 100~{\rm GeV}$.
}}
\label{f9}
\end{center}
\end{figure}

In Fig.~\ref{f9}  we depict the monopole contribution to 
$BR(\tau \to \mu \rho)$ (upper panels) and $BR(\tau \to \mu \pi)$ 
(lower panels), either  in the  $(\mu, \mt_{\tilde{q}})$ plane 
(left panels) or in the $(M_2, \mt_{L_3})$ plane 
(right panels). 
Also for these processes cancellation effects 
are visible (see left panels).
Indeed, box contributions interfere destructively 
with $A^Z_L$ and $C^\ga_L$. 
The  cancellation regions for 
$\tau \to \mu \rho$ and $\tau \to \mu \pi$ are somewhat 
different, also because $C^\ga_L$ only contributes 
to $\tau \to \mu \rho$.
Outside the cancellation regions 
the monopole-induced $BR(\tau \to \mu \rho)$ and 
$BR(\tau \to \mu \pi)$ can exceed  $10^{-9}$. 
We recall that $\tau \to \mu \rho$, at variance with 
$\tau \to \mu \pi$, also gets a dipole contribution. 
Notice that eq.~(\ref{btmuro}) can be written as 
$BR(\tau \to \mu \rho) \simeq BR^{F}_{(D=0)} + BR^{D}_{(F=0)} 
\pm 1.5 \sqrt{BR^{F}_{(D=0)} BR^D_{(F=0)} }$, where 
$BR^{F}_{(D=0)}$ ($BR^{D}_{(F=0)}$) denotes the pure 
monopole (dipole) contribution.
For instance, if we combine a value 
$BR^F_{(D=0)}\sim 3\times 10^{-9}$ with the  maximal 
allowed  $BR^D_{(F=0)}\sim 0.8\times 10^{-9}$ [eq.~(\ref{dom-dgr})], 
we obtain $BR(\tau \to \mu \rho) \sim 6\times 10^{-9} $ or 
$1.5\times 10^{-9}$, depending on the 
interference sign (see also Fig.~\ref{f5}).
We have not shown the corresponding examples for 
$BR(\tau \to \mu \eta)$ and $BR(\tau \to \mu \eta')$, 
since they exhibit a similar pattern to that of 
$BR(\tau \to \mu \pi)$, though the maximal achievable values 
are somewhat smaller.

\begin{figure}[p]
\begin{center}
\psfrag{mu} {\footnotesize $|\mu|$~[TeV]}
\psfrag{m2}{\footnotesize $\begin{array}{c}
|M_2| \\
 {\rm { [TeV]}}
\end{array}
$}
\psfrag{br1}{\tiny $BR(A\rightarrow \mu\tau)~[10^{-5}]$}
\psfrag{br2}{\tiny $BR(\tau\rightarrow \mu\mu\mu)~[10^{-8}]$}
\psfrag{br3}{\tiny $BR(\tau\rightarrow \mu\gamma)~[10^{-7}]$}
\vglue -6.3cm
\hglue -2.0cm
\scalebox{1.5}{
{\mbox{\epsfig{file=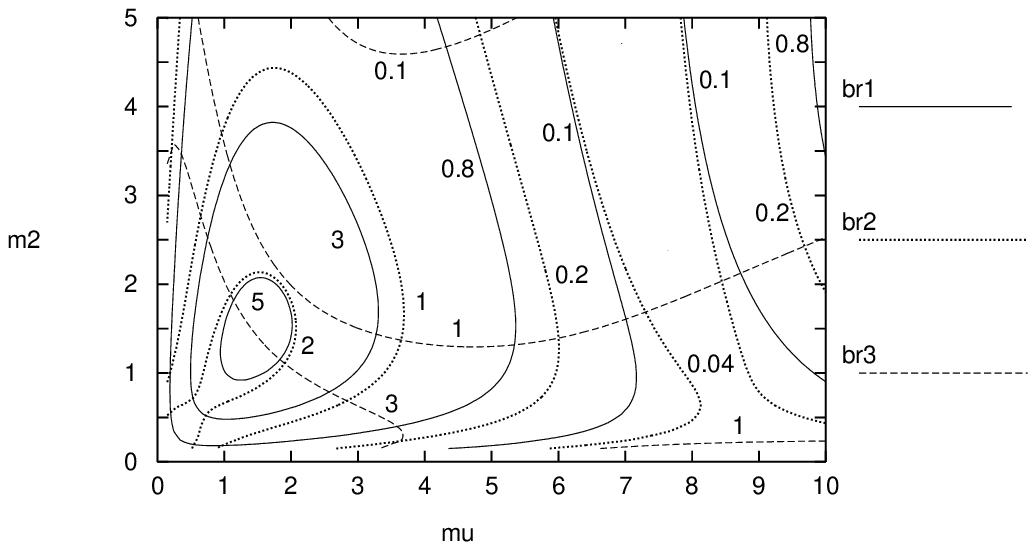}}}}
\vglue -4.2cm
\hglue -2.0cm
\scalebox{1.5}{
{\mbox{\epsfig{file=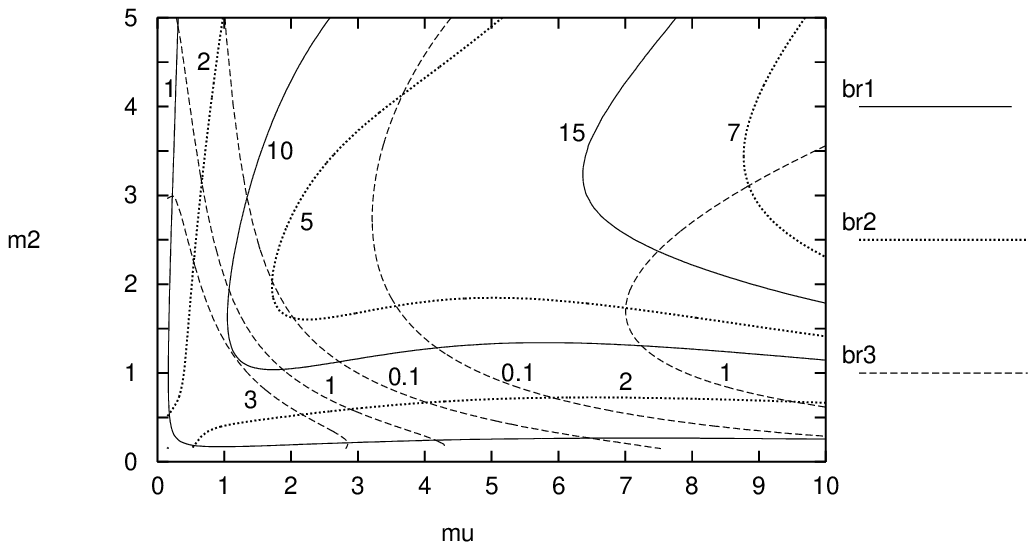}}}}
\vglue -2.cm
\caption{{\ftsz Branching ratios of $\tau \to \mu \ga$ (dashed lines), 
$\tau \to \mu \mu \mu$ (dotted lines) and  $A\to \mu \tau$ (solid lines) 
for  $\tanb =40$, $\theta_L= \pi/4$, 
$\mt_{L_2}= 3~{\rm TeV}$, $\mt_{L_3}= 1~{\rm TeV}$, 
$|M_1|=\frac12 |M_2|$. In the upper (lower) panel ${\rm sign} (M_1 M_2)$ 
is positive (negative) and $\staur =\smur =  2.5~{\rm TeV} ~ (2~{\rm TeV})$. 
We have also fixed $m_A = 100~{\rm GeV}$ (which affects $\tau \to 3 \mu$) 
and $BR(A\to \tau \tau)= 0.1$ (which affects $A\to \mu \tau$). 
}}
\label{f10}
\end{center}
\end{figure}

\subsection{(LFV)$_L$ with large $\tanb$ and 
large masses \label{sllarge}}

Now we discuss the scenario with large $\tanb$. 
In this case we have a strong enhancement of the 
$D^{\ga (b,c)}_L$ dipole coefficients, and Higgs-$\mu$-$\tau$ 
effective operators become relevant. 
Two strategies can be envisaged to deal with the former issue. 
In principle, the parameters can be chosen in such a way 
that cancellations occur (as illustrated in the previous section)  
and keep the total  $D^\ga_L$ dipole  coefficient 
below the bound. 
However, a significant fine tuning  is required,
especially if $M_2$ and $\mt_{L_3}$ are light. 
Alternatively, the individual dipole contributions can be 
kept below the bound by taking large   
mass parameters. This is the option we pursue here. 
In this case, the monopole contributions 
are suppressed, while the Higgs ones are not (for small $m_A$), 
because the $\Delta$ coefficients are 
insensitive to the overall mass scale.  
Hence, at large $\tanb$ and large masses (except for $m_A$)   
the following picture emerges.
\begin{description}
\item
{\it i)} $Z\to \mu \tau$ is strongly suppressed (even including  
$D^Z$ contributions). 
\item 
{\it ii)} The decays $ \tau \to \mu ee$ and 
$\tau \to \mu \rho$ are generically 
dipole dominated, so they are correlated to 
 $\tau \to \mu \ga$ through 
(\ref{dg1}), (\ref{dg3}) [and the bounds   
(\ref{dom-dge}), (\ref{dom-dgr}) hold].
\item
{\it iii)}   
The dominant contributions to $BR(\tau \to \mu \mu \mu)$   
[eq.~(\ref{bt3mu})]  
are induced by the dipole term $D^\ga_L$ and the Higgs-mediated 
term $\delta F^{\mu_L}_R$, proportional 
to  $\Delta_L$  [eq.~(\ref{dflr})]. 
Therefore, $\tau \to \mu \mu \mu$ is correlated to both 
$\tau \to \mu \ga$ and $A\to \mu \tau$ \cite{BR}.
\item
{\it iv)}   
The decay $\tau \to \mu \eta$ is dominated by 
the Higgs-mediated terms $\delta F^{\eta, 8}_L, ~\delta F^{\eta, 0}_L$  
[eqs.~(\ref{dfleta8}), (\ref{dfleta0})]. 
If   $\tau \to \mu \mu \mu$ is mostly induced by  Higgs-exchange, 
then these processes are correlated as in eq.~(\ref{eta3mu}). 
Higgs-exchange is also relevant for the decays  $\tau \to \mu \pi$ 
and  $\tau \to \mu \eta'$
through the contribution $\delta F^{\pi}_L$  [eq.~(\ref{dflpi})] and
 $\delta F^{\eta', 8}_L, ~\delta F^{\eta', 0}_L$ 
[analogous to eqs.~(\ref{dfleta8}), (\ref{dfleta0})], 
respectively. 
\end{description}
 
Let us focus on the processes 
$\tau \to \mu\ga, ~ \tau \to 3\mu, ~A\to \mu \tau$.
The behaviour of these decays is illustrated 
in Fig.~\ref{f10} 
in the plane ($|\mu|, |M_2|$) for $\tanb =40$.  Two examples 
are shown, where ${\rm sign}(M_1 M_2)$ is either positive 
(upper panel) or negative (lower panel). 
The relative sign between $\mu$ and $M_2$ is immaterial for 
large $\tanb$, while that between $M_1$ and $M_2$ matters 
because it determines the interference between $D^{\ga(b)}_L$ and 
$D^{\ga(c)}_L$, as well as that between $\Delta^{(b)}_L$ and 
$\Delta^{(c)}_L$.
In the upper panel  [${\rm sign}(M_1 M_2)>0$] 
the interference is  constructive for  
the  $D^{\ga}_L$ components and destructive for 
the $\Delta_L$ ones. Indeed, we can see a cancellation region 
for $\tau \to 3 \mu$ and $A\to \mu \tau$ \cite{BR}. 
In the lower panel [${\rm sign}(M_1 M_2)<0$]  
the opposite situation occurs and we can notice 
a cancellation region for $\tau \to  \mu\ga$.
In both examples, the contours of  $\tau \to 3 \mu$ and $A\to \mu \tau$ 
follow a very similar pattern and 
are correlated according to eq.~(\ref{crl}). 
This occurs in most of the parameter space shown in Fig.~\ref{f10} 
where  $\tau \to 3 \mu$ is indeed essentially dominated by 
the $\Delta_L$ contribution.
Deviations occur in the regions where the $D^\ga_L$ contribution 
to $\tau \to 3 \mu$  
is comparable to  the $\Delta_L$ one (or larger).
Regarding the numerical values, in the first example 
$BR(A \to \mu \tau)$ and  $BR(\tau \to 3\mu)$ 
can reach $5 \times 10^{-5}$ and $2\times 10^{-8}$,  
respectively.
In the second example 
$BR(A \to \mu \tau)$ and  $BR(\tau \to 3\mu)$ 
can be larger than  
$10^{-4}$ and $5\times 10^{-8}$, respectively.
Notice that in Fig.~\ref{f10} 
we have fixed $m_A= 100~{\rm GeV}$. 
Since the  Higgs-mediated contribution to $BR(\tau \to 3\mu)$  
scales as  $1/m_A^{4}$, the values in both  examples 
would be accordingly reduced for larger $m_A$.
Finally, we recall that $BR(\tau \to \mu \eta)$, 
which can be inferred  
by eq.~(\ref{eta3mu}), is generically 
larger\footnote{
As already emphasized, the value of the ratio (\ref{eta3mu})
is quite sensitive to the parameters $\xi_s, \xi_b$, which 
depend on $\mu$, the gluino mass $M_3$ 
and third  generation squark mass parameters. 
Incidentally, $\xi_b$ also affects 
the total $A$ width (through the main channel $A\to b\ov{b}$) 
and hence   $BR(A\to \tau \tau)$ and $BR(A\to \mu \tau)$.} 
than $BR(\tau \to 3\mu)$ and can approach  
the experimental bound (\ref{e-mueta}). 
In this case $BR(\tau \to \mu \pi), ~ BR(\tau \to \mu \eta')$ can
reach $10^{-9}$ [eqs.~(\ref{pieta}), (\ref{etaetap})].

\section{Large (LFV)$_R$: numerical analysis \label{lfvr}}
In this section we perform a numerical 
analysis in the case of large  (LFV)$_R$, assuming  
vanishing  (LFV)$_L$, {\it i.e.} $\mt^2_{L \mu \tau} = A^L_{\mu \tau}= 0$.   
All operator coefficients depend on 
$\mt_{R_\al}$,  $\theta_R$ and  the bino mass $M_1$.
Some coefficients also depend on additional parameters.
In particular, $A^{Z(a)}_R$ and $D^{\ga (b)}_R$ depend on $\mu$ and 
$\beta$; $D^{\ga (c)}_R$ depends on $\mu$, $\beta$, $\mt_{\tilde{\tau}_L}$, 
$A_\tau, A^R_{\mu\tau}$;   $B^{f}_R$ depends on $\mt_{\tilde{f}}$.

The lightest  eigenvalue of
$\tilde{{\cal M}}^2_R$ is  conventionally chosen to be $\mt^2_{R_3}$ 
(although our formulae in Appendix do not depend on such a choice).
To enhance (LFV)$_R$, in all our numerical 
examples we will take  maximal mixing, $\theta_R=\pi/4$ ({\it i.e.} 
$\mt^2_{R \mu \mu}= \mt^2_{R \tau \tau} $  
in  $\tilde{{\cal M}}^2_R$), and  widely split eigenvalues. 
Again the mass parameters will be varied in such a way that 
charged sparticle masses be $\gsim 100~{\rm GeV}$.   
The parameter $M_1$ will be allowed to reach smaller values 
without being in conflict with LEP bounds on neutralino 
production ($e^+ e^- \to \chi^0_1\chi^0_{2 (3)}$) 
and on non-standard contributions to the invisible $Z$ width 
($Z\to \chi^0_1\chi^0_1$). 
We also recall that in the (LFV)$_R$ case the parameter 
$M_2$ does not play 
a direct role and that we do not impose a specific relation
between $M_1$ and $M_2$, hence small $M_1$ and large $M_2$ 
can coexist.   
It is worthwhile to explore low  $M_1$ values  because 
the monopole coefficients are enhanced.
For simplicity, as assumed in Section~\ref{lfvl}, 
we take a common mass  
$\mt_{\tilde{e}}$ for selectrons and a common mass 
$\mt_{\qt}$ for first and second generation squarks.

\subsection{(LFV)$_R$  with small $\tanb$} 

In analogy with the (LFV)$_L$ case, first we focus on the 
dipole operators and consider a scenario with small $\tanb$. 
The present bound (\ref{e-rad}) on   $\tau \to \mu \ga$ translates into 
$|D^\ga_R| \lsim 5\times 10^{-9}~ {\rm GeV}^{-2}$. 
In Fig.~\ref{fr1} we show contours of $D^{\ga(a)}_R$ 
(left  panel) and $D^{\ga(b)}_R$  (right panel)
in the  plane  $(M_1, \mt_{R_3})$  with $\theta_R =\pi/4$ and 
$\mt_{R_2} = 1~{\rm TeV}$.
 \begin{figure}[ht]
\begin{center}
\psfrag{m1}{\footnotesize $M_1$~[GeV]}
\psfrag{mr3}{\footnotesize $\begin{array}{c}
\tilde{m}_{R_3} \\
{\rm { [GeV]}}\end{array}$}
\psfrag{c1}{\footnotesize 50}
\psfrag{c2}{\footnotesize 30}
\psfrag{c3}{\footnotesize 20}
\psfrag{s1}{\footnotesize 7}
\psfrag{s2}{\footnotesize 5}
\psfrag{s3}{\footnotesize 3}
\psfrag{D}{\small  $D^{\gamma (a)}_R~{[10^{-9}\cdot {\rm GeV}^{-2}]}$}
\psfrag{Db}{\small  $D^{\gamma (b)}_R~{[10^{-9}\cdot {\rm GeV}^{-2}]}$}
\vskip -1.5cm
\hglue -0.8cm
\scalebox{0.77}{
{\mbox{\epsfig{file=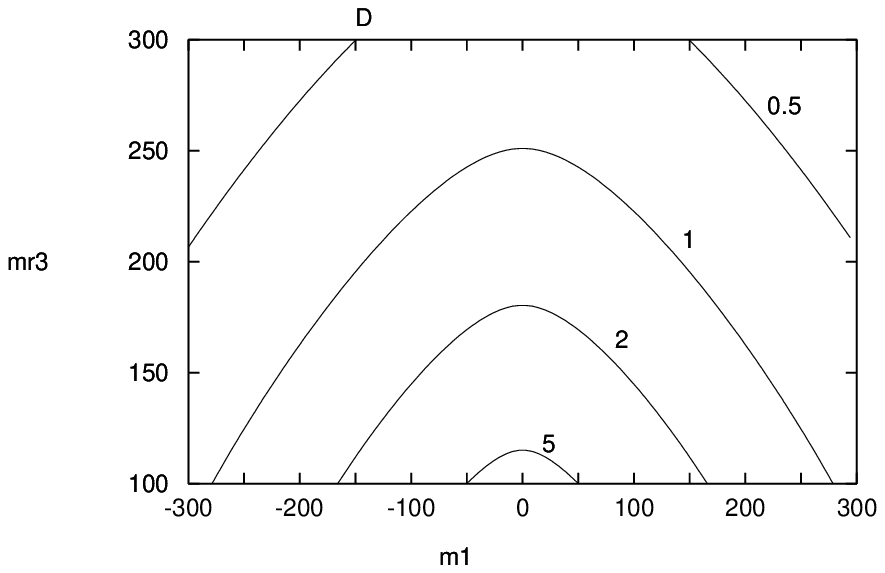}}}
\hglue -3.5cm
{\mbox{
\epsfig{file=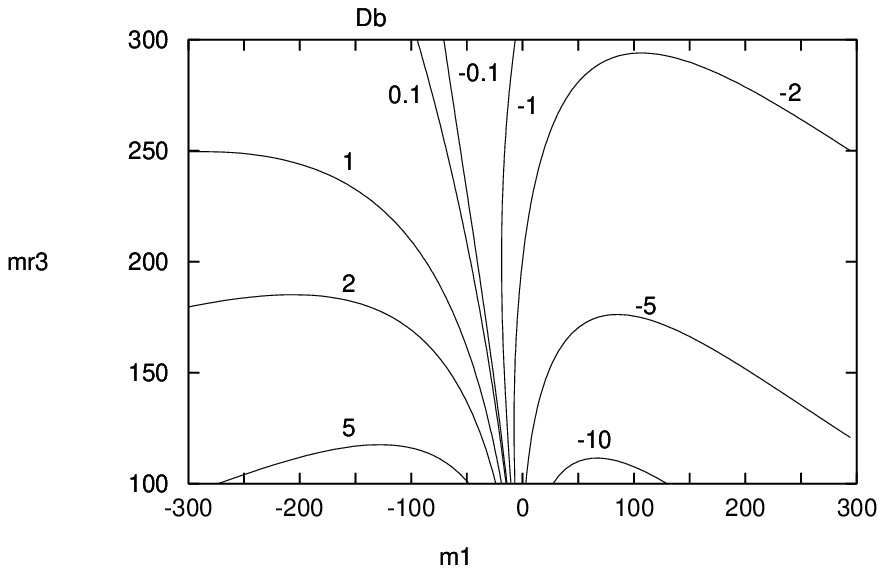}}}}
\vglue -1.2cm
\caption{{\footnotesize Contours of 
$D^{\gamma(a)}_R$ (left panel) and $D^{\gamma(b)}_R$ (right panel) 
for  $\mt_{R_2} = 1~{\rm TeV}$ and $\theta_R=\pi/4$. 
To determine  $D^{\gamma(b)}_R$ we have also fixed 
 $\tanb =3$ and $\mu =150~{\rm GeV}$.
}}
\vglue -0.6cm
\label{fr1}
\end{center}
\end{figure}
The  $D^{\ga(b)}_R$ plot refers to the choice  $\tanb =3$ and 
 $\mu =150~{\rm GeV}$.
We can see that  $D^{\ga(a)}_R$ and $D^{\ga(b)}_R$
are comparable to or even larger than the bound 
in some regions of the parameter space, and can interfere 
with each other destructively or constructively.
Also $D^{\ga (c)}_R$ may or may not exceed the bound, depending on 
the range of the extra parameters $\mt_{\tilde{\tau}_L}$, 
$A_\tau$, $A^R_{\mu\tau}$.  Anyway, 
mutual cancellations involving all three contributions
can bring the total dipole coefficient  
$D^{\ga }_R$ below the bound.
\begin{figure}[ht]
\begin{center}
\psfrag{atau}{\footnotesize $A_\tau$~[TeV]}
\psfrag{ml3}{\footnotesize $\begin{array}{c}
\tilde{m}_{\tilde{\tau}_L} \\
 {\rm { [GeV]}}
\end{array}
$}
\vglue -2.4cm
\scalebox{1.10}{
{\mbox{\epsfig{file=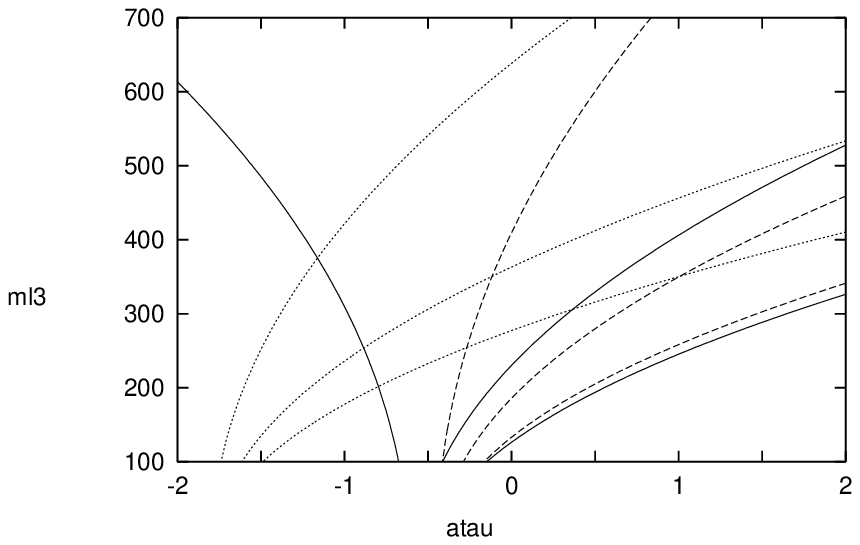}}}}
\vglue -1.8cm
\caption{{\ftsz 
$D^\gamma_R$ contours for $\tanb=3, \theta_R=\pi/4,
\mt_{R_2} =1~{\rm TeV}, 
A^R_{\mu\tau}=0$ and  three choices of $(\mt_{R_3}, \mu, M_1)$ in GeV: 
(100,150,50) (dashed),
(150,200,100) (solid),
(100,600,-50) (dotted). 
For each example the two external lines correspond to  
 $|D^{\ga }_R| = 5 \times 10^{-9}~
{\rm GeV}^{-2}$ and  the middle one  to $D^{\ga }_R=0$.
}}
\vglue -0.6cm
\label{fr2}
\end{center}
\end{figure}
This is illustrated in  Fig.~\ref{fr2}, where we plot   
contours of $D^{\ga }_R$ $(\pm 5\times 10^{-9}~{\rm  GeV}^{-2}, 0)$  
in the plane $(A_\tau, 
\mt_{\tilde{\tau}_L})$ for $\tanb=3, \mt_{R_2} =1~{\rm TeV}, 
A^R_{\mu\tau}=0$ and three choices of $(\mt_{R_3}, \mu, M_1)$.
\begin{figure}[p]
\begin{center}
\psfrag{atau}{\footnotesize $A_\tau$~[GeV]}
\psfrag{br}{\footnotesize $BR$}
\psfrag{ml}{\footnotesize $\tilde{m}_{\tilde{\tau}_L}$~[GeV]
}
\psfrag{t0}{\scriptsize $10^{-10}$}
\psfrag{t1}{\scriptsize $10^{-9}$}
\psfrag{t2}{\scriptsize $10^{-8}$}
\psfrag{t3}{\scriptsize $10^{-7}$}
\psfrag{t4}{\scriptsize $10^{-6}$}
\psfrag{br1}{\footnotesize $\tau\rightarrow \mu\gamma$}
\psfrag{br2}{\footnotesize $Z\rightarrow \mu\tau$}
\psfrag{br3}{\footnotesize $\tau\rightarrow \mu e e$}
\psfrag{br4}{\footnotesize $\tau\rightarrow 3 \mu$}
\psfrag{br5}{\footnotesize $\tau\rightarrow \mu\rho$}
\psfrag{br6}{\footnotesize $\tau\rightarrow \mu\pi$}
\psfrag{exp}{\scriptsize $3\times 10^{-7}$}
\vglue -1.0cm
\scalebox{1.0}{
{\mbox{\epsfig{file=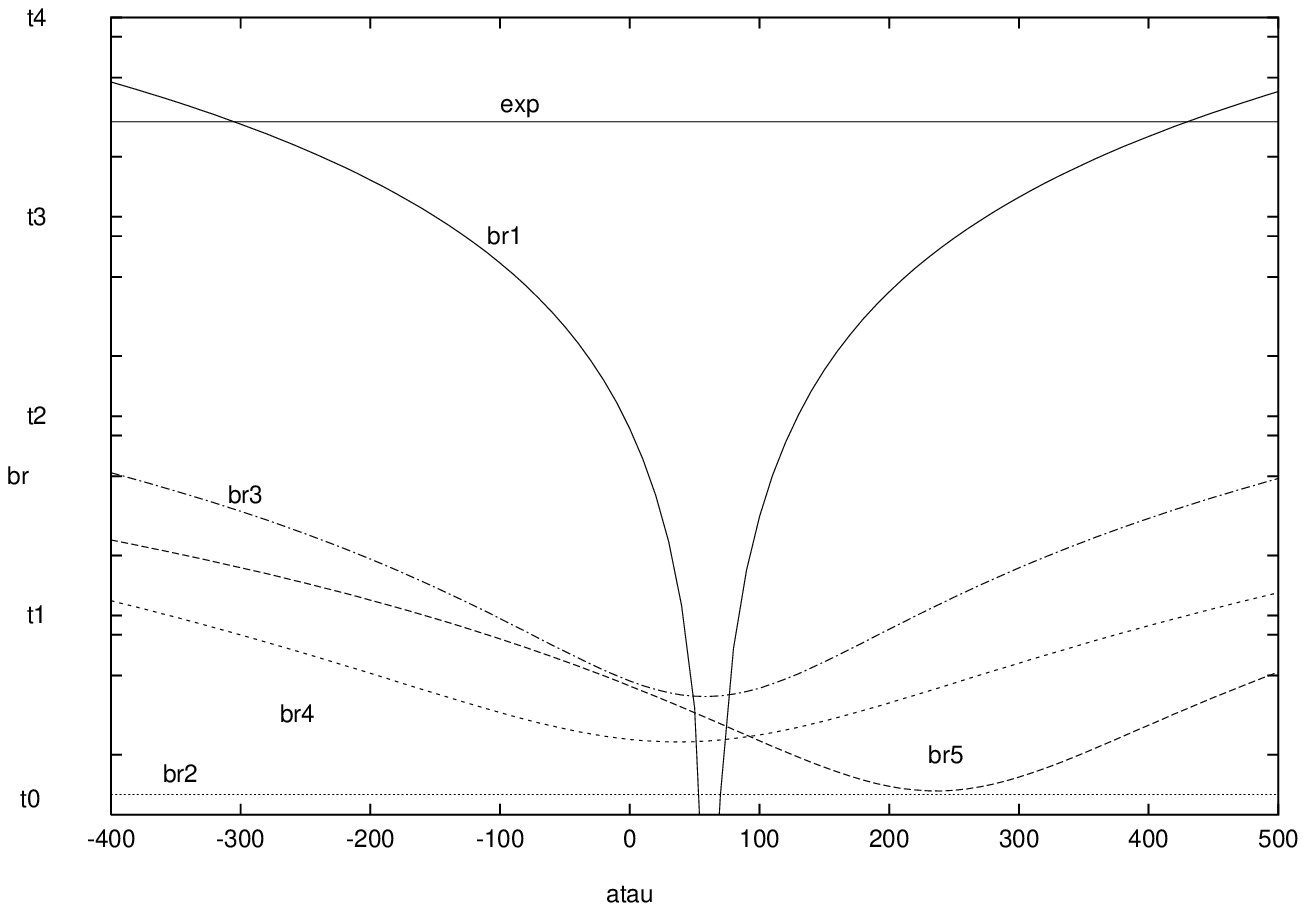}}}}

\vglue 0.8cm
\scalebox{1.0}{
{\mbox{\epsfig{file=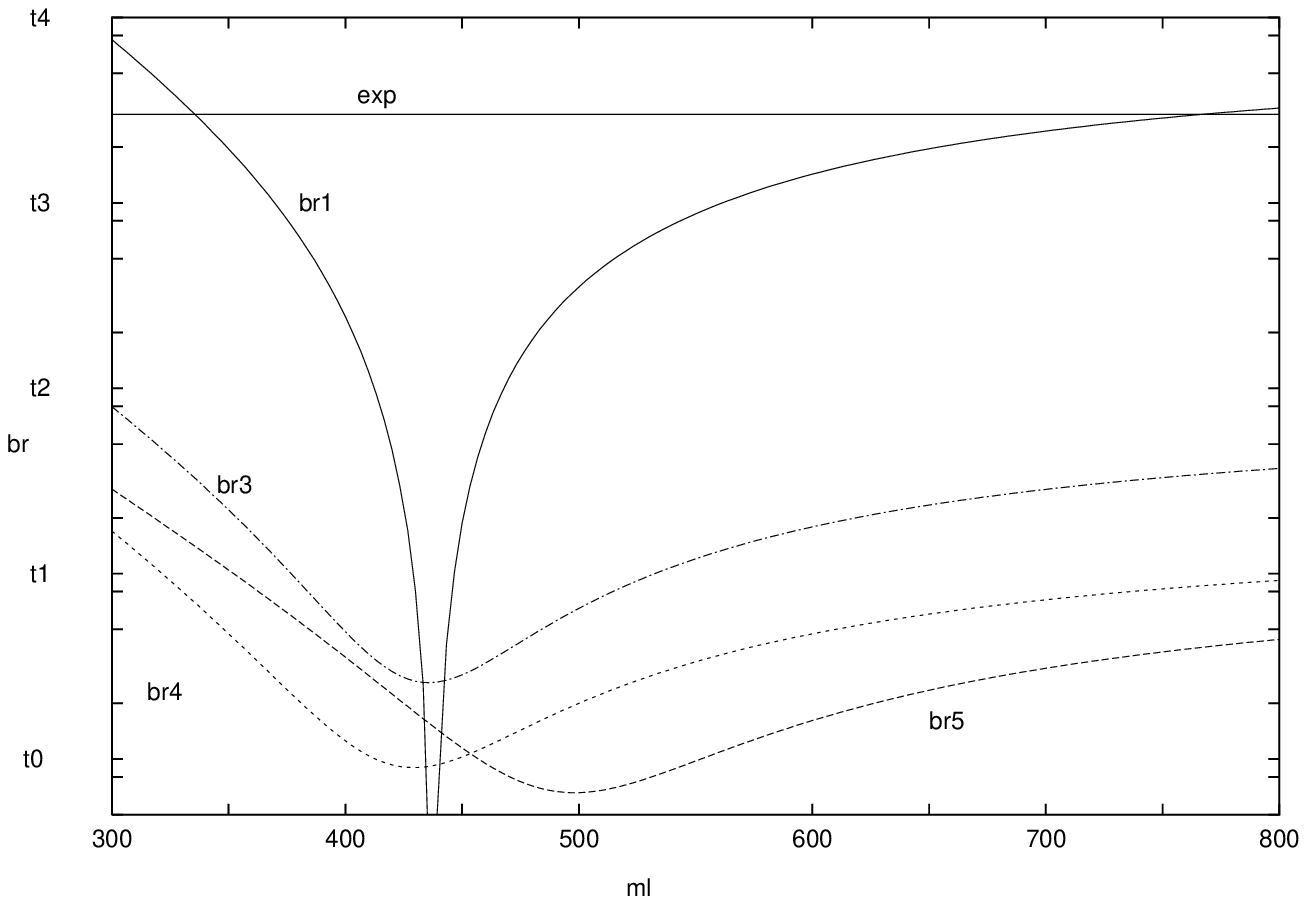}}}}
\vglue 0.5cm
\caption{{\ftsz Branching ratios of  
LFV decays for $\tanb=3$, $\mt_{R_2}= 1~{\rm TeV}$, 
 $\mt_{R_3}= 100~{\rm GeV}$, 
$\theta_R= \pi/4$, $A^R_{\mu \tau}=0$ and 
$\mt_{\tilde{e}}= 100~{\rm GeV}, \mt_{\tilde{q}}= 200~{\rm GeV}$.
In the upper panel, the remaining parameters are: 
$\mu =150~{\rm GeV}$, $M_1=50~{\rm GeV}$, 
$\mt_{\tilde{\tau}_L}= \mt_{\tilde{\mu}_L} = 200~{\rm GeV}$. 
In the lower panel: 
$\mu =800~{\rm GeV}$, $M_1=-50~{\rm GeV}$, 
$A_\tau=0$ and 
$\mt_{\tilde{\mu}_L}=\mt_{\tilde{\tau}_L}$. 
The solid horizontal line indicates the present bound on 
$BR(\tau\to \mu\ga)$. 
In the upper  example, 
$BR(\tau \to \mu \pi)\sim 2\times 10^{-11},~
BR(\tau \to \mu \eta)\sim 2\times 10^{-12},~
BR(\tau \to \mu \eta')\sim 4\times 10^{-11}$  
 and   $BR(Z \to \mu \tau)\sim 10^{-10}$.
 In the lower  example,
$BR(\tau \to \mu \pi)\sim 3\times 10^{-12},~
BR(\tau \to \mu \eta)\sim 4\times 10^{-12},~
BR(\tau \to \mu \eta')\sim  10^{-11}$
and   $BR(Z \to \mu \tau)\sim  2\times 10^{-11}$.  
}}
\label{fr3}
\end{center}
\end{figure}
\begin{figure}[ht]
\begin{center}
\psfrag{mu} {\Large $|\mu|$~[GeV]}
\psfrag{t1}{\large $10^{-11}$}
\psfrag{t2}{\large $10^{-10}$}
\psfrag{t3}{ \large $10^{-9}$}
\psfrag{t4}{\large $10^{-8}$}
\psfrag{m1}{\footnotesize $|M_1|$~[GeV]}
\psfrag{mr3}{\footnotesize $\begin{array}{c}
\tilde{m}_{R_3} \\
{\rm { [GeV]}}\end{array}$}
\psfrag{m25}{\large $|M_1|=25$~GeV}
\psfrag{m50}{\large $50$~GeV}
\psfrag{m100}{\large $100$~GeV}
\psfrag{br1}{\Large $BR(Z\rightarrow \mu\tau)$}
\psfrag{zmu}{\small $BR(Z\rightarrow \mu\tau)~[10^{-9}]$}
\hglue -7.8cm
\scalebox{0.47}{
{\mbox{\epsfig{file=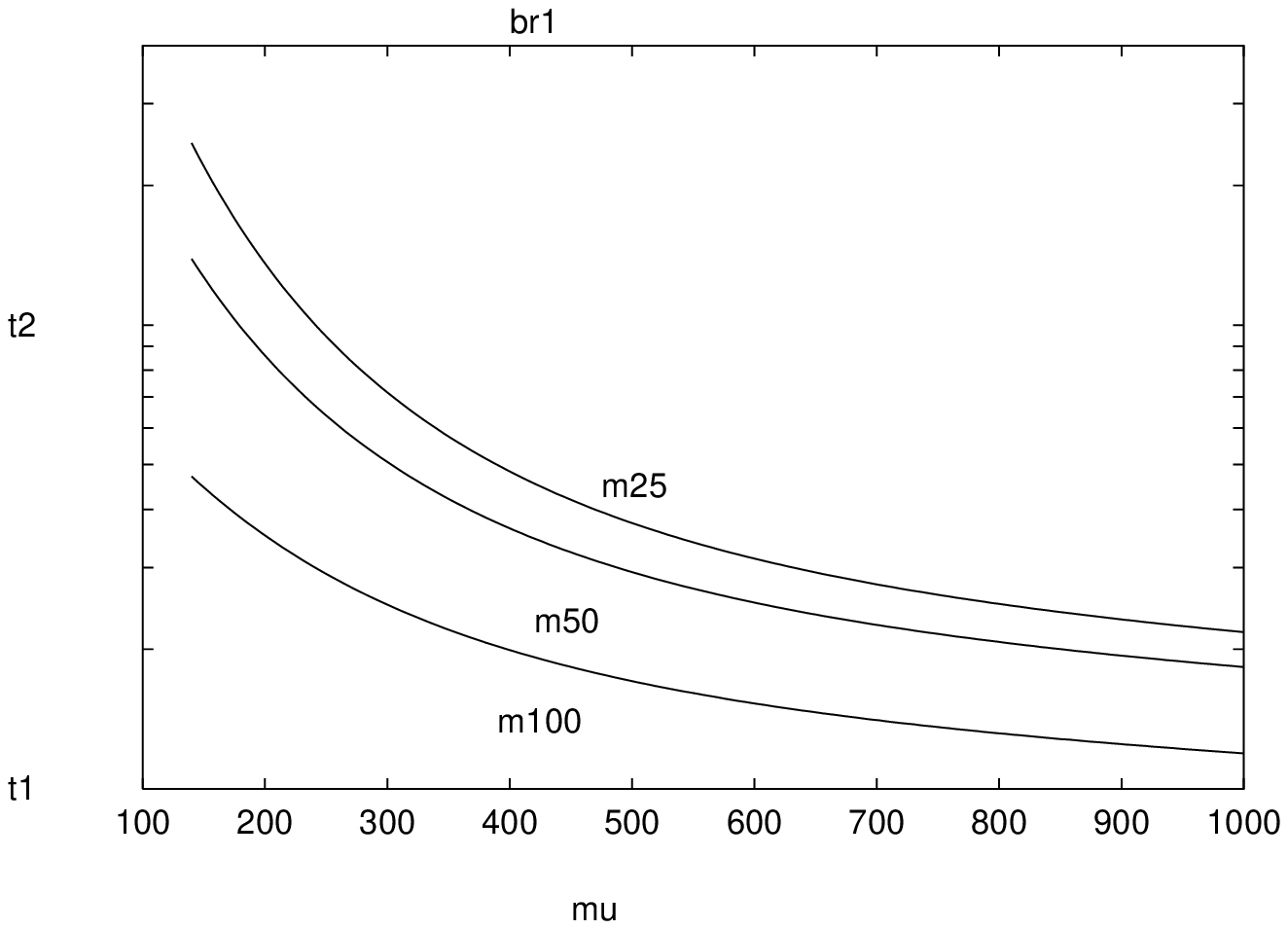}}}}
\vskip -5.5cm
\hglue 6.4cm
\scalebox{0.77}{
{\mbox{\epsfig{file=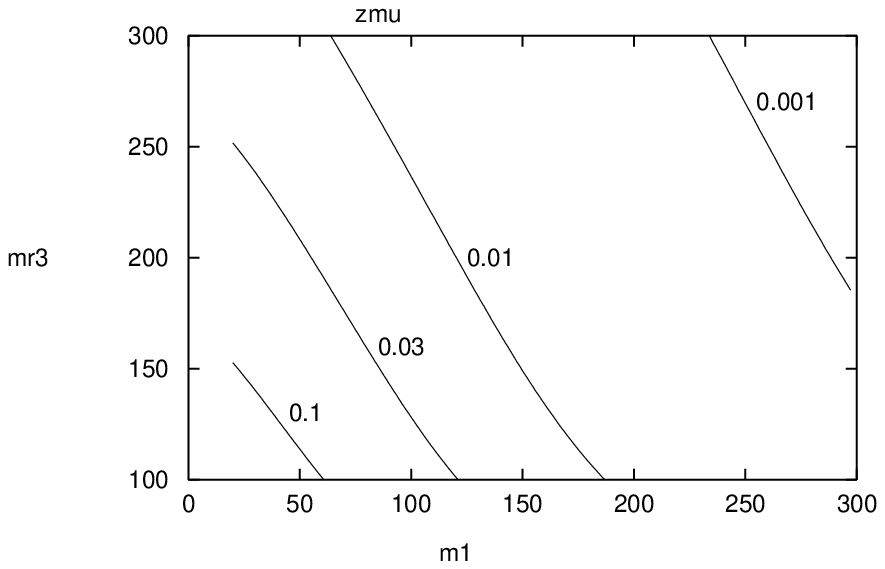}}}}
\vglue -1.2cm
\caption{{\footnotesize $BR(Z\to \mu \tau)$ for $\tanb =3$, $\mt_{R_2}= 
1~{\rm TeV}$, $\theta_R= \pi/4$ and  $\staul=  300~{\rm GeV}$.
In the left panel:   $\mt_{R_3}= 100~{\rm GeV}$ and 
$|M_1|$ as shown.  
In the right  panel: $|\mu| =  150~{\rm GeV}$. 
}}
\label{fr4}
\end{center}
\end{figure}
\begin{figure}[p]
\begin{center}
\psfrag{mu} {\Large $|\mu|$~[GeV]}
\psfrag{t2}{\large $10^{-10}$}
\psfrag{t3}{ \large $10^{-9}$}
\psfrag{m1}{\footnotesize $|M_1|$~[GeV]}
\psfrag{mr3}{\footnotesize $\begin{array}{c}
\tilde{m}_{R_3} \\
{\rm { [GeV]}}\end{array}$}
\psfrag{m25}{\large $|M_1|=25$~GeV}
\psfrag{m50}{\large $50$~GeV}
\psfrag{m100}{\large $100$~GeV}
\psfrag{br1}{\Large $BR(\tau\rightarrow \mu\mu\mu)_{D^\gamma=0}$}
\psfrag{t3mu}{\small $BR(\tau\rightarrow \mu\mu\mu)_{D^\gamma=0}~[10^{-9}]$}
\vglue -2.cm
\hglue -7.8cm
\scalebox{0.47}{
{\mbox{\epsfig{file=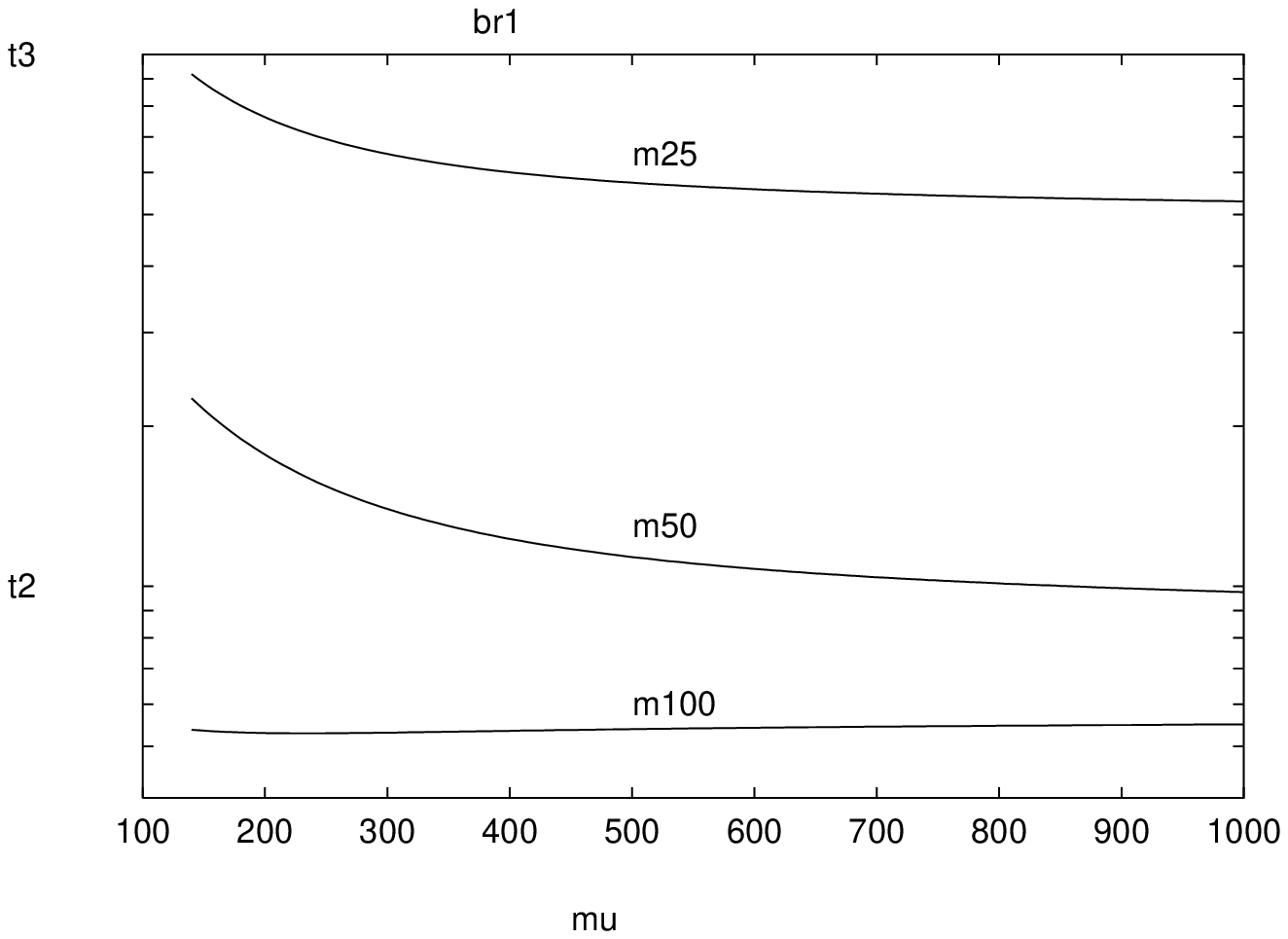}}}}
\vskip -5.5cm
\hglue 6.4cm
\scalebox{0.77}{
{\mbox{\epsfig{file=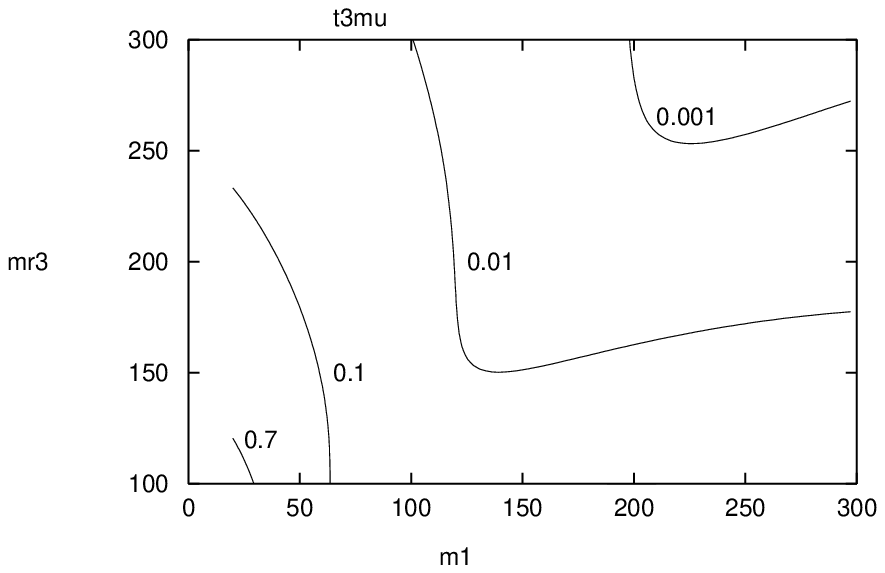}}}}
\vglue -1.2cm
\caption{{
\footnotesize  Monopole contribution to $BR(\tau\to \mu \mu \mu)$ 
for $\tanb =3$, $\mt_{R_2}= 
1~{\rm TeV}$, $\theta_R= \pi/4$ and $\staul= 
 \mt_{\tilde{\mu}_L}= 300~{\rm GeV}$. 
In the left panel:   $\mt_{R_3}= 100~{\rm GeV}$ and 
$|M_1|$ as shown.  
In the right  panel: $|\mu| =  150~{\rm GeV}$.
}}
\label{fr5}
\end{center}
\begin{center}
\psfrag{mu} {\footnotesize  $|\mu|$~[GeV]}
\psfrag{m1}{\footnotesize $|M_1|$~[GeV]}
\psfrag{mr3}{\footnotesize $\begin{array}{c}
\tilde{m}_{R_3} \\
{\rm { [GeV]}}\end{array}$}
\psfrag{sel}{\footnotesize $\begin{array}{c}
\tilde{m}_{\tilde{e}} \\
{\rm { [GeV]}}\end{array}$}
\psfrag{tmue}{\small $BR(\tau\rightarrow \mu e e)_{D^\gamma =0}~[10^{-9}]$}
\vglue -1.cm
\scalebox{0.77}{
{\mbox{\epsfig{file=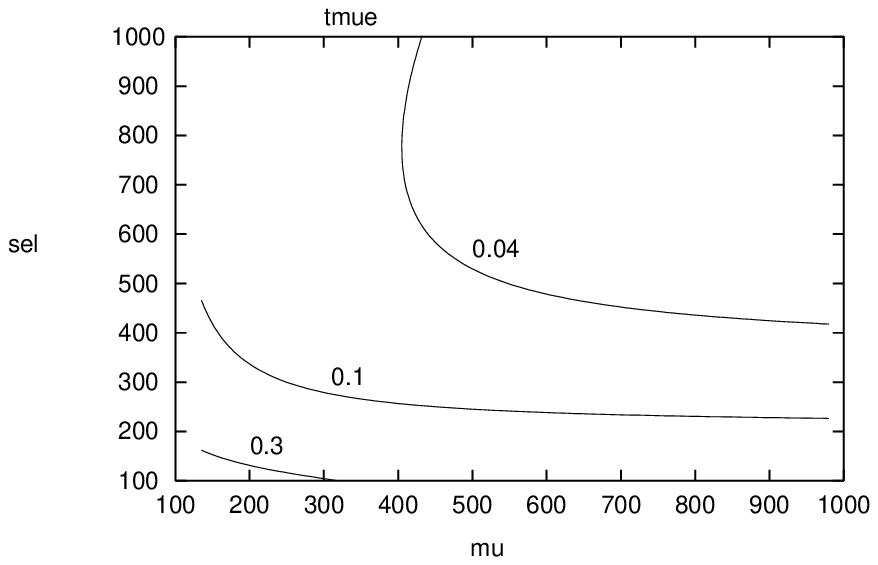}}}
\hglue -3.5cm
{\mbox{\epsfig{file=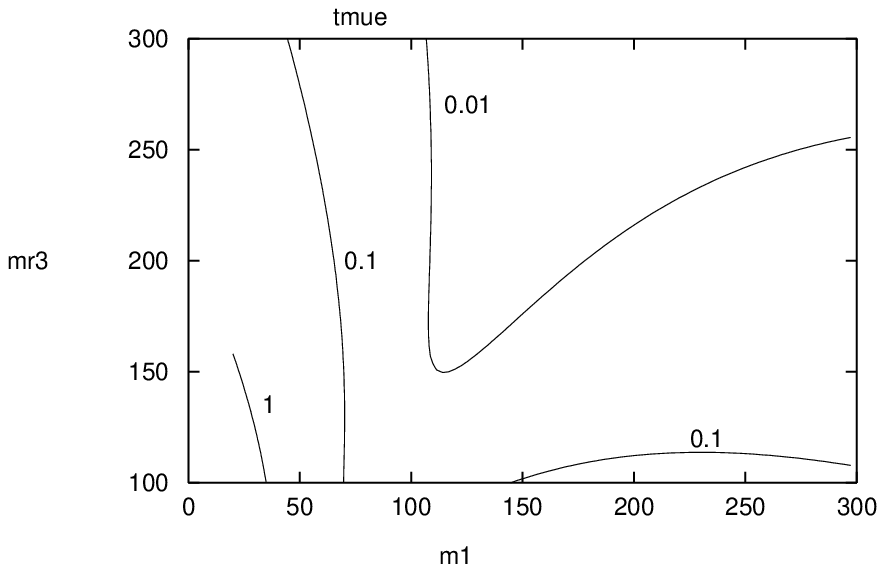}}}}
\vglue -1.2cm
\caption{{\footnotesize 
Monopole contribution to $BR(\tau \to \mu e e)$ 
for $\tanb =3$, $\mt_{R_2}= 
1~{\rm TeV}$, $\theta_R= \pi/4$ and $\staul=  300~{\rm GeV}$. 
In the left panel:   $\mt_{R_3}= 100~{\rm GeV}$, 
$|M_1|= 50~{\rm GeV}$.
In the right  panel: $|\mu| =  150~{\rm GeV}, 
\mt_{\tilde{e}} =100~{\rm GeV}$.
}}
\label{fr6}
\end{center}
\begin{center}
\psfrag{mu} {\footnotesize $|\mu|$~[GeV]}
\psfrag{m1}{\footnotesize $|M_1|$~[GeV]}
\psfrag{mr3}{\footnotesize $\begin{array}{c}
\tilde{m}_{R_3} \\
{\rm { [GeV]}}\end{array}$}
\psfrag{msq}{\footnotesize $\begin{array}{c}
\tilde{m}_{\tilde{q}} \\
{\rm { [GeV]}}\end{array}$}
\psfrag{brrp}{\small $BR(\tau\rightarrow \mu\rho)_{D^\gamma =0}~, ~
BR(\tau\rightarrow \mu\pi)
~[10^{-9}]$}
\vglue -1.cm
\scalebox{0.77}{
{\mbox{\epsfig{file=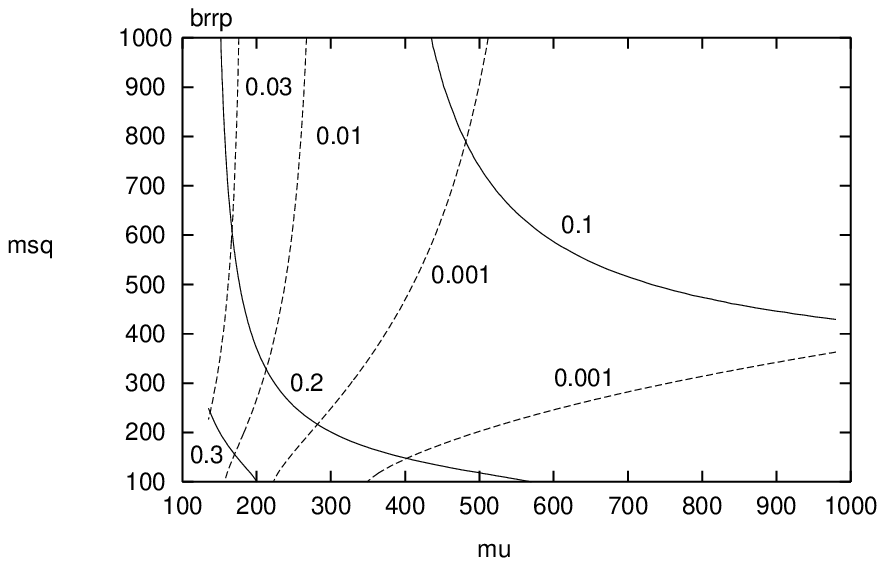}}}
\hglue -3.5cm
{\mbox{\epsfig{file=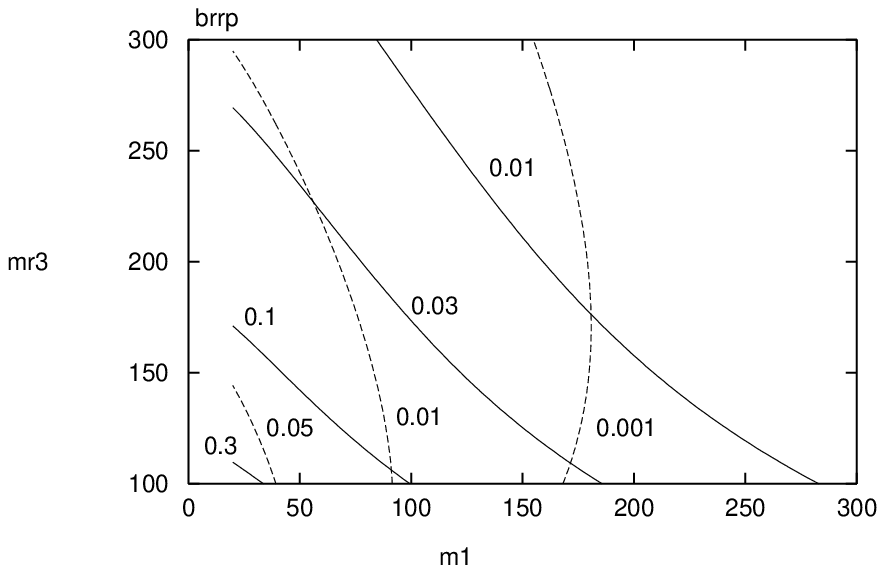}}}}
\vglue -1.2cm
\caption{{\footnotesize 
Monopole contribution to $BR(\tau \to \mu \rho)$ (solid lines)
and $BR(\tau \to \mu \pi)$ (dashed lines), 
for $\tanb =3$, $\mt_{R_2}= 
1~{\rm TeV}$, $\theta_R= \pi/4$ and $\staul = 300~{\rm GeV}$. 
In the left panel:   $\mt_{R_3}= 100~{\rm GeV}, 
|M_1|= 50~{\rm GeV}$.
In the right  panel: $|\mu| =  150~{\rm GeV}, 
\mt_{\tilde{q}} =500~{\rm GeV}$.  
}}
\label{fr7}
\end{center}
\end{figure}
As a general feature the allowed regions are wider 
than in the (LFV)$_L$ examples (see Fig.~\ref{f4}) because 
the individual dipole contributions are now smaller. 
In particular, it is easier to fulfil the bound even with 
vanishing trilinear terms.

In Fig.~\ref{fr3} we show the branching ratios of 
the different LFV processes in two examples, 
where $\mu$ is either  small (upper panel) or large (lower panel). 
In each case 
the dipole $D^{\ga}_R$  contribution is varied (through 
either $A_\tau$ or  $\mt_{\tilde{\tau}_L}$)  
and the monopole ones are essentially  fixed. 
The behaviour of  $BR(\tau \to \mu \mu \mu), 
BR(\tau \to \mu e e), BR(\tau \to \mu \rho)$ 
reflects the interplay between 
the dipole and monopole contributions, 
similarly to  the (LFV)$_L$ case.
Now let us examine in more detail the parameter dependence 
of monopole contributions, considering one process
at a time. 

In Fig.~\ref{fr4} $BR(Z\to \mu \tau)$ is plotted as a function of 
$|\mu|$ (left panel) and in the $(|M_1|, \mt_{R_3})$ plane
for small $\mu$ (right panel).
At variance with the (LFV)$_L$ case, the interference between 
the $A^Z$ and $C^Z$ contributions is now constructive. 
For small $\mu$   we have $|A^Z_R| \gsim  |C^Z_R|$. For large 
$\mu$,  $|A^Z_R|$ is suppressed. [The  $A^{Z(c)}_R$ component 
grows with $\mu$, but it remains small since $\tanb$ is small. 
A similar comment applies to $|D^{Z (c)}_R|$.]
Notice that  $BR(Z\to \mu \tau)$ does not reach $\sim 10^{-9}$, 
even for small mass parameters.  
This is in contrast with the (LFV)$_L$ case.

In Fig.~\ref{fr5} the monopole contribution to 
$BR(\tau \to \mu \mu\mu)$ is shown as a function of 
$|\mu|$ (left panel) and in the $(|M_1|, \mt_{R_3})$ plane (right panel).
For $|M_1|\gsim 50~{\rm GeV}$ we have 
$BR(\tau \to \mu \mu\mu)_{D^\ga =0} \lsim 10^{-10}$, while 
values close  to $10^{-9}$ can only be achieved for  
quite a small $|M_1|$. We recall the pure dipole contribution
to $BR(\tau \to \mu \mu\mu)$ cannot exceed $10^{-9}$ 
[eq.~(\ref{dom-dgm})].

In Fig.~\ref{fr6} the monopole contribution to 
$BR(\tau \to \mu e e)$ is plotted 
in the $(|\mu|, \mt_{\tilde{e}})$ plane
(left panel) and in the $(|M_1|, \mt_{R_3})$ plane (right panel).
Values ${\cal O}(10^{-9})$ can only be reached for small $|M_1|$.
The overall $BR(\tau\to \mu e e)$ can be few times  $10^{-9}$ when 
the maximal allowed dipole contribution is accounted
(see also Fig.~\ref{fr3}).

In Fig.~\ref{fr7} we show the monopole contribution to 
$BR(\tau \to \mu \rho)$ (solid) and $BR(\tau \to \mu \pi)$ 
(dashed), either  in the  $(|\mu|, \mt_{\tilde{q}})$ plane 
(left panel) or in the $(|M_1|, \mt_{R_3})$ plane 
(right panel).
In the left panel 
cancellation effects are visible in $BR(\tau \to \mu \pi)$,
and are due to a destructive interference between 
$A^Z_R$ and  box contributions.
Even outside such an interference region, however,
the $\tau \to \mu \pi$ decay is strongly disfavoured, as 
its $BR$ can hardly reach $10^{-10}$.   
Regarding $\tau \to \mu \rho$, we see that the pure monopole contribution 
can exceed $10^{-10}$ for $|M_1|\lsim 100~{\rm GeV}$. 
Suppose the latter contribution is $ 0.3\times 10^{-9}$ and the pure dipole 
one is $0.8\times 10^{-9}$ [maximal value, see eq.~(\ref{dom-dgr})].
Then we have $BR(\tau \to \mu \rho) \sim 2\times 10^{-9} $ or 
$0.4\times 10^{-9}$, depending on the 
interference sign (see also Fig.~\ref{fr3}). 
As regards $\tau \to \mu \eta$, the $BR$ exhibits a pattern similar to
that of $\tau \to \mu \pi$, but its maximal values are even smaller. 
In the case of $\tau \to \mu \eta'$, the interference between 
$A^Z_R$ and box contributions tends to be constructive, and 
the $BR$ could reach $10^{-10}$.

\subsection{(LFV)$_R$ with large $\tanb$ and large masses\label{srlarge} }
To discuss large  (LFV)$_R$ in the  large $\tanb$ limit 
we follow an approach similar to that adopted in 
Section~\ref{sllarge}. 
In this limit the Higgs-$\mu$-$\tau$ effective operators 
can play an important role.
The enhanced contributions to  the $D^{\ga (b,c)}_R$ dipole
coefficients 
can be kept under control either through mutual cancellations 
(for small mass parameters) or by taking large masses. 
The former option requires some fine tuning, although 
this is milder than in the (LFV)$_L$ case.
Among monopole coefficients, $A^Z_R$ can be slightly larger 
for large $\tanb$ [$A^{Z(a)}_R$ is proportional to $\cos 2\beta$, 
which gets closer to -1, and  $A^{Z(b,c)}_R$ can play some role]. 
Let us consider in more detail   
the second possibility, {\it i.e.} the large mass limit (with 
small $m_A$).
\begin{figure}[ht]
\begin{center}
\psfrag{mu} {\footnotesize $|\mu|$~[TeV]}
\psfrag{m1}{\footnotesize $\begin{array}{c}
|M_1| \\
 {\rm { [TeV]}}
\end{array}
$}
\psfrag{br1}{\tiny $BR(A\rightarrow \mu\tau)~[10^{-5}]$}
\psfrag{br2}{\tiny $BR(\tau\rightarrow \mu\mu\mu)~[10^{-8}]$}
\psfrag{br3}{\tiny $BR(\tau\rightarrow \mu\gamma)~[10^{-7}]$}
\vglue -3.0cm
\hglue -2.2cm
\scalebox{1.5}{
{\mbox{\epsfig{file=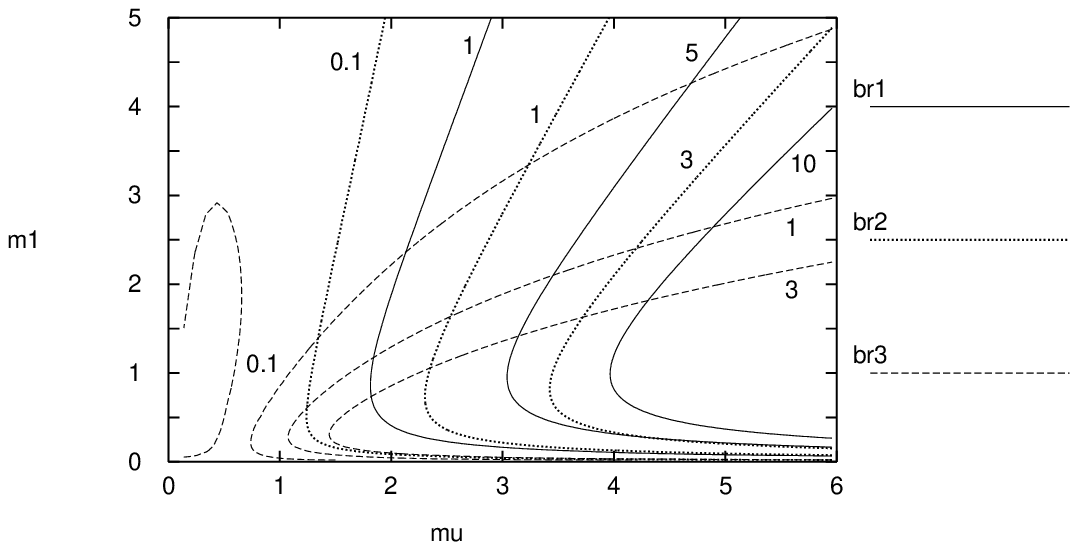}}}}
\vglue -2.0cm
\caption{{\ftsz  Branching ratios of $\tau \to \mu \ga$ (dashed lines), 
$\tau \to \mu \mu \mu$ (dotted lines) and  $A\to \mu \tau$ (solid lines) 
for  $\tanb =40$, $\theta_R= \pi/4$, 
$\mt_{R_2}= 3~{\rm TeV}$, $\mt_{R_3}= 0.7~{\rm TeV}$ and  
$\staul =  0.7~{\rm TeV}$. 
We have also fixed $m_A = 100~{\rm GeV}$  
and $BR(A\to \tau \tau)= 0.1$.
}}
\label{fr8}
\end{center}
\end{figure} 
In this case all monopole contributions 
are suppressed and so is $Z\to \mu \tau$. 
The decays $\tau \to \mu ee$ and 
$\tau \to \mu \rho$ are generically dipole dominated, so the relations 
(\ref{dg1}), (\ref{dg3}) [and the bounds   
(\ref{dom-dge}), (\ref{dom-dgr})] hold.
The dominant contributions to  $BR(\tau \to \mu \mu \mu)$ 
[eq.~(\ref{bt3mu})]  
are induced by the dipole term $D^\ga_R$  and the Higgs-mediated 
term $\delta F^{\mu_R}_L$, proportional 
to  $\Delta_R$  [eq.~(\ref{dflr})]. 
Therefore, $\tau \to \mu \mu \mu$ is correlated to both 
$\tau \to \mu \ga$ and $A\to \mu \tau$ \cite{BR}. 
The processes $\tau \to \mu \eta, \tau \to \mu \eta'$ and  
$\tau \to \mu \pi$ 
are essentially induced by the Higgs-exchange terms  
$\delta F^{\eta, 8}_R, ~\delta F^{\eta, 0}_R,~ 
\delta F^{\eta', 8}_R,~\delta F^{\eta', 0}_R$  
and $\delta F^{\pi}_R$, so they are correlated to one another 
[eqs.~(\ref{etaetap}), (\ref{pieta})], to 
$\tau \to 3 \mu$ [eq.~(\ref{eta3mu})], 
and to $A\to \mu \tau$.
  
In Fig.~\ref{fr8} we show the $BR$s of 
$\tau \to \mu \ga, ~\tau \to 3\mu, ~A\to \mu \tau$ 
in the plane  ($|\mu|, |M_1|$) for $\tanb =40$. 
In the left side of the plot all decays are suppressed, in part because 
of internal cancellations  both in $D^\ga_R$ (visible in the figure) 
and in $\Delta_R$ (not shown here, see Fig.~3 in \cite{BR}).
All $BR$s increase for increasing $\mu$ because $D^{\ga(c)}_R$ and 
$\Delta^{(c)}_R$ dominate. 
The pattern of the contours for $\tau \to \mu \ga$ 
is different from that of $\tau \to 3 \mu$ and  $A\to \mu \tau$,  
whose rates are correlated to each other 
in the regions where $\tau \to 3 \mu$ is  dominated by 
the $\Delta_R$ contribution [see eq.~(\ref{crl})]. 
In particular, there is a region where 
$BR(\tau \to \mu \ga)$ is below the present bound (\ref{e-rad}), 
whereas 
$BR(A \to \mu \tau)$ and $BR(\tau \to 3\mu)$ 
can reach $10^{-4}$ and $5\times 10^{-8}$, respectively. 
Also notice that we have fixed   $m_A= 100~{\rm GeV}$:
$BR(\tau \to 3\mu)$  
is smaller for larger $m_A$, since the 
Higgs-mediated contribution scales as  $1/m_A^{4}$.
The largest values of $BR(\tau \to 3\mu)$ are correlated to 
values of $BR(\tau \to \mu\eta)$ close to its present bound 
(\ref{e-mueta}), and to values of 
$BR(\tau \to \mu\pi)$ and $BR(\tau \to \mu\eta')$ 
of order $10^{-9}$. 

\section{Conclusions}
The observation of LFV processes would be a clear 
evidence of physics beyond the Standard Model. 
In this paper we have focused our attention on LFV in the 
$\mu$-$\tau$ sector by considering several LFV decays, namely  
$\tau \to \mu X$ ($X=\ga,  e^+ e^-, \mu^+ \mu^-, \rho, \pi, \eta, \eta'$), 
$Z\to \mu \tau$ and Higgs boson decays into $\mu \tau$.
First we have presented a model independent 
treatment of such decays, 
then we have analysed them in a general {\it unconstrained} 
MSSM framework, without assuming specific relations 
among mass parameters. In particular, we have allowed slepton mass 
matrices to have large  $\mu$-$\tau$ entries 
(whose origin is {\it unspecified}).
This possibility is not excluded by 
the present experimental bounds 
[essentially that on $BR(\tau\to \mu\ga)$], even 
for a relatively light superparticle spectrum.
It is well known, instead, that the analogous  situation for 
$e$-$\mu$ flavour violation is strongly  
disfavoured by the  stringent bound 
$BR(\mu \to e\ga) < 1.2 \times 10^{-11}$ \cite{mega}. 
We have implicitly assumed that  $\mu$-$e$ flavour 
transitions are adequately suppressed by 
sufficiently small $e$-$\mu$ and $e$-$\tau$ entries in the 
slepton mass matrices. 
Issues such as the generation and the amount of LFV, as well 
as other model-dependent aspects,   
go beyond the scope of our paper and can only be addressed 
in specific frameworks (see {\it e.g.} \cite{BM,HKR,HMTY,lfv}).

We have analysed the behaviour of the various LFV processes in the 
MSSM parameter space in terms of the relevant operator coefficients. 
In particular, we have disentangled the 
 dipole and non-dipole  contributions to the LFV decays 
in order to better appreciate the correlations among them.
We have provided general formulae for the operator coefficients 
in Appendix and presented a numerical analysis with several 
examples in Sections~4 and 5. 
For the sake of brevity, we can try to summarize our results by 
schematically distinguishing a few  different scenarios.
For definiteness, consider the case of large (LFV)$_L$. 
\begin{enumerate}
\item
Each of the contributions to the photon dipole operator 
fulfils the 
bound dictated by $\tau \to \mu \ga$, 
so no mutual cancellations among them need to be invoked. 
Also suppose that MSSM mass parameters have comparable 
magnitudes, with the possible exception of $m_A$.
For convenience, we distinguish the cases of small and 
large $\tanb$. 
{\it i)} For small $\tanb$, mass parameters should be larger 
than a few hundreds GeV to accomplish the  $\tau \to \mu \ga$ 
bound.
This implies that monopole contributions to all branching ratios are 
generically smaller than $10^{-10}$. 
Also the Higgs-mediated contributions are small, 
because  $\tanb$ is small.
Hence $\tau \to \mu P$ ($P=\pi, \eta, \eta'$) and 
$Z\to \mu \tau$ have $BR$s  smaller than  $10^{-10}$.
The processes $\tau \to \mu ee,~ \tau \to 3 \mu,~ \tau \to \mu \rho$
tend to be dipole-dominated 
[eqs.~(\ref{dg1}),~(\ref{dg2}),~(\ref{dg3})], and their 
$BR$s  can be ${\cal O}(10^{-9})$ 
if $BR(\tau \to \mu \ga)$ is close to its present bound 
[see eqs.~(\ref{dom-dge}),~(\ref{dom-dgm}),~(\ref{dom-dgr})].
{\it ii)} For large $\tanb$, 
the mass parameters have to be pushed towards 
the TeV range to accomplish the  $\tau \to \mu \ga$ bound. 
The monopole contributions are even more suppressed 
than in the previous case. 
On the other hand, the Higgs-mediated 
contributions are strongly enhanced and may become
comparable to or even more important 
than the dipole ones (for small $m_A$). Hence the processes 
$\tau \to 3 \mu, ~\tau \to \mu \eta$  
become excellent signatures, together with LFV Higgs boson 
decays, to which they are correlated.
Indeed, one can have $BR(A \to \mu \tau) 
\sim 10^{-4}$ and, for small $m_A$,  
$BR(\tau \to 3\mu)$ close to $10^{-7}$, with   
$BR(\tau \to \mu \eta)$ even larger than the latter one 
[and $BR(\tau \to \mu \pi), ~BR(\tau \to \mu \eta')$ 
possibly reaching $10^{-9}$].   
Regarding the other processes (and also $\tau \to 3 \mu,
~\tau \to \mu P$ for large $m_A$), 
the conclusions drawn in case {\it i)} hold again. 

\item
The dipole operator satisfies the $\tau \to \mu \ga$ bound 
due to cancellations among individual contributions and/or 
because of hierarchies among mass parameters 
({\it e.g.} with large $\mu$), 
like in the examples displayed in Figs.~\ref{f4} and \ref{f5}. 
Such cases are interesting because  $\mt_{L_3}$ and $M_2$ are not
forced to be large, so monopole operators 
could contribute 
to each branching ratio at the level of $10^{-9}$ (or more),
unless cancellations occur among different
monopole components (which indeed could happen, as we
have shown). Such contributions can bring $BR(\tau \to \mu P)$  
($P=\pi, \eta, \eta'$) and 
$BR(Z\to \mu \tau)$ to values ${\cal O}(10^{-9})$,
and an interesting interplay between dipole and 
monopole contributions can 
take place in the other processes. 
This breaks the direct correlation 
[eqs.~(\ref{dg1}),~(\ref{dg2}),~(\ref{dg3})] 
between  $\tau \to \mu \ga$ and 
$\tau \to  \mu ee, ~\tau \to  3\mu, ~ 
\tau \to \mu \rho$. 
For each of the latter  processes, 
the $BR$  can be smaller or larger than its maximal 
$D^\ga$-dominated value, and can become a few $\times 10^{-9}$.
Such values  can be reached even in the case of 
strongly suppressed dipole contributions. 
In this extreme case $\tau \to  \mu ee, ~\tau \to  3\mu, ~ 
\tau \to \mu \rho$ could be as sensitive as $\tau \to \mu \ga$ 
to supersymmetric LFV effects.
All these comments apply to the small $\tan\beta$ case.
If we insist on having small $\mt_{L_3}$ and $M_2$ 
with large $\tan\beta$, a strong fine-tuning is required 
to keep the total $D^\ga$ below the bound (through
cancellations).
In this special situation, the Higgs-mediated contributions
could push $BR(\tau \to 3\mu)$ and 
$BR(\tau \to \mu \eta)$ to values ${\cal O}(10^{-7})$
for small  $m_A$, and $Z$-dipole contributions could help 
$BR(Z\to \mu \tau)$ to reach $10^{-9}$.
  
\end{enumerate}

The case of large (LFV)$_R$ can be summarized along similar lines, 
although some differences with respect to (LFV)$_L$ are present 
(see Section~\ref{lfvr} for details).
For instance, it is easier to fulfil the $\tau \to \mu \ga$ 
bound on the dipole operator with small mass parameters,
especially at small $\tanb$. 
Monopole contributions are typically 
smaller than in the (LFV)$_L$ case, unless 
$M_1$ is stretched to very small values. 
Barring this possibility, and assuming that  $\tau \to \mu \ga$ is
close to its present bound, the only processes whose
$BR$s can be ${\cal O}(10^{-9})$ at small $\tanb$ are 
$\tau \to \mu ee,~ \tau \to 3 \mu,~ \tau \to \mu \rho$,
mainly through dipole contributions.  
At large  $\tanb$, also Higgs-mediated contributions can play 
an important role. If $\mu$ is larger than other mass 
parameters and $m_A$ is small, $BR(\tau \to 3\mu)$ and 
$BR(\tau \to \mu \eta)$ can reach values ${\cal O}(10^{-7})$.

As a final remark, we recall that future 
collider experiments will give crucial
information (or constraints) on the MSSM 
parameters, including those related to LFV. 
Indeed, supersymmetric $\mu$-$\tau$ 
LFV effects can also show 
up, already at the tree level, 
in the decays of superparticles directly 
produced at  LHC or leptonic colliders \cite{collider}. 
This important search will be complementary to 
that of the loop-induced LFV decays discussed in our work.

\vspace{0.5cm}

\noindent
{\large \bf Acknowledgments}\\
\noindent
This work was partially supported by MIUR and 
by the European Union under the contracts 
HPRN-CT-2000-00148 (Across the Energy Frontier) and
HPRN-CT-2000-00149 (Collider Physics).

\vspace{0.3cm}
\noindent
{\it This article is dedicated to the memory of Anna's parents.}

\vspace{1cm}

\appendix
\section{Operator coefficients in the MSSM}

\subsection{MSSM Lagrangian and conventions \label{a1}}
We recall here the lagrangian terms relevant for our computations, 
and in this way  also establish our conventions and notations.
The gauge bosons we need are the neutral ones (photon and $Z$).
The covariant kinetic term for a generic Weyl fermion $\psi$ 
is $i \ov{\psi}\smub D_{\mu} \psi$, where $\psi$ can be  
 either a matter fermion  
($f, f^c$) or a gaugino ($\tilde{B}^0, \tilde{W}^0, \tilde{W}^{\pm}$) 
or a higgsino ($\tilde{H}^0_{1}, \tilde{H}^0_{2}, 
\tilde{H}^{-}_1, \tilde{H}^{+}_2$), and 
the covariant derivative reads as: 
\be{dcov}
D_{\mu} \psi = \left(\partial_\mu + i e Q_\psi A_\mu + i g_Z (T^3_\psi -  
  Q_\psi  s^2_W) Z_\mu\right ) \psi , 
\ee 
with $g_Z = g/c_W$  
($c_W =\cos\theta_W, s_W = \sin\theta_W$, where $\theta_W$ is 
the weak mixing angle).
Electric charge ($Q_\psi$) is related 
to weak isospin ($T^3_\psi$) and hypercharge ($Y_\psi$) through 
$ Q_\psi= T^3_\psi + Y_\psi$.
Similarly, for a generic complex scalar $\phi$ ($ \tilde{f}=\fl, 
\tilde{f}^c=\fr^*, {H}^0_{1},{H}^0_{2},{H}^{-}_1,{H}^{+}_2$), 
we have  $(D^{\mu} \phi)^* (D_{\mu} \phi)$.
The gaugino and higgsino mass terms are:
\be{masse}
-\frac12 M_1 \tilde{B}^0 \tilde{B}^0 -\frac12 M_2 \tilde{W}^0 \tilde{W}^0 
 - M_2 \tilde{W}^{+} \tilde{W}^{-} - \mu ( \tilde{H}^0_{1}  \tilde{H}^0_{2} 
-  \tilde{H}^{-}_1\tilde{H}^{+}_2) +{\rm h.c.} . 
\ee
Notice that our sign convention for the $\mu$ parameter is
{\em opposite} to the one commonly used. The gaugino masses and the 
$\mu$ parameter will be taken  real in the diagrammatic 
computations.
The relevant fermion-sfermion-gaugino interaction terms are:
\beqn{fsg}
&&-g'\sqrt{2} \left(Y_f \tilde{f}^* f +Y_{f^c} \tilde{f}^{c *} f^c -\frac12
H^{0*}_1 \tilde{H}^0_{1} +\frac12 H^{0*}_2 \tilde{H}^0_{2} \right)\tilde{B}^0 
\nonumber \\ 
&& -g\sqrt{2} \left(T^3_f  \tilde{f}^* f + \frac12
H^{0*}_1 \tilde{H}^0_{1} -\frac12 H^{0*}_2 \tilde{H}^0_{2} \right)\tilde{W}^0 
\nonumber \\ 
&& -g \left(\tilde{f}^*_u f_d + H^{0*}_1 \tilde{H}^{-}_1 \right)\tilde{W}^+ 
-g \left(\tilde{f}^*_d f_u + H^{0*}_2 \tilde{H}^{+}_2 \right)\tilde{W}^- 
+ {\rm h.c.}
\eea
where $f_{u,d} (\tilde{f}_{u,d})$ are up- or down-type components of 
fermion (sfermion) doublets. 
The $h_\tau$-Yukawa interactions we use are:
\be{yuk}
-h_\tau H^0_1 \tau^c\tau - h_\tau\tilde{\tau}^c\tilde{H}^0_1\tau 
- h_\tau {\tau}^c( \tilde{H}^0_1\tilde{\tau} -
\tilde{H}^-_1 \tilde{\nu}_\tau) +{\rm h.c.} . 
\ee
Analogous expressions hold for other Yukawa couplings (in particular,  
$h_\mu$). 

The (soft-breaking) scalar mass terms have 
the generic form $-\mt^2_f |\ft|^2 - 
\mt^2_{\fc} |\fct|^2$ (equivalently denoted as $-\mt^2_{\fl} |\fl|^2 - 
\mt^2_{\fr} |\fr|^2$). We assume that 
such terms are flavour conserving, except for those in the  
smuon-stau sector, which require a more detailed
description. In the left sector the relevant terms are:
\be{l23}
- \tilde{L}^\dagger_\mu \mt^2_{L\mu \mu} \tilde{L}_\mu 
-\tilde{L}^\dagger_\tau \mt^2_{L\tau \tau} \tilde{L}_\tau
-(\tilde{L}^\dagger_\mu \mt^2_{L \mu \tau} \tilde{L}_\tau +{\rm h.c.}) , 
\ee
where  $\tilde{L}_\mu = (\tilde{\nu}_\mu , \tilde{\mu}_L)^T, 
\tilde{L}_\tau = (\tilde{\nu}_\tau , \tilde{\tau}_L)^T$.
In the right sector the relevant terms are:
\be{r23}
- \tilde{\mu}^*_R \mt^2_{R\mu \mu} \tilde{\mu}_R 
-\tilde{\tau}^*_R \mt^2_{R\tau \tau} \tilde{\tau}_R
- (\tilde{\mu}^*_R \mt^2_{R\mu \tau} \tilde{\tau}_R +{\rm h.c.} ) . 
\ee
Finally, in the left-right sector  the terms with neutral Higgs 
bosons are:
\be{lr23}
- h_\mu (\mu H^{0*}_2 + A_\mu H^{0}_1) 
\tilde{\mu}^*_R\tilde{\mu}_L  
- h_\tau (\mu H^{0*}_2 + A_\tau H^{0}_1) 
\tilde{\tau}^*_R\tilde{\tau}_L  
-h_\tau A^L_{\mu\tau} H^{0}_1 \tilde{\tau}^*_R\tilde{\mu}_L-
h_\tau A^R_{\mu\tau} H^{0}_1 \tilde{\mu}^*_R\tilde{\tau}_L + {\rm h.c.} , 
\ee
where $A_\mu, A_\tau, A^{L,R}_{\mu\tau}$ are mass parameters, which 
will be taken real in the diagrammatic computations.

The mass parameters in eqs.~(\ref{l23}-\ref{r23}) are the entries of the 
following mass matrices:
\be{matri}
\tilde{{\cal M}}^2_L =
\pmatrix{
\tilde{m}^2_{L \mu \mu} & \tilde{m}^2_{L \mu \tau} \cr
\tilde{m}^2_{L \mu \tau} & \tilde{m}^2_{L \tau \tau} }, \, \,~~~~~~
\tilde{{\cal M}}^2_R = \pmatrix{
\tilde{m}^2_{R \mu \mu} & \tilde{m}^2_{R \mu \tau} \cr
\tilde{m}^2_{R \mu \tau} & \tilde{m}^2_{R \tau \tau} } , 
\end{equation}
where we have now assumed $\tilde{m}^2_{L \mu \tau}, \tilde{m}^2_{R \mu \tau}$ 
to be real.
The flavour states $\tilde{L}_\mu$, $\tilde{L}_\tau$ 
are related to the 
$\tilde{{\cal M}}^2_L$ eigenstates 
$\tilde{L}_2 = 
(\tilde{\nu}_2 , \lt_{L_2})^T, 
\tilde{L}_3 = 
(\tilde{\nu}_3 , \lt_{L_3})^T$ 
by the relations $\tilde{L}_\mu = c_L \tilde{L}_2 - s_L \tilde{L}_3, ~
\tilde{L}_\tau = s_L \tilde{L}_2 + c_L \tilde{L}_3$, where 
$c_L = \cos\theta_L, s_L = \sin\theta_L$. 
Analogous relations hold for the right-handed 
sleptons:   $\tilde{\mu}_{R} = c_R \lt_{R_2} - s_R \lt_{R_3}, ~
\tilde{\tau}_{R} = s_R \lt_{R_2} + c_R \lt_{R_3}$, 
where $\lt_{R_2}$, $\lt_{R_3}$ are the eigenstates of 
$\tilde{{\cal M}}^2_R$ and  
$c_R = \cos\theta_R, s_R = \sin\theta_R$.
The mixing parameters satisfy the following relations:
\be{mix}
s_L c_L = 
\frac{\tilde{m}^2_{L \mu \tau}}{\tilde{m}^2_{L_2}-\tilde{m}^2_{L_3} }~ , 
~~~~~~~s_R c_R = 
\frac{\tilde{m}^2_{R \mu \tau}}{\tilde{m}^2_{R_2}-\tilde{m}^2_{R_3} } ~, 
\ee
where $\tilde{m}^2_{L_\al}$ and $\tilde{m}^2_{R_\al}$ ($\al =2,3$) are 
the eigenvalues of $\tilde{{\cal M}}^2_L$ and $\tilde{{\cal M}}^2_R$, 
respectively.

The mass and interaction terms presented above are sufficient 
to perform diagrammatic calculations of the operator 
coefficients. As anticipated in Section~\ref{gao}, electroweak
breaking effects (Higgs insertions) will be treated
at lowest order. 
In the next sections, for each coefficient we 
show the relevant diagrams and display the analytical 
results. 
In diagrams with gauge boson insertions in a fermionic line 
we use a shorthand notation, explained in Fig.~\ref{short}. 

\begin{figure}[htb]
\begin{center}
\begin{picture}(200,35)(0,-15)
\ArrowLine(20,0)(0,0)
\ArrowLine(40,0)(20,0)
\Photon(20,0)(20,20){2}{3}
\GCirc(20,0){2}{0.5}
\Text(8,8)[]{\ftsz{$\tilde{W}^+$}}
\Text(34,8)[]{\ftsz{$\tilde{W}^+$}}
\Text(50,0)[]{$=$}
\ArrowLine(80,0)(60,0)
\ArrowLine(100,0)(80,0)
\Photon(80,0)(80,20){2}{3}
\Text(68,8)[]{\ftsz{$\tilde{W}^+$}}
\Text(94,8)[]{\ftsz{$\tilde{W}^+$}}
\Text(110,0)[]{$+$}
\ArrowLine(140,0)(120,0)
\ArrowLine(140,0)(160,0)
\ArrowLine(160,0)(180,0)
\ArrowLine(200,0)(180,0)
\Photon(160,0)(160,20){2}{3}
\Text(128,8)[]{\ftsz{$\tilde{W}^+$}}
\Text(148,8)[]{\ftsz{$\tilde{W}^-$}}
\Text(174,8)[]{\ftsz{$\tilde{W}^-$}}
\Text(194,8)[]{\ftsz{$\tilde{W}^+$}}
\end{picture}
\hspace{0.8 cm}
\begin{picture}(200,35)(0,-15)
\ArrowLine(0,0)(20,0)
\ArrowLine(40,0)(20,0)
\Photon(20,0)(20,20){2}{3}
\GCirc(20,0){2}{0.5}
\Text(8,8)[]{\ftsz{$\tilde{W}^-$}}
\Text(34,8)[]{\ftsz{$\tilde{W}^+$}}
\Text(50,0)[]{$=$}
\ArrowLine(60,0)(80,0)
\ArrowLine(100,0)(80,0)
\ArrowLine(120,0)(100,0)
\Photon(100,0)(100,20){2}{3}
\Text(68,8)[]{\ftsz{$\tilde{W}^-$}}
\Text(88,8)[]{\ftsz{$\tilde{W}^+$}}
\Text(114,8)[]{\ftsz{$\tilde{W}^+$}}
\Text(130,0)[]{$+$}
\ArrowLine(140,0)(160,0)
\ArrowLine(160,0)(180,0)
\ArrowLine(200,0)(180,0)
\Photon(160,0)(160,20){2}{3}
\Text(148,8)[]{\ftsz{$\tilde{W}^-$}}
\Text(174,8)[]{\ftsz{$\tilde{W}^-$}}
\Text(194,8)[]{\ftsz{$\tilde{W}^+$}}
\end{picture}
\end{center}
\begin{center}
\begin{picture}(200,35)(0,-15)
\ArrowLine(0,0)(20,0)
\ArrowLine(20,0)(40,0)
\Photon(20,0)(20,20){2}{3}
\GCirc(20,0){2}{0.5}
\Text(10,8)[]{\ftsz{$\tilde{H}^-_1$}}
\Text(34,8)[]{\ftsz{$\tilde{H}^-_1$}}
\Text(50,0)[]{$=$}
\ArrowLine(60,0)(80,0)
\ArrowLine(80,0)(100,0)
\Photon(80,0)(80,20){2}{3}
\Text(70,8)[]{\ftsz{$\tilde{H}^-_1$}}
\Text(94,8)[]{\ftsz{$\tilde{H}^-_1$}}
\Text(110,0)[]{$+$}
\ArrowLine(120,0)(140,0)
\ArrowLine(160,0)(140,0)
\ArrowLine(180,0)(160,0)
\ArrowLine(180,0)(200,0)
\Photon(160,0)(160,20){2}{3}
\Text(130,8)[]{\ftsz{$\tilde{H}^-_1$}}
\Text(146,8)[]{\ftsz{$\tilde{H}^+_2$}}
\Text(174,8)[]{\ftsz{$\tilde{H}^+_2$}}
\Text(194,8)[]{\ftsz{$\tilde{H}^-_1$}}
\end{picture}
\hspace{0.8 cm}
\begin{picture}(200,35)(0,-15)
\ArrowLine(0,0)(20,0)
\ArrowLine(40,0)(20,0)
\Photon(20,0)(20,20){2}{3}
\GCirc(20,0){2}{0.5}
\Text(10,8)[]{\ftsz{$\tilde{H}^-_1$}}
\Text(34,8)[]{\ftsz{$\tilde{H}^+_2$}}
\Text(8,-10)[]{\ftsz{$( \tilde{H}^0_1$}}
\Text(34,-10)[]{\ftsz{$\tilde{H}^0_2 )$}}
\Text(50,0)[]{$=$}
\ArrowLine(60,0)(80,0)
\ArrowLine(100,0)(80,0)
\ArrowLine(120,0)(100,0)
\Photon(100,0)(100,20){2}{3}
\Text(70,8)[]{\ftsz{$\tilde{H}^-_1$}}
\Text(86,8)[]{\ftsz{$\tilde{H}^+_2$}}
\Text(114,8)[]{\ftsz{$\tilde{H}^+_2$}}
\Text(68,-10)[]{\ftsz{$(\tilde{H}^0_1$}}
\Text(88,-10)[]{\ftsz{$\tilde{H}^0_2$}}
\Text(114,-10)[]{\ftsz{$\tilde{H}^0_2)$}}
\Text(130,0)[]{$+$}
\ArrowLine(140,0)(160,0)
\ArrowLine(160,0)(180,0)
\ArrowLine(200,0)(180,0)
\Photon(160,0)(160,20){2}{3}
\Text(150,8)[]{\ftsz{$\tilde{H}^-_1$}}
\Text(174,8)[]{\ftsz{$\tilde{H}^-_1$}}
\Text(194,8)[]{\ftsz{$\tilde{H}^+_2$}}
\Text(148,-10)[]{\ftsz{$(\tilde{H}^0_1$}}
\Text(174,-10)[]{\ftsz{$\tilde{H}^0_1$}}
\Text(194,-10)[]{\ftsz{$\tilde{H}^0_2)$}}
\end{picture}
\end{center}
\caption{\ftsz Shorthand notation used in some diagrams.}
\label{short}
\end{figure}

\subsection{Loop integrals}
The results of our diagrammatic computations will be expressed in terms 
of the following standard loop integrals
\bea
I_N(m_1^2,\ldots,m_N^2) & \equiv & \! {i \over \pi^2}  \!
\int  \! {{\rm d}^4 k \over (k^2 -m_1^2) \ldots  (k^2 -m_N^2)}
=(-1)^{N+1}  \! \int_0^{\infty}  \! \! \!
{s \, {\rm d} s \over (s +m_1^2) \ldots  (s+m_N^2)}
\nonumber\\
& &
\\
J_N(m_1^2,\ldots,m_N^2) & \equiv & \! {i \over \pi^2} \! 
\int  \! {k^2 \, {\rm d}^4 k \over (k^2 -m_1^2) \ldots (k^2 -m_N^2)}
=(-1)^N  \! \int_0^{\infty}  \! \! \!
{s^2 \, {\rm d} s \over (s +m_1^2) \ldots  (s+m_N^2)}
\nonumber\\
& &
\\
K_N(m_1^2,\ldots,m_N^2) & \equiv &  \! {i \over \pi^2}  \!
\int  \! {k^4 \,  {\rm d}^4 k \over (k^2 -m_1^2) \ldots (k^2 -m_N^2)}
=(-1)^{N+1}  \! \int_0^{\infty}  \! \! \!
{s^3 \, {\rm d} s \over (s +m_1^2) \ldots  (s+m_N^2)} .
\nonumber\\
& &
\eea

It is worth recalling that the expression of a diagrammatic
computation in terms of these loop integrals is not unique, 
since such functions are not independent (this fact is also reflected
in the arbitrariness in the parametrization of diagram momenta).
Several identities relating the above functions can be derived,
either by simple manipulations of the integrands ({\it i.e.} the
propagators) or by using obvious scaling properties (see {\it e.g.} 
\cite{moroi}).
Examples of the  former kind are 
\beqn{rel1}
I_N(m_1^2,\ldots,m_N^2) &= &\frac{1}{m_1^2 - m_N^2}\left[
I_{N-1}(m_1^2,\ldots,m_{N-1}^2)- 
I_{N-1}(m_2^2,\ldots,m_N^2) \right]  \\
J_N(m_1^2,\ldots,m_N^2) &= &I_{N-1}(m_1^2,\ldots,m_{N-1}^2)+ m_{N}^2
I_N(m_1^2,\ldots,m_N^2) . \label{rel2}
\eea
An example of the latter kind is (for $N>2$) 
\be{rel3}
\frac{\rm d}{{\rm d} \xi} \left[\xi^{N-2} I_N(\xi m_1^2,\ldots,\xi m_N^2)
\right]_{\xi =1} = 0 .
\ee
By combining  identities of both kinds, new ones can be obtained.
For instance, from eqs.~(\ref{rel2}) and (\ref{rel3}) we get
\beqn{rel4}
\sum_{i=1}^{N-2} J_{N+1}(m_1^2,\ldots,m_i^2, m_i^2, \ldots,m_N^2) &=& 
-m_{N-1}^2 I_{N+1}(m_1^2,\ldots,m_{N-1}^2,m_{N-1}^2,m_N^2) \nonumber \\
&& -m_{N}^2 I_{N+1}(m_1^2,\ldots,m_{N-1}^2,m_{N}^2,m_N^2) .
\eea
Specializing this equation to the case $N=4$ and $m_3^2 = m_4^2$  
gives a particularly useful relation
\be{rel5}
 J_{5}(m_1^2,m_1^2,m_2^2,m_3^2,m_3^2)+ J_{5}(m_1^2,m_2^2,m_2^2,m_3^2,m_3^2)
= -2 m_3^2 I_{5}(m_1^2,m_2^2,m_3^2,m_3^2,m_3^2) .
\ee

All such identities can be used  to cast a given 
expression into a convenient form. One can even reduce all loop 
functions to a simple one\footnote{
For example $I_3$, which has the explicit form 
$I_3 (x,y,z)=(xy \log\frac{x}{y}  +yz \log\frac{y}{z} 
+ zx \log\frac{z}{x})/[(x-y) (z-y)(z-x)]$.},
although the latter operation may not be the 
best choice for numerical evaluations.

\subsection{Contributions to $A^Z_{L,R}$ \label{a3}}
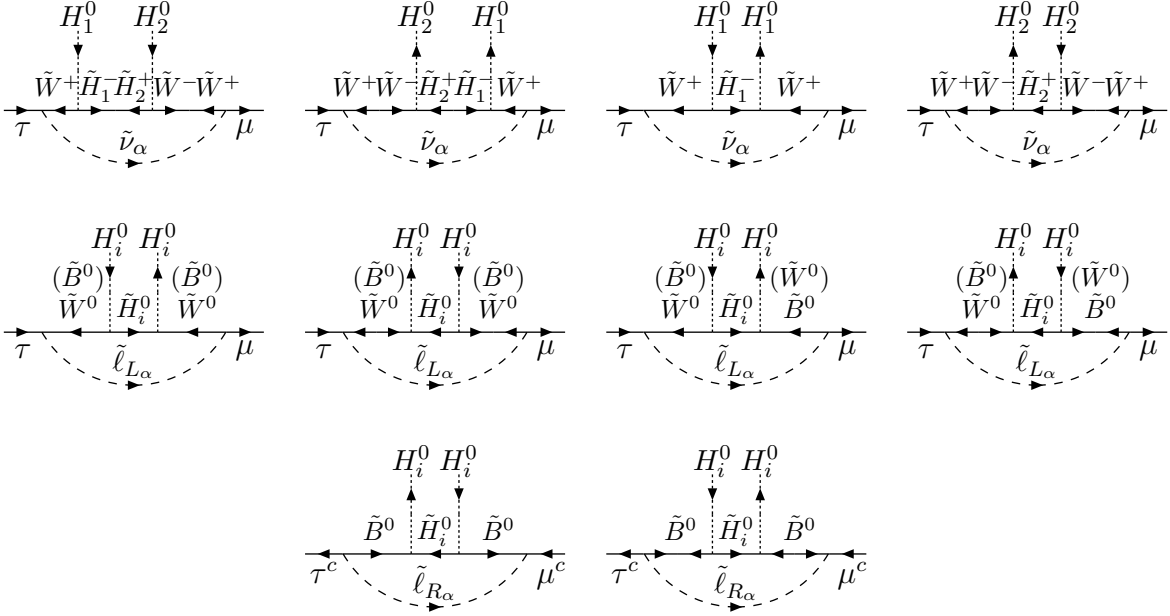
\begin{figure}[htb]
\begin{center}
\begin{picture}(110,70)(-55,-30)
\ArrowLine(-49,0)(-35,0)
\ArrowLine(-21,0)(-35,0)
\ArrowLine(-21,0)(-7,0)
\ArrowLine(7,0)(-7,0)
\ArrowLine(7,0)(21,0)
\ArrowLine(35,0)(21,0)
\ArrowLine(35,0)(49,0)
\DashArrowLine(-21,30)(-21,15){1}
\DashLine(-21,15)(-21,0){1}
\DashArrowLine(7,30)(7,15){1}
\DashLine(7,15)(7,0){1}
\DashArrowArc(0,20)(40,210,330){3}
\Text(-42,-7)[]{$\tau$}
\Text(42,-7)[]{$\mu$}
\Text(-29,10)[]{\ftsz{$\tilde{W}^+$}}
\Text(-13,9)[]{\ftsz{$\tilde{H}_1^-$}}
\Text(0,9)[]{\ftsz{$\tilde{H}_2^+$}}
\Text(16,10)[]{\ftsz{$\tilde{W}^-$}}
\Text(32,10)[]{\ftsz{$\tilde{W}^+$}}
\Text(-21,32)[b]{\small{$H_1^0$}}
\Text(7,32)[b]{\small{$H_2^0$}}
\Text(0,-12)[]{\small{$\tilde{\nu}_{\alpha}$}}
\end{picture}
\begin{picture}(110,70)(-55,-30)
\ArrowLine(-49,0)(-35,0)
\ArrowLine(-21,0)(-35,0)
\ArrowLine(-21,0)(-7,0)
\ArrowLine(7,0)(-7,0)
\ArrowLine(7,0)(21,0)
\ArrowLine(35,0)(21,0)
\ArrowLine(35,0)(49,0)
\DashArrowLine(-7,15)(-7,30){1}
\DashLine(-7,15)(-7,0){1}
\DashArrowLine(21,15)(21,30){1}
\DashLine(21,15)(21,0){1}
\DashArrowArc(0,20)(40,210,330){3}
\Text(-42,-7)[]{$\tau$}
\Text(42,-7)[]{$\mu$}
\Text(-30,10)[]{\ftsz{$\tilde{W}^+$}}
\Text(-14,10)[]{\ftsz{$\tilde{W}^-$}}
\Text(1,9)[]{\ftsz{$\tilde{H}_2^+$}}
\Text(15,9)[]{\ftsz{$\tilde{H}_1^-$}}
\Text(32,10)[]{\ftsz{$\tilde{W}^+$}}
\Text(-7,32)[b]{\small{$H_2^0$}}
\Text(21,32)[b]{\small{$H_1^0$}}
\Text(0,-12)[]{\small{$\tilde{\nu}_{\alpha}$}} 
\end{picture}
\begin{picture}(110,70)(-55,-30)
\ArrowLine(-49,0)(-35,0)
\ArrowLine(-9,0)(-35,0)
\ArrowLine(-9,0)(9,0)
\ArrowLine(35,0)(9,0)
\ArrowLine(35,0)(49,0)
\DashArrowLine(-9,30)(-9,15){1}
\DashLine(-9,15)(-9,0){1}
\DashArrowLine(9,15)(9,30){1}
\DashLine(9,15)(9,0){1}
\DashArrowArc(0,20)(40,210,330){3}
\Text(-42,-7)[]{$\tau$}
\Text(42,-7)[]{$\mu$}
\Text(-21,10)[]{\ftsz{$\tilde{W}^+$}}
\Text(0,9)[]{\ftsz{$\tilde{H}_1^-$}}
\Text(24,10)[]{\ftsz{$\tilde{W}^+$}}
\Text(-9,32)[b]{\small{$H_1^0$}}
\Text(9,32)[b]{\small{$H_1^0$}}
\Text(0,-12)[]{\small{$\tilde{\nu}_{\alpha}$}}
\end{picture}
\begin{picture}(110,70)(-55,-30)
\ArrowLine(-49,0)(-35,0)
\ArrowLine(-22,0)(-35,0)
\ArrowLine(-22,0)(-9,0)
\ArrowLine(9,0)(-9,0)
\ArrowLine(9,0)(22,0)
\ArrowLine(35,0)(22,0)
\ArrowLine(35,0)(49,0)
\DashArrowLine(-9,15)(-9,30){1}
\DashLine(-9,15)(-9,0){1}
\DashArrowLine(9,30)(9,15){1}
\DashLine(9,15)(9,0){1}
\DashArrowArc(0,20)(40,210,330){3}
\Text(-42,-7)[]{$\tau$}
\Text(42,-7)[]{$\mu$}
\Text(-32,10)[]{\ftsz{$\tilde{W}^+$}}
\Text(-16,10)[]{\ftsz{$\tilde{W}^-$}}
\Text(0,9)[]{\ftsz{$\tilde{H}_2^+$}}
\Text(18,10)[]{\ftsz{$\tilde{W}^-$}}
\Text(34,10)[]{\ftsz{$\tilde{W}^+$}}
\Text(-9,32)[b]{\small{$H_2^0$}}
\Text(9,32)[b]{\small{$H_2^0$}}
\Text(0,-12)[]{\small{$\tilde{\nu}_{\alpha}$}}
\end{picture}
\end{center}
\begin{center}
\begin{picture}(110,70)(-55,-30)
\ArrowLine(-49,0)(-35,0)
\ArrowLine(-9,0)(-35,0)
\ArrowLine(-9,0)(9,0)
\ArrowLine(35,0)(9,0)
\ArrowLine(35,0)(49,0)
\DashArrowLine(-9,30)(-9,15){1}
\DashLine(-9,15)(-9,0){1}
\DashArrowLine(9,15)(9,30){1}
\DashLine(9,15)(9,0){1}
\DashArrowArc(0,20)(40,210,330){3}
\Text(-42,-7)[]{$\tau$}
\Text(42,-7)[]{$\mu$}
\Text(-21,10)[]{\ftsz{$\tilde{W}^0$}}
\Text(-21,21)[]{\ftsz{$(\tilde{B}^0)$}}
\Text(0,9)[]{\ftsz{$\tilde{H}_i^0$}}
\Text(24,10)[]{\ftsz{$\tilde{W}^0$}}
\Text(24,21)[]{\ftsz{$(\tilde{B}^0)$}}
\Text(-9,32)[b]{\small{$H_i^0$}}
\Text(9,32)[b]{\small{$H_i^0$}}
\Text(0,-11)[]{\small{$\tilde{\ell}_{L_\alpha}$}}
\end{picture}
\begin{picture}(110,70)(-55,-30)
\ArrowLine(-49,0)(-35,0)
\ArrowLine(-22,0)(-35,0)
\ArrowLine(-22,0)(-9,0)
\ArrowLine(9,0)(-9,0)
\ArrowLine(9,0)(22,0)
\ArrowLine(35,0)(22,0)
\ArrowLine(35,0)(49,0)
\DashArrowLine(-9,15)(-9,30){1}
\DashLine(-9,15)(-9,0){1}
\DashArrowLine(9,30)(9,15){1}
\DashLine(9,15)(9,0){1}
\DashArrowArc(0,20)(40,210,330){3}
\Text(-42,-7)[]{$\tau$}
\Text(42,-7)[]{$\mu$}
\Text(-21,10)[]{\ftsz{$\tilde{W}^0$}}
\Text(-21,21)[]{\ftsz{$(\tilde{B}^0)$}}
\Text(0,9)[]{\ftsz{$\tilde{H}_i^0$}}
\Text(24,10)[]{\ftsz{$\tilde{W}^0$}}
\Text(24,21)[]{\ftsz{$(\tilde{B}^0)$}}
\Text(-9,32)[b]{\small{$H_i^0$}}
\Text(9,32)[b]{\small{$H_i^0$}}
\Text(0,-11)[]{\small{$\tilde{\ell}_{L_\alpha}$}}
\end{picture}
\begin{picture}(110,70)(-55,-30)
\ArrowLine(-49,0)(-35,0)
\ArrowLine(-9,0)(-35,0)
\ArrowLine(-9,0)(9,0)
\ArrowLine(35,0)(9,0)
\ArrowLine(35,0)(49,0)
\DashArrowLine(-9,30)(-9,15){1}
\DashLine(-9,15)(-9,0){1}
\DashArrowLine(9,15)(9,30){1}
\DashLine(9,15)(9,0){1}
\DashArrowArc(0,20)(40,210,330){3}
\Text(-42,-7)[]{$\tau$}
\Text(42,-7)[]{$\mu$}
\Text(-21,10)[]{\ftsz{$\tilde{W}^0$}}
\Text(-21,21)[]{\ftsz{$(\tilde{B}^0)$}}
\Text(0,9)[]{\ftsz{$\tilde{H}_i^0$}}
\Text(24,10)[]{\ftsz{$\tilde{B}^0$}}
\Text(24,21)[]{\ftsz{$(\tilde{W}^0)$}}
\Text(-9,32)[b]{\small{$H_i^0$}}
\Text(9,32)[b]{\small{$H_i^0$}}
\Text(0,-11)[]{\small{$\tilde{\ell}_{L_\alpha}$}}
\end{picture}
\begin{picture}(110,70)(-55,-30)
\ArrowLine(-49,0)(-35,0)
\ArrowLine(-22,0)(-35,0)
\ArrowLine(-22,0)(-9,0)
\ArrowLine(9,0)(-9,0)
\ArrowLine(9,0)(22,0)
\ArrowLine(35,0)(22,0)
\ArrowLine(35,0)(49,0)
\DashArrowLine(-9,15)(-9,30){1}
\DashLine(-9,15)(-9,0){1}
\DashArrowLine(9,30)(9,15){1}
\DashLine(9,15)(9,0){1}
\DashArrowArc(0,20)(40,210,330){3}
\Text(-42,-7)[]{$\tau$}
\Text(42,-7)[]{$\mu$}
\Text(-21,10)[]{\ftsz{$\tilde{W}^0$}}
\Text(-21,21)[]{\ftsz{$(\tilde{B}^0)$}}
\Text(0,9)[]{\ftsz{$\tilde{H}_i^0$}}
\Text(24,10)[]{\ftsz{$\tilde{B}^0$}}
\Text(24,21)[]{\ftsz{$(\tilde{W}^0)$}}
\Text(-9,32)[b]{\small{$H_i^0$}}
\Text(9,32)[b]{\small{$H_i^0$}}
\Text(0,-11)[]{\small{$\tilde{\ell}_{L_\alpha}$}}
\end{picture}
\end{center}
\begin{center}
\begin{picture}(110,70)(-55,-30)
\ArrowLine(-35,0)(-49,0)
\ArrowLine(-35,0)(-9,0)
\ArrowLine(9,0)(-9,0)
\ArrowLine(9,0)(35,0)
\ArrowLine(49,0)(35,0)
\DashArrowLine(-9,15)(-9,30){1}
\DashLine(-9,15)(-9,0){1}
\DashArrowLine(9,30)(9,15){1}
\DashLine(9,15)(9,0){1}
\DashArrowArc(0,20)(40,210,330){3}
\Text(-42,-7)[]{$\tau^c$}
\Text(44,-8)[]{$\mu^c$}
\Text(-21,10)[]{\ftsz{$\tilde{B}^0$}}
\Text(0,9)[]{\ftsz{$\tilde{H}_i^0$}}
\Text(24,10)[]{\ftsz{$\tilde{B}^0$}}
\Text(-9,32)[b]{\small{$H_i^0$}}
\Text(9,32)[b]{\small{$H_i^0$}}
\Text(0,-11)[]{\small{$\tilde{\ell}_{R_\alpha}$}}
\end{picture}
\begin{picture}(110,70)(-55,-30)
\ArrowLine(-35,0)(-49,0)
\ArrowLine(-35,0)(-22,0)
\ArrowLine(-9,0)(-22,0)
\ArrowLine(-9,0)(9,0)
\ArrowLine(22,0)(9,0)
\ArrowLine(22,0)(35,0)
\ArrowLine(49,0)(35,0)
\DashArrowLine(-9,30)(-9,15){1}
\DashLine(-9,15)(-9,0){1}
\DashArrowLine(9,15)(9,30){1}
\DashLine(9,15)(9,0){1}
\DashArrowArc(0,20)(40,210,330){3}
\Text(-42,-7)[]{$\tau^c$}
\Text(44,-8)[]{$\mu^c$}
\Text(-21,10)[]{\ftsz{$\tilde{B}^0$}}
\Text(0,9)[]{\ftsz{$\tilde{H}_i^0$}}
\Text(24,10)[]{\ftsz{$\tilde{B}^0$}}
\Text(-9,32)[b]{\small{$H_i^0$}}
\Text(9,32)[b]{\small{$H_i^0$}}
\Text(0,-11)[]{\small{$\tilde{\ell}_{R_\alpha}$}}
\end{picture}
\end{center}
\caption{\ftsz Diagrams that contribute to $A^{Z (a)}_L$
(first and second rows) and $A^{Z (a)}_R$ (third row).}
\label{alra}
\end{figure}

The coefficients $A^Z_{L,R}$ receive 
contributions of three different types:
$A^Z_{L,R} = A^{Z(a)}_{L,R}  +A^{Z(b)}_{L,R}  +A^{Z(c)}_{L,R}$.
The diagrams corresponding to $A^{Z(a)}_{L,R} $ are depicted in 
Fig.~\ref{alra} and give:
\bea
{ A^{Z(a)}_L \over s_L c_L}  & = & 
{g^2 c_W^2 \over 16 \pi^2} \cdot {1 \over 8}  \left[
-(2+3 c_{2 \beta})\left( \mu^2 J_5(M_2^2,M_2^2,\mu^2,\mu^2,\mt^2_{L_2})
+ 2 J_4(M_2^2,M_2^2,\mu^2,\mt^2_{L_2}) \right)
\right.
\nonumber \\
& & \left.
-(2-3 c_{2 \beta}) M_2^2 \left(\mu^2 I_5(M_2^2,M_2^2,\mu^2,\mu^2,\mt^2_{L_2}) 
- I_4(M_2^2,M_2^2,\mu^2,\mt^2_{L_2}) \right) \right.
\nonumber \\
& & \left.
+ 4 s_{2 \beta} \mu M_2 \, J_5(M_2^2,M_2^2,\mu^2,\mu^2,\mt^2_{L_2}) \right]
\nonumber \\
& + & 
{ g'^2 c_W^2  \over 16 \pi^2} \cdot {1\over 4} \, c_{2 \beta} \left[
\mu^2 J_5(M_1^2,M_2^2,\mu^2,\mu^2,\mt^2_{L_2})
+ 2 J_4(M_1^2,M_2^2,\mu^2,\mt^2_{L_2}) \right.
\nonumber \\
& & \left.
- M_1 M_2 \left( \mu^2 I_5(M_1^2,M_2^2,\mu^2,\mu^2,\mt^2_{L_2}) 
- I_4(M_1^2,M_2^2,\mu^2,\mt^2_{L_2}) \right) \right]
\nonumber \\
& + & 
{ g'^2 s_W^2 \over 16 \pi^2} \cdot {1\over 8} \, c_{2 \beta} \left[
- \mu^2 J_5(M_1^2,M_1^2,\mu^2,\mu^2,\mt^2_{L_2})
- 2 J_4(M_1^2,M_1^2,\mu^2,\mt^2_{L_2}) \right.
\nonumber \\
& & \left.
+ M_1^2 \left( \mu^2 I_5(M_1^2,M_1^2,\mu^2,\mu^2,\mt^2_{L_2}) 
-  I_4(M_1^2,M_1^2,\mu^2,\mt^2_{L_2}) \right) \right]
\nonumber \\
& - &
(L_2 \to L_3)
\\
{ A^{Z (a)}_R \over s_R c_R}  & = & 
{ g'^2 s_W^2 \over 16 \pi^2} \cdot {1\over 2} \, c_{2 \beta} \left[
\mu^2 J_5(M_1^2,M_1^2,\mu^2,\mu^2,\mt^2_{R_2})
+ 2 J_4(M_1^2,M_1^2,\mu^2,\mt^2_{R_2}) \right.
\nonumber \\
& & \left.
- M_1^2 \left( \mu^2 I_5(M_1^2,M_1^2,\mu^2,\mu^2,\mt^2_{R_2}) 
- I_4(M_1^2,M_1^2,\mu^2,\mt^2_{R_2}) \right) \right]
\nonumber \\
& - &
( R_2 \to R_3)
\eea

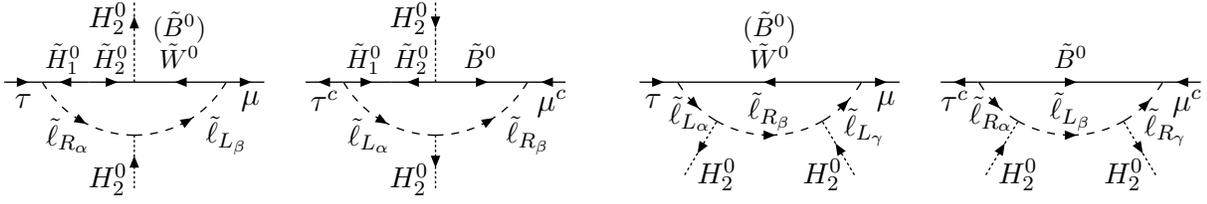
\begin{figure}[htb]
\begin{center}
\begin{picture}(110,80)(-55,-40)
\ArrowLine(-49,0)(-35,0)
\ArrowLine(-17,0)(-35,0)
\ArrowLine(-17,0)(0,0)
\ArrowLine(35,0)(0,0)
\ArrowLine(35,0)(49,0)
\DashArrowLine(0,15)(0,30){1}
\DashLine(0,15)(0,0){1}
\DashArrowLine(0,-40)(0,-20){1}
\DashArrowArc(0,20)(40,210,270){3}
\DashArrowArc(0,20)(40,270,330){3}
\Text(-42,-7)[]{$\tau$}
\Text(44,-7)[]{$\mu$}
\Text(-27,9)[]{\ftsz{$\tilde{H}^0_1$}}
\Text(-9,9)[]{\ftsz{$\tilde{H}^0_2$}}
\Text(17,10)[]{\ftsz{$\tilde{W}^0$}}
\Text(17,21)[]{\ftsz{$(\tilde{B}^0)$}}
\Text(-10,30)[t]{\small{$H_2^0$}}
\Text(-10,-40)[b]{\small{$H_2^0$}}
\Text(-25,-20)[]{\small{$\tilde{\ell}_{R_\alpha}$}}
\Text(35,-20)[]{\small{$\tilde{\ell}_{L_\beta}$}}
\end{picture}
\begin{picture}(110,80)(-55,-40)
\ArrowLine(-35,0)(-49,0)
\ArrowLine(-35,0)(-17,0)
\ArrowLine(0,0)(-17,0)
\ArrowLine(0,0)(35,0)
\ArrowLine(49,0)(35,0)
\DashArrowLine(0,30)(0,15){1}
\DashLine(0,15)(0,0){1}
\DashArrowLine(0,-20)(0,-40){1}
\DashArrowArc(0,20)(40,210,270){3}
\DashArrowArc(0,20)(40,270,330){3}
\Text(-42,-7)[]{$\tau^c$}
\Text(44,-8)[]{$\mu^c$}
\Text(-27,9)[]{\ftsz{$\tilde{H}^0_1$}}
\Text(-9,9)[]{\ftsz{$\tilde{H}^0_2$}}
\Text(17,10)[]{\ftsz{$\tilde{B}^0$}}
\Text(-10,30)[t]{\small{$H_2^0$}}
\Text(-10,-40)[b]{\small{$H_2^0$}}
\Text(-25,-20)[]{\small{$\tilde{\ell}_{L_\alpha}$}}
\Text(35,-20)[]{\small{$\tilde{\ell}_{R_\beta}$}}
\end{picture}
\hspace{0.3 cm}
\begin{picture}(110,80)(-55,-40)
\ArrowLine(-49,0)(-35,0)
\ArrowLine(35,0)(-35,0)
\ArrowLine(35,0)(49,0)
\DashArrowLine(-20,-15)(-32,-35){1}
\DashArrowLine(32,-35)(20,-15){1}
\DashArrowArc(0,20)(40,210,240){3}
\DashArrowArc(0,20)(40,240,300){3}
\DashArrowArc(0,20)(40,300,330){3}
\Text(-44,-7)[]{$\tau$}
\Text(44,-7)[]{$\mu$}
\Text(0,10)[]{\ftsz{$\tilde{W}^0$}}
\Text(0,21)[]{\ftsz{$(\tilde{B}^0)$}}
\Text(-20,-35)[]{\small{$H_2^0$}}
\Text(20,-35)[]{\small{$H_2^0$}}
\Text(-30,-12)[]{\small{$\tilde{\ell}_{L_\alpha}$}}
\Text(36,-17)[]{\small{$\tilde{\ell}_{L_\gamma}$}}
\Text(0,-11)[]{\small{$\tilde{\ell}_{R_\beta}$}}
\end{picture}
\begin{picture}(110,80)(-55,-40)
\ArrowLine(-35,0)(-49,0)
\ArrowLine(-35,0)(35,0)
\ArrowLine(49,0)(35,0)
\DashArrowLine(-32,-35)(-20,-15){1}
\DashArrowLine(20,-15)(32,-35){1}
\DashArrowArc(0,20)(40,210,240){3}
\DashArrowArc(0,20)(40,240,300){3}
\DashArrowArc(0,20)(40,300,330){3}
\Text(-44,-7)[]{$\tau^c$}
\Text(44,-8)[]{$\mu^c$}
\Text(0,10)[]{\ftsz{$\tilde{B}^0$}}
\Text(-20,-35)[]{\small{$H_2^0$}}
\Text(20,-35)[]{\small{$H_2^0$}}
\Text(-30,-12)[]{\small{$\tilde{\ell}_{R_\alpha}$}}
\Text(36,-17)[]{\small{$\tilde{\ell}_{R_\gamma}$}}
\Text(0,-11)[]{\small{$\tilde{\ell}_{L_\beta}$}}
\end{picture}
\end{center}
\caption{\ftsz Diagrams that contribute to $A^{Z (b)}_{L,R}$
(left side) and $A^{Z (c)}_{L,R}$ (right side).}
\label{alrbc}
\end{figure}

The diagrams corresponding to $A^{Z(b)}_{L,R}$ (Fig.~\ref{alrbc})
give:
\bea
{ A^{Z(b)}_L \over s_L c_L}  & = & 
{ h_{\tau}^2 c_W^2 \over 16 \pi^2}\cdot {1 \over 4} \,
s_{\beta}^2 \mu^2 \left[
s_R^2 \left( J_5(M_2^2,\mu^2,\mt^2_{R_2},\mt^2_{R_2},\mt^2_{L_2}) 
+ J_5(M_2^2,\mu^2,\mu^2,\mt^2_{R_2},\mt^2_{L_2})\right)\right.
\nonumber \\
& & \left. +
c_R^2 \left( J_5(M_2^2,\mu^2,\mt^2_{R_3},\mt^2_{R_3},\mt^2_{L_2}) 
+ J_5(M_2^2,\mu^2,\mu^2,\mt^2_{R_3},\mt^2_{L_2})\right)\right]
\nonumber \\
& + & 
{ h_{\tau}^2 s_W^2 \over 16 \pi^2}\cdot {1 \over 4} \,
s_{\beta}^2 \mu^2 \left[
-s_R^2 \left( J_5(M_1^2,\mu^2,\mt^2_{R_2},\mt^2_{R_2},\mt^2_{L_2}) 
+ J_5(M_1^2,\mu^2,\mu^2,\mt^2_{R_2},\mt^2_{L_2})\right)\right.
\nonumber \\
& & \left. 
-c_R^2 \left( J_5(M_1^2,\mu^2,\mt^2_{R_3},\mt^2_{R_3},\mt^2_{L_2}) 
+ J_5(M_1^2,\mu^2,\mu^2,\mt^2_{R_3},\mt^2_{L_2})\right)\right]
\nonumber \\
& - & (L_2 \to L_3)
\\
{ A^{Z (b)}_R \over s_R c_R}  & = & 
{ h_{\tau}^2 s_W^2 \over 16 \pi^2}\cdot {1 \over 2} \, 
s_{\beta}^2  \mu^2  \left[
-s_L^2 \left( J_5(M_1^2,\mu^2,\mt^2_{L_2},\mt^2_{L_2},\mt^2_{R_2}) 
+ J_5(M_1^2,\mu^2,\mu^2,\mt^2_{L_2},\mt^2_{R_2})\right)\right.
\nonumber \\
& & \left. 
-c_L^2 \left( J_5(M_1^2,\mu^2,\mt^2_{L_3},\mt^2_{L_3},\mt^2_{R_2}) 
+ J_5(M_1^2,\mu^2,\mu^2,\mt^2_{L_3},\mt^2_{R_2})\right)\right]
\nonumber \\
& - & (R_2 \to R_3)
\eea

The diagrams corresponding to $A^{Z(c)}_{L,R} $ (Fig.~\ref{alrbc}) give: 
\bea
{ A^{Z(c)}_L \over s_L c_L}  & = & 
{ h_{\tau}^2 c_W^2 \over 16 \pi^2} \cdot {1 \over 4}\, 
s_{\beta}^2 \mu^2 \left[ s_L^2 \left( 
s_R^2 J_5(M_2^2,\mt^2_{R_2},\mt^2_{R_2},\mt^2_{L_2},\mt^2_{L_2}) 
\right. \right.
\nonumber \\
& & \left. \left.
+ c_R^2 J_5(M_2^2,\mt^2_{R_3},\mt^2_{R_3},\mt^2_{L_2},\mt^2_{L_2}) 
\right) 
- c_L^2 \left( 
s_R^2 J_5(M_2^2,\mt^2_{R_2},\mt^2_{R_2},\mt^2_{L_3},\mt^2_{L_3}) 
\right. \right.
\nonumber \\
& & \left. \left.
+ c_R^2 J_5(M_2^2,\mt^2_{R_3},\mt^2_{R_3},\mt^2_{L_3},\mt^2_{L_3}) 
\right) 
- (s_L^2-c_L^2) \left( 
s_R^2 J_5(M_2^2,\mt^2_{R_2},\mt^2_{R_2},\mt^2_{L_2},\mt^2_{L_3}) 
\right. \right.
\nonumber \\
& &  \left. \left.
+ c_R^2 J_5(M_2^2,\mt^2_{R_3},\mt^2_{R_3},\mt^2_{L_2},\mt^2_{L_3}) 
\right) \right]
\nonumber \\
& + & { h_{\tau}^2 s_W^2 \over 16 \pi^2} \cdot {1 \over 4} \, 
s_{\beta}^2 \mu^2 \left[ s_L^2 \left( 
s_R^2 J_5(M_1^2,\mt^2_{R_2},\mt^2_{R_2},\mt^2_{L_2},\mt^2_{L_2}) 
\right. \right.
\nonumber \\
& & \left. \left.
+ c_R^2 J_5(M_1^2,\mt^2_{R_3},\mt^2_{R_3},\mt^2_{L_2},\mt^2_{L_2}) 
\right) 
- c_L^2 \left( 
s_R^2 J_5(M_1^2,\mt^2_{R_2},\mt^2_{R_2},\mt^2_{L_3},\mt^2_{L_3}) 
\right. \right.
\nonumber \\
& & \left. \left.
+ c_R^2 J_5(M_1^2,\mt^2_{R_3},\mt^2_{R_3},\mt^2_{L_3},\mt^2_{L_3}) 
\right) 
- (s_L^2-c_L^2) \left( 
s_R^2 J_5(M_1^2,\mt^2_{R_2},\mt^2_{R_2},\mt^2_{L_2},\mt^2_{L_3}) 
\right. \right.
\nonumber \\
& &  \left. \left.
+ c_R^2 J_5(M_1^2,\mt^2_{R_3},\mt^2_{R_3},\mt^2_{L_2},\mt^2_{L_3}) 
\right) \right]
\\
{ A^{Z (c)}_R \over s_R c_R}  & = & 
{ h_{\tau}^2 s_W^2 \over 16 \pi^2} 
s_{\beta}^2 \mu^2 \left[ - s_R^2 \left( 
s_L^2 J_5(M_1^2,\mt^2_{L_2},\mt^2_{L_2},\mt^2_{R_2},\mt^2_{R_2}) 
\right. \right.
\nonumber \\
& & \left. \left.
+ c_L^2 J_5(M_1^2,\mt^2_{L_3},\mt^2_{L_3},\mt^2_{R_2},\mt^2_{R_2}) 
\right) 
+ c_R^2 \left( 
s_L^2 J_5(M_1^2,\mt^2_{L_2},\mt^2_{L_2},\mt^2_{R_3},\mt^2_{R_3}) 
\right. \right.
\nonumber \\
& & \left. \left.
+ c_L^2 J_5(M_1^2,\mt^2_{L_3},\mt^2_{L_3},\mt^2_{R_3},\mt^2_{R_3}) 
\right) 
+ (s_R^2-c_R^2) \left( 
s_L^2 J_5(M_1^2,\mt^2_{L_2},\mt^2_{L_2},\mt^2_{R_2},\mt^2_{R_3}) 
\right. \right.
\nonumber \\
& &  \left. \left.
+ c_L^2 J_5(M_1^2,\mt^2_{L_3},\mt^2_{L_3},\mt^2_{R_2},\mt^2_{R_3}) 
\right) \right]
\eea

\subsection{ Contributions to $C^{\ga}_{L,R}$}

The diagrams corresponding to $C^{\ga}_{L,R} $ are depicted in 
Fig.~\ref{cga} and give:
\bea
{C^{\ga}_L  \over s_L c_L}& = & 
{g^2 \over 16 \pi^2} \cdot {1 \over 12}  \left[
- K_5(M_2^2,\mt^2_{L_2},\mt^2_{L_2},\mt^2_{L_2},\mt^2_{L_2})
\right.
\nonumber \\
& & \left.
- 4 K_5(M_2^2,M_2^2,M_2^2,M_2^2,\mt^2_{L_2}) 
+ 6 M_2^2 J_5(M_2^2,M_2^2,M_2^2,M_2^2,\mt^2_{L_2}) \right]
\nonumber \\
& + & 
{g'^2 \over 16 \pi^2} \cdot {1 \over 12} 
[- K_5(M_1^2,\mt^2_{L_2},\mt^2_{L_2},\mt^2_{L_2},\mt^2_{L_2})]
- (L_2 \to L_3)
\\
{C^{\ga}_R  \over s_R c_R} & = & 
{g'^2 \over 16 \pi^2} \cdot {1 \over 3} \, 
[- K_5(M_1^2,\mt^2_{R_2},\mt^2_{R_2},\mt^2_{R_2},\mt^2_{R_2})]
- (R_2 \to R_3)
\eea
%

\begin{figure}[htb]
\begin{center}
\begin{picture}(110,60)(-55,-30)
\ArrowLine(-49,0)(-35,0)
\ArrowLine(0,0)(-35,0)
\ArrowLine(35,0)(0,0)
\ArrowLine(35,0)(49,0)
\Photon(0,0)(0,25){2}{3}
\GCirc(0,0){2}{0.5}
\DashArrowArc(0,20)(40,210,330){3}
\Text(-42,-7)[]{$\tau$}
\Text(42,-7)[]{$\mu$}
\Text(-17,10)[]{\ftsz{$\tilde{W}^+$}}
\Text(17,10)[]{\ftsz{$\tilde{W}^+$}}
\Text(7,25)[]{$\gamma$}
\Text(0,-12)[]{\small{$\tilde{\nu}_{\alpha}$}}
\end{picture}
\begin{picture}(110,60)(-55,-30)
\ArrowLine(-49,0)(-35,0)
\ArrowLine(35,0)(-35,0)
\ArrowLine(35,0)(49,0)
\DashArrowArc(0,20)(40,210,270){3}
\DashArrowArc(0,20)(40,270,330){3}
\Photon(0,-20)(0,-40){2}{3}
\Text(-44,-7)[]{$\tau$}
\Text(44,-7)[]{$\mu$}
\Text(0,10)[]{\ftsz{$\tilde{W}^0$}}
\Text(0,21)[]{\ftsz{$(\tilde{B}^0)$}}
\Text(7,-40)[]{$\gamma$}
\Text(0,-11)[]{\small{$\tilde{\ell}_{L_\alpha}$}}
\end{picture}
\hspace{0.3 cm}
\begin{picture}(110,60)(-55,-30)
\ArrowLine(-35,0)(-49,0)
\ArrowLine(-35,0)(35,0)
\ArrowLine(49,0)(35,0)
\DashArrowArc(0,20)(40,210,270){3}
\DashArrowArc(0,20)(40,270,330){3}
\Photon(0,-20)(0,-40){2}{3}
\Text(-44,-7)[]{$\tau^c$}
\Text(44,-8)[]{$\mu^c$}
\Text(0,10)[]{\ftsz{$\tilde{B}^0$}}
\Text(7,-40)[]{$\gamma$}
\Text(0,-11)[]{\small{$\tilde{\ell}_{R_\alpha}$}}
\end{picture}
\end{center}
\caption{\ftsz Diagrams that contribute to $C^{\gamma}_{L,R}$.}
\label{cga}
\end{figure}

\subsection{Contributions to $C^Z_{L,R}$}

The diagrams corresponding to $C^{Z}_{L,R} $ are depicted in 
Fig.~\ref{cz} and give:
\bea
{C^Z_L  \over s_L c_L} & = & 
{g^2  \over 16 \pi^2}  \cdot {1 \over 24} \left[  
(1 + 2 s_W^2) 
K_5(M_2^2,\mt^2_{L_2},\mt^2_{L_2},\mt^2_{L_2},\mt^2_{L_2})
\right.
\nonumber \\
& & \left.
- 4 c_W^2 \left( 2 K_5(M_2^2,M_2^2,M_2^2,M_2^2,\mt^2_{L_2}) 
- 3 M_2^2 J_5(M_2^2,M_2^2,M_2^2,M_2^2,\mt^2_{L_2}) \right) \right]
\nonumber \\
& + & 
{g'^2 \over 16 \pi^2}  \cdot {1 \over 24} \, 
( -1 + 2 s_W^2 ) 
K_5(M_1^2,\mt^2_{L_2},\mt^2_{L_2},\mt^2_{L_2},\mt^2_{L_2})
- (L_2 \to L_3)
\\
{C^Z_R \over s_R c_R} & = & 
{g'^2 \over 16 \pi^2}  \cdot {1 \over 3} \,
s_W^2  K_5(M_1^2,\mt^2_{R_2},\mt^2_{R_2},\mt^2_{R_2},\mt^2_{R_2})
- (R_2 \to R_3)
\eea
%

\begin{figure}[htb]
\begin{center}
\begin{picture}(110,60)(-55,-30)
\ArrowLine(-49,0)(-35,0)
\ArrowLine(0,0)(-35,0)
\ArrowLine(35,0)(0,0)
\ArrowLine(35,0)(49,0)
\Photon(0,0)(0,25){2}{3}
\GCirc(0,0){2}{0.5}
\DashArrowArc(0,20)(40,210,330){3}
\Text(-42,-7)[]{$\tau$}
\Text(42,-7)[]{$\mu$}
\Text(-17,10)[]{\ftsz{$\tilde{W}^+$}}
\Text(17,10)[]{\ftsz{$\tilde{W}^+$}}
\Text(8,25)[]{$Z$}
\Text(0,-12)[]{\small{$\tilde{\nu}_{\alpha}$}}
\end{picture}
\begin{picture}(110,60)(-55,-30)
\ArrowLine(-49,0)(-35,0)
\ArrowLine(35,0)(-35,0)
\ArrowLine(35,0)(49,0)
\DashArrowArc(0,20)(40,210,270){3}
\DashArrowArc(0,20)(40,270,330){3}
\Photon(0,-20)(0,-40){2}{3}
\Text(-44,-7)[]{$\tau$}
\Text(44,-7)[]{$\mu$}
\Text(0,10)[]{\ftsz{$\tilde{W}^+$}}
\Text(7,-40)[]{$Z$}
\Text(0,-12)[]{\small{$\tilde{\nu}_{\alpha}$}}
\end{picture}
\begin{picture}(110,60)(-55,-30)
\ArrowLine(-49,0)(-35,0)
\ArrowLine(35,0)(-35,0)
\ArrowLine(35,0)(49,0)
\DashArrowArc(0,20)(40,210,270){3}
\DashArrowArc(0,20)(40,270,330){3}
\Photon(0,-20)(0,-40){2}{3}
\Text(-44,-7)[]{$\tau$}
\Text(44,-7)[]{$\mu$}
\Text(0,10)[]{\ftsz{$\tilde{W}^0$}}
\Text(0,21)[]{\ftsz{$(\tilde{B}^0)$}}
\Text(7,-40)[]{$Z$}
\Text(0,-11)[]{\small{$\tilde{\ell}_{L_\alpha}$}}
\end{picture}
\hspace{0.3 cm}
\begin{picture}(110,60)(-55,-30)
\ArrowLine(-35,0)(-49,0)
\ArrowLine(-35,0)(35,0)
\ArrowLine(49,0)(35,0)
\DashArrowArc(0,20)(40,210,270){3}
\DashArrowArc(0,20)(40,270,330){3}
\Photon(0,-20)(0,-40){2}{3}
\Text(-44,-7)[]{$\tau^c$}
\Text(44,-8)[]{$\mu^c$}
\Text(0,10)[]{\ftsz{$\tilde{B}^0$}}
\Text(7,-40)[]{$Z$}
\Text(0,-11)[]{\small{$\tilde{\ell}_{R_\alpha}$}}
\end{picture}
\end{center}
\caption{\ftsz Diagrams that contribute to $C^Z_{L,R}$.}
\label{cz}
\end{figure}

\subsection{ Contributions to $D^{\ga}_{L,R}$} 
The coefficients $D^\ga_{L,R}$ receive 
contributions of three different types:
$D^\ga_{L,R} = D^{\ga(a)}_{L,R}  +D^{\ga(b)}_{L,R} +D^{\ga(c)}_{L,R}$.

The diagrams corresponding to $D^{\ga(a)}_{L,R} $ are depicted in 
Fig.~\ref{dgalra} and give:
\bea
{ D^{\ga (a)}_L  \over s_L c_L} & = & 
{g^2 \over 16 \pi^2} \cdot {1 \over 4}  \left[ 
\mt^2_{L_2} J_5(M_2^2,\mt^2_{L_2},\mt^2_{L_2},\mt^2_{L_2},\mt^2_{L_2})
-2 M_2^2 J_5(M_2^2,M_2^2,M_2^2,M_2^2,\mt^2_{L_2}) \right]
\nonumber \\
& + & 
{g'^2 \over 16 \pi^2} \cdot {1 \over 4}  
\mt^2_{L_2} J_5(M_1^2,\mt^2_{L_2},\mt^2_{L_2},\mt^2_{L_2},\mt^2_{L_2})
- (L_2 \to L_3)
\\
{ D^{\ga (a)}_R  \over s_R c_R} & = & 
{g'^2 \over 16 \pi^2}
\mt^2_{R_2} J_5(M_1^2,\mt^2_{R_2},\mt^2_{R_2},\mt^2_{R_2},\mt^2_{R_2})
- (R_2 \to R_3)
\eea

%
\begin{figure}[htb]
\begin{center}
\begin{picture}(120,60)(-65,-30)
\ArrowLine(-49,0)(-63,0)
\ArrowLine(-49,0)(-35,0)
\ArrowLine(0,0)(-35,0)
\ArrowLine(35,0)(0,0)
\ArrowLine(35,0)(49,0)
\DashArrowLine(-49,0)(-49,20){1}
\Photon(0,0)(0,25){2}{3}
\GCirc(0,0){2}{0.5}
\DashArrowArc(0,20)(40,210,330){3}
\Text(-56,-7)[]{$\tau^c$}
\Text(-42,-8)[]{$\tau$}
\Text(42,-7)[]{$\mu$}
\Text(-49,25)[]{\small{$H_1^0$}}
\Text(-17,10)[]{\ftsz{$\tilde{W}^+$}}
\Text(17,10)[]{\ftsz{$\tilde{W}^+$}}
\Text(7,25)[]{$\gamma$}
\Text(0,-12)[]{\small{$\tilde{\nu}_{\alpha}$}}
\end{picture}
\begin{picture}(120,60)(-65,-30)
\ArrowLine(-49,0)(-63,0)
\ArrowLine(-49,0)(-35,0)
\ArrowLine(35,0)(-35,0)
\ArrowLine(35,0)(49,0)
\DashArrowLine(-49,0)(-49,20){1}
\DashArrowArc(0,20)(40,210,270){3}
\DashArrowArc(0,20)(40,270,330){3}
\Photon(0,-20)(0,-40){2}{3}
\Text(-56,-7)[]{$\tau^c$}
\Text(-42,-8)[]{$\tau$}
\Text(44,-7)[]{$\mu$}
\Text(-49,25)[]{\small{$H_1^0$}}
\Text(0,10)[]{\ftsz{$\tilde{W}^0$}}
\Text(0,21)[]{\ftsz{$(\tilde{B}^0)$}}
\Text(7,-40)[]{$\gamma$}
\Text(0,-11)[]{\small{$\tilde{\ell}_{L_\alpha}$}}
\end{picture}
\hspace{0.3 cm}
\begin{picture}(120,60)(-65,-30)
\ArrowLine(-63,0)(-49,0)
\ArrowLine(-35,0)(-49,0)
\ArrowLine(-35,0)(35,0)
\ArrowLine(49,0)(35,0)
\DashArrowLine(-49,20)(-49,0){1}
\DashArrowArc(0,20)(40,210,270){3}
\DashArrowArc(0,20)(40,270,330){3}
\Photon(0,-20)(0,-40){2}{3}
\Text(-56,-8)[]{$\tau$}
\Text(-42,-7)[]{$\tau^c$}
\Text(44,-8)[]{$\mu^c$}
\Text(-49,25)[]{\small{$H_1^0$}}
\Text(0,10)[]{\ftsz{$\tilde{B}^0$}}
\Text(7,-40)[]{$\gamma$}
\Text(0,-11)[]{\small{$\tilde{\ell}_{R_\alpha}$}}
\end{picture}
\end{center}
\caption{\ftsz Diagrams that contribute to $D^{\gamma (a)}_{L,R}$.}
\label{dgalra}
\end{figure}

The diagrams corresponding to $D^{\ga(b)}_{L,R} $ are depicted in 
Fig.~\ref{dgalrb} and give:
\bea
{ D^{\ga (b)}_L  \over s_L c_L} & = & 
{g^2 \over 16 \pi^2}  \cdot {1 \over 2} \left[ 
- \mt^2_{L_2} 
J_5(M_2^2,\mu^2,\mt^2_{L_2},\mt^2_{L_2},\mt^2_{L_2})
- 2 \mt^4_{L_2} I_5(M_2^2,\mu^2,\mt^2_{L_2},\mt^2_{L_2},\mt^2_{L_2})
\right.
\nonumber \\
& & \left.
+ M_2 \mu \tan\beta \left(   \mt^2_{L_2} 
I_5(M_2^2,\mu^2,\mt^2_{L_2},\mt^2_{L_2},\mt^2_{L_2})
+ 2 J_5(M_2^2,M_2^2,\mu^2,\mu^2,\mt^2_{L_2}) \right. \right.
\nonumber \\
& & \left. \left.
+ 2 J_5(M_2^2,M_2^2,M_2^2,\mu^2,\mt^2_{L_2})
+ 2 J_5(M_2^2,\mu^2,\mu^2,\mu^2,\mt^2_{L_2})
\right) \right]
\nonumber \\
& + & 
{g'^2 \over  16 \pi^2} \, {\mt^2_{L_2}  \over 2}  
\left[ J_5(M_1^2,\mu^2,\mt^2_{L_2},\mt^2_{L_2},\mt^2_{L_2}) 
- M_1 \mu \tan\beta \,
I_5(M_1^2,\mu^2,\mt^2_{L_2},\mt^2_{L_2},\mt^2_{L_2}) \right]
\nonumber \\
& - & (L_2 \to L_3)
\\
{ D^{\ga (b)}_R  \over s_R c_R} & = & 
{g'^2 \over  16 \pi^2} \mt^2_{R_2} 
\left[ - J_5(M_1^2,\mu^2,\mt^2_{R_2},\mt^2_{R_2},\mt^2_{R_2}) 
+ M_1 \mu \tan\beta \,
I_5(M_1^2,\mu^2,\mt^2_{R_2},\mt^2_{R_2},\mt^2_{R_2}) \right]
\nonumber \\
& - & (R_2 \to R_3)
\eea

\begin{figure}[htb]
\begin{center}
\begin{picture}(110,60)(-55,-30)
\ArrowLine(-35,0)(-49,0)
\ArrowLine(-35,0)(0,0)
\ArrowLine(20,0)(0,0)
\ArrowLine(35,0)(20,0)
\ArrowLine(35,0)(49,0)
\DashLine(0,0)(0,15){1}
\DashArrowLine(0,15)(0,30){1}
\Photon(20,0)(20,30){2}{3}
\GCirc(20,0){2}{0.5}
\DashArrowArc(0,20)(40,210,330){3}
\Text(-42,-8)[]{$\tau^c$}
\Text(42,-7)[]{$\mu$}
\Text(-10,25)[]{\small{$H_1^0$}}
\Text(-15,8)[]{\ftsz{$\tilde{H}^-_1$}}
\Text(10,9)[]{\ftsz{$\tilde{W}^+$}}
\Text(32,9)[]{\ftsz{$\tilde{W}^+$}}
\Text(28,25)[]{$\gamma$}
\Text(0,-12)[]{\small{$\tilde{\nu}_{\alpha}$}}
\end{picture}
\begin{picture}(110,60)(-55,-30)
\ArrowLine(-35,0)(-49,0)
\ArrowLine(-35,0)(-18,0)
\ArrowLine(-18,0)(0,0)
\ArrowLine(35,0)(0,0)
\ArrowLine(35,0)(49,0)
\DashLine(0,0)(0,15){1}
\DashArrowLine(0,15)(0,30){1}
\Photon(-20,0)(-20,30){2}{3}
\GCirc(-20,0){2}{0.5}
\DashArrowArc(0,20)(40,210,330){3}
\Text(-42,-8)[]{$\tau^c$}
\Text(42,-7)[]{$\mu$}
\Text(10,25)[]{\small{$H_1^0$}}
\Text(-29,8)[]{\ftsz{$\tilde{H}^-_1$}}
\Text(-9,8)[]{\ftsz{$\tilde{H}^-_1$}}
\Text(20,9)[]{\ftsz{$\tilde{W}^+$}}
\Text(-28,25)[]{$\gamma$}
\Text(0,-12)[]{\small{$\tilde{\nu}_{\alpha}$}}
\end{picture}
\begin{picture}(110,60)(-55,-30)
\ArrowLine(-35,0)(-49,0)
\ArrowLine(-35,0)(-18,0)
\ArrowLine(0,0)(-18,0)
\ArrowLine(0,0)(20,0)
\ArrowLine(35,0)(20,0)
\ArrowLine(35,0)(49,0)
\DashLine(0,0)(0,15){1}
\DashArrowLine(0,30)(0,15){1}
\Photon(20,0)(20,30){2}{3}
\GCirc(20,0){2}{0.5}
\DashArrowArc(0,20)(40,210,330){3}
\Text(-42,-8)[]{$\tau^c$}
\Text(42,-7)[]{$\mu$}
\Text(-10,25)[]{\small{$H_2^0$}}
\Text(-28,8)[]{\ftsz{$\tilde{H}^-_1$}}
\Text(-13,8)[]{\ftsz{$\tilde{H}^+_2$}}
\Text(10,9)[]{\ftsz{$\tilde{W}^-$}}
\Text(32,9)[]{\ftsz{$\tilde{W}^+$}}
\Text(28,25)[]{$\gamma$}
\Text(0,-12)[]{\small{$\tilde{\nu}_{\alpha}$}}
\end{picture}
\begin{picture}(110,60)(-55,-30)
\ArrowLine(-35,0)(-49,0)
\ArrowLine(-35,0)(-20,0)
\ArrowLine(0,0)(-20,0)
\ArrowLine(0,0)(20,0)
\ArrowLine(35,0)(20,0)
\ArrowLine(35,0)(49,0)
\DashLine(0,0)(0,15){1}
\DashArrowLine(0,30)(0,15){1}
\Photon(-20,0)(-20,30){2}{3}
\GCirc(-20,0){2}{0.5}
\DashArrowArc(0,20)(40,210,330){3}
\Text(-42,-8)[]{$\tau^c$}
\Text(42,-7)[]{$\mu$}
\Text(10,25)[]{\small{$H_2^0$}}
\Text(-30,8)[]{\ftsz{$\tilde{H}^-_1$}}
\Text(-8,8)[]{\ftsz{$\tilde{H}^+_2$}}
\Text(12,9)[]{\ftsz{$\tilde{W}^-$}}
\Text(30,9)[]{\ftsz{$\tilde{W}^+$}}
\Text(-28,25)[]{$\gamma$}
\Text(0,-12)[]{\small{$\tilde{\nu}_{\alpha}$}}
\end{picture}
\end{center}
\begin{center}
\begin{picture}(110,70)(-55,-40)
\ArrowLine(-35,0)(-49,0)
\ArrowLine(-35,0)(0,0)
\ArrowLine(35,0)(0,0)
\ArrowLine(35,0)(49,0)
\DashLine(0,0)(0,15){1}
\DashArrowLine(0,15)(0,30){1}
\Photon(0,-20)(0,-40){2}{3}
\DashArrowArc(0,20)(40,210,270){3}
\DashArrowArc(0,20)(40,270,330){3}
\Text(-42,-8)[]{$\tau^c$}
\Text(42,-7)[]{$\mu$}
\Text(-10,25)[]{\small{$H_1^0$}}
\Text(-15,8)[]{\ftsz{$\tilde{H}^0_1$}}
\Text(18,9)[]{\ftsz{$\tilde{W}^0$}}
\Text(18,20)[]{\ftsz{$(\tilde{B}^0)$}}
\Text(8,-40)[]{$\gamma$}
\Text(0,-11)[]{\small{$\tilde{\ell}_{L_\alpha}$}}
\end{picture}
\begin{picture}(110,70)(-55,-40)
\ArrowLine(-35,0)(-49,0)
\ArrowLine(-35,0)(-18,0)
\ArrowLine(0,0)(-18,0)
\ArrowLine(0,0)(18,0)
\ArrowLine(35,0)(18,0)
\ArrowLine(35,0)(49,0)
\DashLine(0,0)(0,15){1}
\DashArrowLine(0,30)(0,15){1}
\Photon(0,-20)(0,-40){2}{3}
\DashArrowArc(0,20)(40,210,270){3}
\DashArrowArc(0,20)(40,270,330){3}
\Text(-42,-8)[]{$\tau^c$}
\Text(42,-7)[]{$\mu$}
\Text(-10,25)[]{\small{$H_2^0$}}
\Text(-27,8)[]{\ftsz{$\tilde{H}^0_1$}}
\Text(-8,8)[]{\ftsz{$\tilde{H}^0_2$}}
\Text(18,9)[]{\ftsz{$\tilde{W}^0$}}
\Text(18,20)[]{\ftsz{$(\tilde{B}^0)$}}
\Text(8,-40)[]{$\gamma$}
\Text(0,-11)[]{\small{$\tilde{\ell}_{L_\alpha}$}}
\end{picture}
\hspace{0.3 cm}
\begin{picture}(110,70)(-55,-40)
\ArrowLine(-49,0)(-35,0)
\ArrowLine(0,0)(-35,0)
\ArrowLine(0,0)(35,0)
\ArrowLine(49,0)(35,0)
\DashLine(0,0)(0,15){1}
\DashArrowLine(0,30)(0,15){1}
\Photon(0,-20)(0,-40){2}{3}
\DashArrowArc(0,20)(40,210,270){3}
\DashArrowArc(0,20)(40,270,330){3}
\Text(-42,-7)[]{$\tau$}
\Text(44,-8)[]{$\mu^c$}
\Text(-10,25)[]{\small{$H_1^0$}}
\Text(-15,8)[]{\ftsz{$\tilde{H}^0_1$}}
\Text(18,9)[]{\ftsz{$\tilde{B}^0$}}
\Text(8,-40)[]{$\gamma$}
\Text(0,-11)[]{\small{$\tilde{\ell}_{R_\alpha}$}}
\end{picture}
\begin{picture}(110,70)(-55,-40)
\ArrowLine(-49,0)(-35,0)
\ArrowLine(-18,0)(-35,0)
\ArrowLine(-18,0)(0,0)
\ArrowLine(18,0)(0,0)
\ArrowLine(18,0)(35,0)
\ArrowLine(49,0)(35,0)
\DashLine(0,0)(0,15){1}
\DashArrowLine(0,15)(0,30){1}
\Photon(0,-20)(0,-40){2}{3}
\DashArrowArc(0,20)(40,210,270){3}
\DashArrowArc(0,20)(40,270,330){3}
\Text(-42,-7)[]{$\tau$}
\Text(44,-8)[]{$\mu^c$}
\Text(-10,25)[]{\small{$H_2^0$}}
\Text(-26,8)[]{\ftsz{$\tilde{H}^0_1$}}
\Text(-9,8)[]{\ftsz{$\tilde{H}^0_2$}}
\Text(18,9)[]{\ftsz{$\tilde{B}^0$}}
\Text(8,-40)[]{$\gamma$}
\Text(0,-11)[]{\small{$\tilde{\ell}_{R_\alpha}$}}
\end{picture}
\end{center}
\caption{\ftsz Diagrams that contribute to $D^{\gamma (b)}_{L,R}$.}
\label{dgalrb}
\end{figure}

The diagrams corresponding to $D^{\ga(c)}_{L,R} $ are depicted in 
Fig.~\ref{dgalrc} and give:
\bea
D^{\ga (c)}_L & = & 
- {g'^2 \over 16 \pi^2} M_1^3 \, \times
\nonumber \\
& & \left\{
\left[ s_L c_L \left( s_R^2 [A_{\tau} + \mu \tan\beta ]
+ s_R c_R A^R_{\mu \tau} \right) + c_L^2 s_R^2 A^L_{\mu \tau} \right]
I_5(M_1^2,M_1^2,M_1^2,\mt^2_{L_2},\mt^2_{R_2}) \right.
\nonumber \\
& & \left. \!\!\! 
- \left[ s_L c_L \left( s_R^2 [A_{\tau} + \mu \tan\beta ]
+ s_R c_R A^R_{\mu \tau} \right) - s_L^2 s_R^2 A^L_{\mu \tau} \right]
I_5(M_1^2,M_1^2,M_1^2,\mt^2_{L_3},\mt^2_{R_2}) \right.
\nonumber \\
& &  \left. \!\!\! 
+ \left[ s_L c_L \left( c_R^2 [A_{\tau} + \mu \tan\beta ]
- s_R c_R A^R_{\mu \tau} \right) + c_L^2 c_R^2 A^L_{\mu \tau} \right]
I_5(M_1^2,M_1^2,M_1^2,\mt^2_{L_2},\mt^2_{R_3}) \right.
\nonumber \\
& &  \left.  \!\! \!
- \left[ s_L c_L \left( c_R^2 [A_{\tau} + \mu \tan\beta ]
- s_R c_R A^R_{\mu \tau} \right) - s_L^2 c_R^2 A^L_{\mu \tau} \right]
I_5(M_1^2,M_1^2,M_1^2,\mt^2_{L_3},\mt^2_{R_3}) \right\}
\nonumber \\
& & \label{a6dc}\\
D^{\ga (c)}_R  & = &
- {g'^2 \over 16 \pi^2} M_1^3  \, \times
\nonumber \\
& & \left\{
\left[ s_R c_R \left( s_L^2 [A_{\tau} + \mu \tan\beta]
+ s_L c_L A^L_{\mu \tau} \right) + c_R^2 s_L^2 A^R_{\mu \tau} \right]
I_5(M_1^2,M_1^2,M_1^2,\mt^2_{L_2},\mt^2_{R_2}) \right.
\nonumber \\
& & \left. \!\!\! 
- \left[ s_R c_R \left( s_L^2 [A_{\tau} + \mu \tan\beta ]
+ s_L c_L A^L_{\mu \tau} \right) - s_R^2 s_L^2 A^R_{\mu \tau} \right]
I_5(M_1^2,M_1^2,M_1^2,\mt^2_{L_2},\mt^2_{R_3}) \right.
\nonumber \\
& & \left. \!\!\! 
+ \left[ s_R c_R \left( c_L^2 [A_{\tau} + \mu \tan\beta ]
- s_L c_L A^L_{\mu \tau} \right) + c_R^2 c_L^2 A^R_{\mu \tau} \right]
I_5(M_1^2,M_1^2,M_1^2,\mt^2_{L_3},\mt^2_{R_2}) \right.
\nonumber \\
& & \left. \!\!\! 
- \left[ s_R c_R \left( c_L^2 [A_{\tau} + \mu \tan\beta ]
- s_L c_L A^L_{\mu \tau} \right) - s_R^2 c_L^2 A^R_{\mu \tau} \right]
I_5(M_1^2,M_1^2,M_1^2,\mt^2_{L_3},\mt^2_{R_3}) \right\}
\nonumber \\
& & 
\eea
\vspace{-2cm}

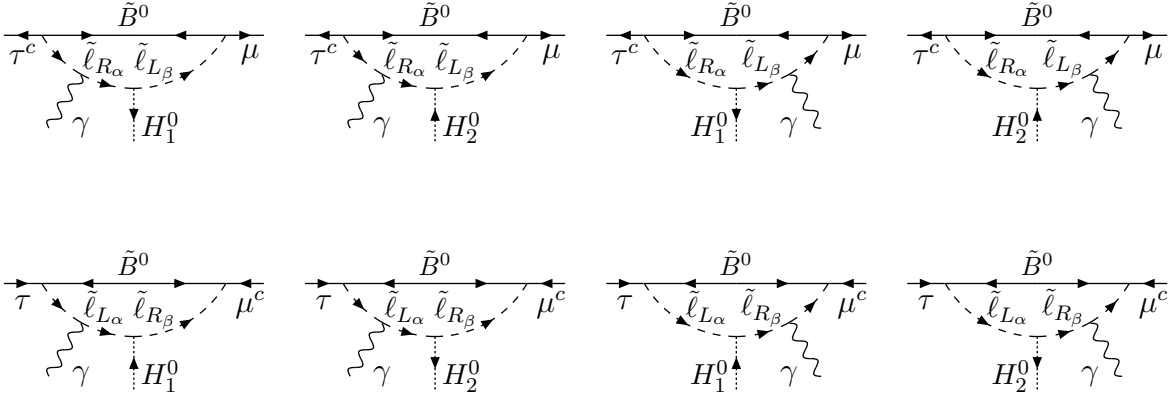
\begin{figure}[htb]
\begin{center}
\begin{picture}(110,80)(-55,-40)
\ArrowLine(-35,0)(-49,0)
\ArrowLine(-35,0)(0,0)
\ArrowLine(35,0)(0,0)
\ArrowLine(35,0)(49,0)
\Photon(-32,-35)(-20,-15){2}{3}
\DashArrowLine(0,-20)(0,-40){1}
\DashArrowArc(0,20)(40,210,240){3}
\DashArrowArc(0,20)(40,240,270){3}
\DashArrowArc(0,20)(40,270,330){3}
\Text(-42,-7)[]{$\tau^c$}
\Text(44,-7)[]{$\mu$}
\Text(0,8)[]{\ftsz{$\tilde{B}^0$}}
\Text(-20,-35)[]{$\gamma$}
\Text(10,-35)[]{\small{$H_1^0$}}
\Text(-11,-9)[]{\small{$\tilde{\ell}_{R_\alpha}$}}
\Text(8,-10)[]{\small{$\tilde{\ell}_{L_\beta}$}}
\end{picture}
\begin{picture}(110,80)(-55,-40)
\ArrowLine(-35,0)(-49,0)
\ArrowLine(-35,0)(0,0)
\ArrowLine(35,0)(0,0)
\ArrowLine(35,0)(49,0)
\Photon(-32,-35)(-20,-15){2}{3}
\DashArrowLine(0,-40)(0,-20){1}
\DashArrowArc(0,20)(40,210,240){3}
\DashArrowArc(0,20)(40,240,270){3}
\DashArrowArc(0,20)(40,270,330){3}
\Text(-42,-7)[]{$\tau^c$}
\Text(44,-7)[]{$\mu$}
\Text(0,8)[]{\ftsz{$\tilde{B}^0$}}
\Text(-20,-35)[]{$\gamma$}
\Text(10,-35)[]{\small{$H_2^0$}}
\Text(-11,-9)[]{\small{$\tilde{\ell}_{R_\alpha}$}}
\Text(8,-10)[]{\small{$\tilde{\ell}_{L_\beta}$}}
\end{picture}
\begin{picture}(110,80)(-55,-40)
\ArrowLine(-35,0)(-49,0)
\ArrowLine(-35,0)(0,0)
\ArrowLine(35,0)(0,0)
\ArrowLine(35,0)(49,0)
\Photon(32,-35)(20,-15){2}{3}
\DashArrowLine(0,-20)(0,-40){1}
\DashArrowArc(0,20)(40,210,270){3}
\DashArrowArc(0,20)(40,270,300){3}
\DashArrowArc(0,20)(40,300,330){3}
\Text(-42,-7)[]{$\tau^c$}
\Text(44,-7)[]{$\mu$}
\Text(0,8)[]{\ftsz{$\tilde{B}^0$}}
\Text(20,-35)[]{$\gamma$}
\Text(-10,-35)[]{\small{$H_1^0$}}
\Text(-11,-9)[]{\small{$\tilde{\ell}_{R_\alpha}$}}
\Text(10,-9)[]{\small{$\tilde{\ell}_{L_\beta}$}}
\end{picture}
\begin{picture}(110,80)(-55,-40)
\ArrowLine(-35,0)(-49,0)
\ArrowLine(-35,0)(0,0)
\ArrowLine(35,0)(0,0)
\ArrowLine(35,0)(49,0)
\Photon(32,-35)(20,-15){2}{3}
\DashArrowLine(0,-40)(0,-20){1}
\DashArrowArc(0,20)(40,210,270){3}
\DashArrowArc(0,20)(40,270,300){3}
\DashArrowArc(0,20)(40,300,330){3}
\Text(-42,-7)[]{$\tau^c$}
\Text(44,-7)[]{$\mu$}
\Text(0,8)[]{\ftsz{$\tilde{B}^0$}}
\Text(20,-35)[]{$\gamma$}
\Text(-10,-35)[]{\small{$H_2^0$}}
\Text(-11,-9)[]{\small{$\tilde{\ell}_{R_\alpha}$}}
\Text(10,-9)[]{\small{$\tilde{\ell}_{L_\beta}$}}
\end{picture}
\end{center}
\begin{center}
\begin{picture}(110,80)(-55,-40)
\ArrowLine(-49,0)(-35,0)
\ArrowLine(0,0)(-35,0)
\ArrowLine(0,0)(35,0)
\ArrowLine(49,0)(35,0)
\Photon(-32,-35)(-20,-15){2}{3}
\DashArrowLine(0,-40)(0,-20){1}
\DashArrowArc(0,20)(40,210,240){3}
\DashArrowArc(0,20)(40,240,270){3}
\DashArrowArc(0,20)(40,270,330){3}
\Text(-42,-7)[]{$\tau$}
\Text(44,-8)[]{$\mu^c$}
\Text(0,8)[]{\ftsz{$\tilde{B}^0$}}
\Text(-20,-35)[]{$\gamma$}
\Text(10,-35)[]{\small{$H_1^0$}}
\Text(-11,-9)[]{\small{$\tilde{\ell}_{L_\alpha}$}}
\Text(8,-10)[]{\small{$\tilde{\ell}_{R_\beta}$}}
\end{picture}
\begin{picture}(110,80)(-55,-40)
\ArrowLine(-49,0)(-35,0)
\ArrowLine(0,0)(-35,0)
\ArrowLine(0,0)(35,0)
\ArrowLine(49,0)(35,0)
\Photon(-32,-35)(-20,-15){2}{3}
\DashArrowLine(0,-20)(0,-40){1}
\DashArrowArc(0,20)(40,210,240){3}
\DashArrowArc(0,20)(40,240,270){3}
\DashArrowArc(0,20)(40,270,330){3}
\Text(-42,-7)[]{$\tau$}
\Text(44,-8)[]{$\mu^c$}
\Text(0,8)[]{\ftsz{$\tilde{B}^0$}}
\Text(-20,-35)[]{$\gamma$}
\Text(10,-35)[]{\small{$H_2^0$}}
\Text(-11,-9)[]{\small{$\tilde{\ell}_{L_\alpha}$}}
\Text(8,-10)[]{\small{$\tilde{\ell}_{R_\beta}$}}
\end{picture}
\begin{picture}(110,80)(-55,-40)
\ArrowLine(-49,0)(-35,0)
\ArrowLine(0,0)(-35,0)
\ArrowLine(0,0)(35,0)
\ArrowLine(49,0)(35,0)
\Photon(32,-35)(20,-15){2}{3}
\DashArrowLine(0,-40)(0,-20){1}
\DashArrowArc(0,20)(40,210,270){3}
\DashArrowArc(0,20)(40,270,300){3}
\DashArrowArc(0,20)(40,300,330){3}
\Text(-42,-7)[]{$\tau$}
\Text(44,-8)[]{$\mu^c$}
\Text(0,8)[]{\ftsz{$\tilde{B}^0$}}
\Text(20,-35)[]{$\gamma$}
\Text(-10,-35)[]{\small{$H_1^0$}}
\Text(-11,-9)[]{\small{$\tilde{\ell}_{L_\alpha}$}}
\Text(10,-9)[]{\small{$\tilde{\ell}_{R_\beta}$}}
\end{picture}
\begin{picture}(110,80)(-55,-40)
\ArrowLine(-49,0)(-35,0)
\ArrowLine(0,0)(-35,0)
\ArrowLine(0,0)(35,0)
\ArrowLine(49,0)(35,0)
\Photon(32,-35)(20,-15){2}{3}
\DashArrowLine(0,-20)(0,-40){1}
\DashArrowArc(0,20)(40,210,270){3}
\DashArrowArc(0,20)(40,270,300){3}
\DashArrowArc(0,20)(40,300,330){3}
\Text(-42,-7)[]{$\tau$}
\Text(44,-8)[]{$\mu^c$}
\Text(0,8)[]{\ftsz{$\tilde{B}^0$}}
\Text(20,-35)[]{$\gamma$}
\Text(-10,-35)[]{\small{$H_2^0$}}
\Text(-11,-9)[]{\small{$\tilde{\ell}_{L_\alpha}$}}
\Text(10,-9)[]{\small{$\tilde{\ell}_{R_\beta}$}}
\end{picture}
\end{center}
\caption{\ftsz Diagrams that contribute to $D^{\gamma (c)}_L$ (first row)
and $D^{\gamma (c)}_R$ (second row).}
\label{dgalrc}
\end{figure}

\subsection{ Contributions to $D^{Z}_{L,R}$}

The leading contributions to $D^Z_{L,R}$ are of 
 two types:
$D^Z_{L,R} = D^{Z(b)}_{L,R}  +D^{Z(c)}_{L,R}$.
The diagrams corresponding to $D^{Z(a)}_{L,R} $ are depicted in 
Fig.~\ref{dzlrb} and give:
\bea
{ D^{Z(b)}_L  \over s_L c_L} & = & 
{g^2 \over 16 \pi^2} \cdot {1 \over 8} \, 
M_2 \mu \tan\beta \, \left[
-2 ( 1 + 2 s_W^2) \mt^2_{L_2} 
I_5(M_2^2,\mu^2,\mt^2_{L_2},\mt^2_{L_2},\mt^2_{L_2}) \right. 
\nonumber \\
& & \left. + 2 ( 1- 4 s_W^2 )
J_5(M_2^2,\mu^2,\mu^2,\mu^2,\mt^2_{L_2}) 
+ 8 c_W^2 J_5(M_2^2,M_2^2,M_2^2,\mu^2,\mt^2_{L_2}) \right.
\nonumber \\
& & \left.
+ ( 5 - 8 s_W^2 ) 
J_5(M_2^2,M_2^2,\mu^2,\mu^2,\mt^2_{L_2})
\right]
\nonumber \\
& + & 
{g'^2 \over 16 \pi^2} \cdot {1 \over 8} \,  
M_1 \mu \tan\beta \, \left[ 
-2 ( 1 - 2 s_W^2 ) \mt^2_{L_2} 
I_5(M_1^2,\mu^2,\mt^2_{L_2},\mt^2_{L_2},\mt^2_{L_2}) \right.
\nonumber \\
& & \left. 
+ 2 J_5(M_1^2,\mu^2,\mu^2,\mu^2,\mt^2_{L_2})
+ J_5(M_1^2,M_1^2,\mu^2,\mu^2,\mt^2_{L_2}) \right]
- (L_2 \to L_3)
\\
{ D^{Z(b)}_R  \over s_R c_R} & = & 
{g'^2 \over 16 \pi^2} \cdot {1 \over 4} \,   
M_1 \mu \tan\beta \, \left[ 
- 4 s_W^2 \mt^2_{R_2} 
I_5(M_1^2,\mu^2,\mt^2_{R_2},\mt^2_{R_2},\mt^2_{R_2}) \right.
\nonumber \\
& & \left. 
+ 2 J_5(M_1^2,\mu^2,\mu^2,\mu^2,\mt^2_{R_2})
+ J_5(M_1^2,M_1^2,\mu^2,\mu^2,\mt^2_{R_2}) \right]
- (R_2 \to R_3)
\eea
\begin{figure}[htb]
\begin{center}
\begin{picture}(110,60)(-55,-30)
\ArrowLine(-35,0)(-49,0)
\ArrowLine(-35,0)(-18,0)
\ArrowLine(0,0)(-18,0)
\ArrowLine(0,0)(20,0)
\ArrowLine(35,0)(20,0)
\ArrowLine(35,0)(49,0)
\DashLine(0,0)(0,15){1}
\DashArrowLine(0,30)(0,15){1}
\Photon(20,0)(20,30){2}{3}
\GCirc(20,0){2}{0.5}
\DashArrowArc(0,20)(40,210,330){3}
\Text(-42,-8)[]{$\tau^c$}
\Text(42,-7)[]{$\mu$}
\Text(-10,25)[]{\small{$H_2^0$}}
\Text(-27,8)[]{\ftsz{$\tilde{H}^-_1$}}
\Text(-13,8)[]{\ftsz{$\tilde{H}^+_2$}}
\Text(10,9)[]{\ftsz{$\tilde{W}^-$}}
\Text(32,9)[]{\ftsz{$\tilde{W}^+$}}
\Text(28,25)[]{$Z$}
\Text(0,-12)[]{\small{$\tilde{\nu}_{\alpha}$}}
\end{picture}
\begin{picture}(110,60)(-55,-30)
\ArrowLine(-35,0)(-49,0)
\ArrowLine(-35,0)(-20,0)
\ArrowLine(0,0)(-20,0)
\ArrowLine(0,0)(20,0)
\ArrowLine(35,0)(20,0)
\ArrowLine(35,0)(49,0)
\DashLine(0,0)(0,15){1}
\DashArrowLine(0,30)(0,15){1}
\Photon(-20,0)(-20,30){2}{3}
\GCirc(-20,0){2}{0.5}
\DashArrowArc(0,20)(40,210,330){3}
\Text(-42,-8)[]{$\tau^c$}
\Text(42,-7)[]{$\mu$}
\Text(-9,25)[]{\small{$H_2^0$}}
\Text(-29,8)[]{\ftsz{$\tilde{H}^-_1$}}
\Text(-8,8)[]{\ftsz{$\tilde{H}^+_2$}}
\Text(12,9)[]{\ftsz{$\tilde{W}^-$}}
\Text(30,9)[]{\ftsz{$\tilde{W}^+$}}
\Text(-28,25)[]{$Z$}
\Text(0,-12)[]{\small{$\tilde{\nu}_{\alpha}$}}
\end{picture}
\begin{picture}(110,60)(-55,-30)
\ArrowLine(-35,0)(-49,0)
\ArrowLine(-35,0)(-20,0)
\ArrowLine(0,0)(-20,0)
\ArrowLine(0,0)(20,0)
\ArrowLine(35,0)(20,0)
\ArrowLine(35,0)(49,0)
\DashLine(0,0)(0,15){1}
\DashArrowLine(0,30)(0,15){1}
\Photon(-20,0)(-20,30){2}{3}
\GCirc(-20,0){2}{0.5}
\DashArrowArc(0,20)(40,210,330){3}
\Text(-42,-8)[]{$\tau^c$}
\Text(42,-7)[]{$\mu$}
\Text(-9,25)[]{\small{$H_2^0$}}
\Text(-30,8)[]{\ftsz{$\tilde{H}^0_1$}}
\Text(-8,8)[]{\ftsz{$\tilde{H}^0_2$}}
\Text(22,9)[]{\ftsz{$\tilde{W}^0$}}
\Text(22,20)[]{\ftsz{$(\tilde{B}^0)$}}
\Text(-28,25)[]{$Z$}
\Text(0,-11)[]{\small{$\tilde{\ell}_{L_\alpha}$}}
\end{picture}
\end{center}
\begin{center}
\begin{picture}(110,70)(-55,-40)
\ArrowLine(-35,0)(-49,0)
\ArrowLine(-35,0)(-18,0)
\ArrowLine(0,0)(-18,0)
\ArrowLine(0,0)(18,0)
\ArrowLine(35,0)(18,0)
\ArrowLine(35,0)(49,0)
\DashLine(0,0)(0,15){1}
\DashArrowLine(0,30)(0,15){1}
\Photon(0,-20)(0,-40){2}{3}
\DashArrowArc(0,20)(40,210,270){3}
\DashArrowArc(0,20)(40,270,330){3}
\Text(-42,-8)[]{$\tau^c$}
\Text(42,-7)[]{$\mu$}
\Text(-10,25)[]{\small{$H_2^0$}}
\Text(-28,8)[]{\ftsz{$\tilde{H}^0_1$}}
\Text(-14,8)[]{\ftsz{$\tilde{H}^0_2$}}
\Text(18,9)[]{\ftsz{$\tilde{W}^0$}}
\Text(18,20)[]{\ftsz{$(\tilde{B}^0)$}}
\Text(8,-40)[]{$Z$}
\Text(0,-11)[]{\small{$\tilde{\ell}_{L_\alpha}$}}
\end{picture}
\begin{picture}(110,70)(-55,-40)
\ArrowLine(-35,0)(-49,0)
\ArrowLine(-35,0)(-18,0)
\ArrowLine(0,0)(-18,0)
\ArrowLine(0,0)(18,0)
\ArrowLine(35,0)(18,0)
\ArrowLine(35,0)(49,0)
\DashLine(0,0)(0,15){1}
\DashArrowLine(0,30)(0,15){1}
\Photon(0,-20)(0,-40){2}{3}
\DashArrowArc(0,20)(40,210,270){3}
\DashArrowArc(0,20)(40,270,330){3}
\Text(-42,-8)[]{$\tau^c$}
\Text(42,-7)[]{$\mu$}
\Text(-10,25)[]{\small{$H_2^0$}}
\Text(-27,8)[]{\ftsz{$\tilde{H}^-_1$}}
\Text(-13,8)[]{\ftsz{$\tilde{H}^+_2$}}
\Text(12,9)[]{\ftsz{$\tilde{W}^-$}}
\Text(30,9)[]{\ftsz{$\tilde{W}^+$}}
\Text(8,-40)[]{$Z$}
\Text(0,-12)[]{\small{$\tilde{\nu}_{\alpha}$}}
\end{picture}
\hspace{0.3 cm}
\begin{picture}(110,70)(-55,-40)
\ArrowLine(-49,0)(-35,0)
\ArrowLine(-18,0)(-35,0)
\ArrowLine(-18,0)(0,0)
\ArrowLine(18,0)(0,0)
\ArrowLine(18,0)(35,0)
\ArrowLine(49,0)(35,0)
\DashLine(0,0)(0,15){1}
\DashArrowLine(0,15)(0,30){1}
\Photon(0,-20)(0,-40){2}{3}
\DashArrowArc(0,20)(40,210,270){3}
\DashArrowArc(0,20)(40,270,330){3}
\Text(-42,-7)[]{$\tau$}
\Text(44,-8)[]{$\mu^c$}
\Text(-10,25)[]{\small{$H_2^0$}}
\Text(-26,8)[]{\ftsz{$\tilde{H}^0_1$}}
\Text(-9,8)[]{\ftsz{$\tilde{H}^0_2$}}
\Text(18,9)[]{\ftsz{$\tilde{B}^0$}}
\Text(8,-40)[]{$Z$}
\Text(0,-11)[]{\small{$\tilde{\ell}_{R_\alpha}$}}
\end{picture}
\begin{picture}(110,70)(-55,-40)
\ArrowLine(-49,0)(-35,0)
\ArrowLine(-20,0)(-35,0)
\ArrowLine(-20,0)(0,0)
\ArrowLine(20,0)(0,0)
\ArrowLine(20,0)(35,0)
\ArrowLine(49,0)(35,0)
\DashLine(0,0)(0,15){1}
\DashArrowLine(0,15)(0,30){1}
\Photon(-20,0)(-20,30){2}{3}
\GCirc(-20,0){2}{0.5}
\DashArrowArc(0,20)(40,210,330){3}
\Text(-42,-7)[]{$\tau$}
\Text(44,-8)[]{$\mu^c$}
\Text(-9,25)[]{\small{$H_2^0$}}
\Text(-32,8)[]{\ftsz{$\tilde{H}^0_1$}}
\Text(-9,8)[]{\ftsz{$\tilde{H}^0_2$}}
\Text(22,9)[]{\ftsz{$\tilde{B}^0$}}
\Text(-28,25)[]{$Z$}
\Text(0,-11)[]{\small{$\tilde{\ell}_{R_\alpha}$}}
\end{picture}
\end{center}
\caption{\ftsz Diagrams that contribute to $D^{Z (b)}_{L,R}$. }
\label{dzlrb}
\end{figure}

The diagrams corresponding to $D^{Z(c)}_{L,R} $ are depicted in 
Fig.~\ref{dzlrc} and give:
\bea
{ D^{Z (c)}_L  \over s_L c_L} & = & 
{g'^2 \over 16 \pi^2}\cdot {1 \over 4} \, 
M_1 \mu \tan\beta \, \left[
(1 - 2 s_W^2 )
\left( s_R^2 J_5(M_1^2,M_1^2,\mt^2_{L_2},\mt^2_{L_2},\mt^2_{R_2})
\right. \right.
\nonumber \\
& & \left. \left.
+ c_R^2  J_5(M_1^2,M_1^2,\mt^2_{L_2},\mt^2_{L_2},\mt^2_{R_3}) \right)
-2 s_W^2 
\left(s_R^2 J_5(M_1^2,M_1^2,\mt^2_{L_2},\mt^2_{R_2},\mt^2_{R_2})
\right. \right.
\nonumber \\
& & \left. \left.
+ c_R^2  J_5(M_1^2,M_1^2,\mt^2_{L_2},\mt^2_{R_3},\mt^2_{R_3}) \right)
\right] - (L_2 \to L_3)
\\
{ D^{Z (c)}_R  \over s_R c_R} & = &
{g'^2 \over 16 \pi^2}\cdot {1 \over 4} \, 
M_1 \mu \tan\beta \, \left[
(1 - 2 s_W^2 )
\left( s_L^2 J_5(M_1^2,M_1^2,\mt^2_{L_2},\mt^2_{L_2},\mt^2_{R_2})
\right. \right.
\nonumber \\
& & \left. \left.
+ c_L^2  J_5(M_1^2,M_1^2,\mt^2_{L_3},\mt^2_{L_3},\mt^2_{R_2}) \right)
- 2 s_W^2 
\left(s_L^2 J_5(M_1^2,M_1^2,\mt^2_{L_2},\mt^2_{R_2},\mt^2_{R_2})
\right. \right.
\nonumber \\
& & \left. \left.
+ c_L^2  J_5(M_1^2,M_1^2,\mt^2_{L_3},\mt^2_{R_2},\mt^2_{R_2}) \right)
\right] - (R_2 \to R_3)
\eea

\begin{figure}[htb]
\vspace{-1cm}
\begin{center}
\begin{picture}(110,80)(-55,-40)
\ArrowLine(-35,0)(-49,0)
\ArrowLine(-35,0)(0,0)
\ArrowLine(35,0)(0,0)
\ArrowLine(35,0)(49,0)
\Photon(-32,-35)(-20,-15){2}{3}
\DashArrowLine(0,-40)(0,-20){1}
\DashArrowArc(0,20)(40,210,240){3}
\DashArrowArc(0,20)(40,240,270){3}
\DashArrowArc(0,20)(40,270,330){3}
\Text(-42,-7)[]{$\tau^c$}
\Text(44,-7)[]{$\mu$}
\Text(0,8)[]{\ftsz{$\tilde{B}^0$}}
\Text(-22,-35)[]{$Z$}
\Text(10,-35)[]{\small{$H_2^0$}}
\Text(-11,-9)[]{\small{$\tilde{\ell}_{R_\alpha}$}}
\Text(8,-10)[]{\small{$\tilde{\ell}_{L_\beta}$}}
\end{picture}
\begin{picture}(110,80)(-55,-40)
\ArrowLine(-35,0)(-49,0)
\ArrowLine(-35,0)(0,0)
\ArrowLine(35,0)(0,0)
\ArrowLine(35,0)(49,0)
\Photon(32,-35)(20,-15){2}{3}
\DashArrowLine(0,-40)(0,-20){1}
\DashArrowArc(0,20)(40,210,270){3}
\DashArrowArc(0,20)(40,270,300){3}
\DashArrowArc(0,20)(40,300,330){3}
\Text(-42,-7)[]{$\tau^c$}
\Text(44,-7)[]{$\mu$}
\Text(0,8)[]{\ftsz{$\tilde{B}^0$}}
\Text(20,-35)[]{$Z$}
\Text(-10,-35)[]{\small{$H_2^0$}}
\Text(-11,-9)[]{\small{$\tilde{\ell}_{R_\alpha}$}}
\Text(10,-9)[]{\small{$\tilde{\ell}_{L_\beta}$}}
\end{picture}
\hspace{0.3 cm}
\begin{picture}(110,80)(-55,-40)
\ArrowLine(-49,0)(-35,0)
\ArrowLine(0,0)(-35,0)
\ArrowLine(0,0)(35,0)
\ArrowLine(49,0)(35,0)
\Photon(-32,-35)(-20,-15){2}{3}
\DashArrowLine(0,-20)(0,-40){1}
\DashArrowArc(0,20)(40,210,240){3}
\DashArrowArc(0,20)(40,240,270){3}
\DashArrowArc(0,20)(40,270,330){3}
\Text(-42,-7)[]{$\tau$}
\Text(44,-7)[]{$\mu^c$}
\Text(0,8)[]{\ftsz{$\tilde{B}^0$}}
\Text(-22,-35)[]{$Z$}
\Text(10,-35)[]{\small{$H_2^0$}}
\Text(-11,-9)[]{\small{$\tilde{\ell}_{L_\alpha}$}}
\Text(8,-10)[]{\small{$\tilde{\ell}_{R_\beta}$}}
\end{picture}
\begin{picture}(110,80)(-55,-40)
\ArrowLine(-49,0)(-35,0)
\ArrowLine(0,0)(-35,0)
\ArrowLine(0,0)(35,0)
\ArrowLine(49,0)(35,0)
\Photon(32,-35)(20,-15){2}{3}
\DashArrowLine(0,-20)(0,-40){1}
\DashArrowArc(0,20)(40,210,270){3}
\DashArrowArc(0,20)(40,270,300){3}
\DashArrowArc(0,20)(40,300,330){3}
\Text(-42,-7)[]{$\tau$}
\Text(44,-8)[]{$\mu^c$}
\Text(0,8)[]{\ftsz{$\tilde{B}^0$}}
\Text(20,-35)[]{$Z$}
\Text(-10,-35)[]{\small{$H_2^0$}}
\Text(-11,-9)[]{\small{$\tilde{\ell}_{L_\alpha}$}}
\Text(10,-9)[]{\small{$\tilde{\ell}_{R_\beta}$}}
\end{picture}
\end{center}
\caption{\ftsz Diagrams that contribute to $D^{Z (c)}_{L,R}$.}
\label{dzlrc}
\end{figure}

\subsection{\large \bf Contributions to $B^{f_{L,R}}_{L,R}$ }

Contributions to   $B^{f_{L,R}}_{L,R}$ arise 
from the diagrams in Fig.~\ref{blrf}.
%

\begin{figure}[htb]
\begin{center}
\begin{picture}(110,60)(-55,-30)
\ArrowLine(-45,15)(-25,15)
\ArrowLine(-25,-15)(-25,15)
\ArrowLine(-25,-15)(-45,-15)
\ArrowLine(25,15)(45,15)
\ArrowLine(25,15)(25,-15)
\ArrowLine(45,-15)(25,-15)
\DashArrowLine(-25,15)(25,15){3}
\DashArrowLine(25,-15)(-25,-15){3}
\Text(-42,23)[]{$\tau$}
\Text(42,23)[]{$\mu$}
\Text(-42,-24)[]{$f^d$}
\Text(42,-24)[]{$f^d$}
\Text(-14,0)[]{\ftsz{$\tilde{W}^+$}}
\Text(14,0)[]{\ftsz{$\tilde{W}^+$}}
\Text(0,25)[]{\small{$\tilde{\nu}_{\alpha}$}}
\Text(0,-25)[]{\small{$\tilde{f}^u_L$}}
\end{picture}
\begin{picture}(110,60)(-55,-30)
\ArrowLine(-45,15)(-25,15)
\ArrowLine(-25,0)(-25,15)
\ArrowLine(-25,0)(-25,-15)
\ArrowLine(-45,-15)(-25,-15)
\ArrowLine(25,15)(45,15)
\ArrowLine(25,15)(25,0)
\ArrowLine(25,-15)(25,0)
\ArrowLine(25,-15)(45,-15)
\DashArrowLine(-25,15)(25,15){3}
\DashArrowLine(-25,-15)(25,-15){3}
\Text(-42,23)[]{$\tau$}
\Text(42,23)[]{$\mu$}
\Text(-42,-24)[]{$f^u$}
\Text(42,-24)[]{$f^u$}
\Text(-14,7)[]{\ftsz{$\tilde{W}^+$}}
\Text(-14,-7)[]{\ftsz{$\tilde{W}^-$}}
\Text(14,7)[]{\ftsz{$\tilde{W}^+$}}
\Text(14,-7)[]{\ftsz{$\tilde{W}^-$}}
\Text(0,25)[]{\small{$\tilde{\nu}_{\alpha}$}}
\Text(0,-25)[]{\small{$\tilde{f}^d_L$}}
\end{picture}
\begin{picture}(110,60)(-55,-30)
\ArrowLine(-45,15)(-25,15)
\ArrowLine(-25,-15)(-25,15)
\ArrowLine(-25,-15)(-45,-15)
\ArrowLine(25,15)(45,15)
\ArrowLine(25,15)(25,-15)
\ArrowLine(45,-15)(25,-15)
\DashArrowLine(-25,15)(25,15){3}
\DashArrowLine(25,-15)(-25,-15){3}
\Text(-42,23)[]{$\tau$}
\Text(42,23)[]{$\mu$}
\Text(-42,-24)[]{$f$}
\Text(42,-24)[]{$f$}
\Text(-37,-1)[]{\ftsz{$(\tilde{B}^0)$}}
\Text(37,-1)[]{\ftsz{$(\tilde{B}^0)$}}
\Text(-15,0)[]{\ftsz{$\tilde{W}^0$}}
\Text(15,0)[]{\ftsz{$\tilde{W}^0$}}
\Text(0,25)[]{\small{$\tilde{\ell}_{L_\alpha}$}}
\Text(0,-25)[]{\small{$\tilde{f}_L$}}
\end{picture}
\begin{picture}(110,60)(-55,-30)
\ArrowLine(-45,15)(-25,15)
\ArrowLine(-25,0)(-25,15)
\ArrowLine(-25,0)(-25,-15)
\ArrowLine(-45,-15)(-25,-15)
\ArrowLine(25,15)(45,15)
\ArrowLine(25,15)(25,0)
\ArrowLine(25,-15)(25,0)
\ArrowLine(25,-15)(45,-15)
\DashArrowLine(-25,15)(25,15){3}
\DashArrowLine(-25,-15)(25,-15){3}
\Text(-42,23)[]{$\tau$}
\Text(42,23)[]{$\mu$}
\Text(-42,-24)[]{$f$}
\Text(42,-24)[]{$f$}
\Text(-37,-1)[]{\ftsz{$(\tilde{B}^0)$}}
\Text(37,-1)[]{\ftsz{$(\tilde{B}^0)$}}
\Text(-15,0)[]{\ftsz{$\tilde{W}^0$}}
\Text(15,0)[]{\ftsz{$\tilde{W}^0$}}
\Text(0,25)[]{\small{$\tilde{\ell}_{L_\alpha}$}}
\Text(0,-25)[]{\small{$\tilde{f}_L$}}
\end{picture}
\end{center}
\begin{center}
\begin{picture}(110,60)(-55,-30)
\ArrowLine(-45,15)(-25,15)
\ArrowLine(-25,-15)(-25,15)
\ArrowLine(-25,-15)(-45,-15)
\ArrowLine(25,15)(45,15)
\ArrowLine(25,15)(25,-15)
\ArrowLine(45,-15)(25,-15)
\DashArrowLine(-25,15)(25,15){3}
\DashArrowLine(25,-15)(-25,-15){3}
\Text(-42,23)[]{$\tau$}
\Text(42,23)[]{$\mu$}
\Text(-42,-24)[]{$f$}
\Text(42,-24)[]{$f$}
\Text(-37,-1)[]{\ftsz{$(\tilde{B}^0)$}}
\Text(38,-1)[]{\ftsz{$(\tilde{W}^0)$}}
\Text(-15,0)[]{\ftsz{$\tilde{W}^0$}}
\Text(15,0)[]{\ftsz{$\tilde{B}^0$}}
\Text(0,25)[]{\small{$\tilde{\ell}_{L_\alpha}$}}
\Text(0,-25)[]{\small{$\tilde{f}_L$}}
\end{picture}
\begin{picture}(110,60)(-55,-30)
\ArrowLine(-45,15)(-25,15)
\ArrowLine(-25,0)(-25,15)
\ArrowLine(-25,0)(-25,-15)
\ArrowLine(-45,-15)(-25,-15)
\ArrowLine(25,15)(45,15)
\ArrowLine(25,15)(25,0)
\ArrowLine(25,-15)(25,0)
\ArrowLine(25,-15)(45,-15)
\DashArrowLine(-25,15)(25,15){3}
\DashArrowLine(-25,-15)(25,-15){3}
\Text(-42,23)[]{$\tau$}
\Text(42,23)[]{$\mu$}
\Text(-42,-23)[]{$f$}
\Text(42,-23)[]{$f$}
\Text(-37,-1)[]{\ftsz{$(\tilde{B}^0)$}}
\Text(38,-1)[]{\ftsz{$(\tilde{W}^0)$}}
\Text(-15,0)[]{\ftsz{$\tilde{W}^0$}}
\Text(15,0)[]{\ftsz{$\tilde{B}^0$}}
\Text(0,25)[]{\small{$\tilde{\ell}_{L_\alpha}$}}
\Text(0,-25)[]{\small{$\tilde{f}_L$}}
\end{picture}
\begin{picture}(110,60)(-55,-30)
\ArrowLine(-45,15)(-25,15)
\ArrowLine(-25,-15)(-25,15)
\ArrowLine(-25,-15)(-45,-15)
\ArrowLine(25,15)(45,15)
\ArrowLine(25,15)(25,-15)
\ArrowLine(45,-15)(25,-15)
\DashArrowLine(-25,15)(25,15){3}
\DashArrowLine(-25,-15)(25,-15){3}
\Text(-42,23)[]{$\tau$}
\Text(42,23)[]{$\mu$}
\Text(-42,-24)[]{$f^c$}
\Text(42,-24)[]{$f^c$}
\Text(-15,0)[]{\ftsz{$\tilde{B}^0$}}
\Text(15,0)[]{\ftsz{$\tilde{B}^0$}}
\Text(0,25)[]{\small{$\tilde{\ell}_{L_\alpha}$}}
\Text(0,-25)[]{\small{$\tilde{f}_R$}}
\end{picture}
\begin{picture}(110,60)(-55,-30)
\ArrowLine(-45,15)(-25,15)
\ArrowLine(-25,0)(-25,15)
\ArrowLine(-25,0)(-25,-15)
\ArrowLine(-45,-15)(-25,-15)
\ArrowLine(25,15)(45,15)
\ArrowLine(25,15)(25,0)
\ArrowLine(25,-15)(25,0)
\ArrowLine(25,-15)(45,-15)
\DashArrowLine(-25,15)(25,15){3}
\DashArrowLine(25,-15)(-25,-15){3}
\Text(-42,23)[]{$\tau$}
\Text(42,23)[]{$\mu$}
\Text(-42,-24)[]{$f^c$}
\Text(42,-24)[]{$f^c$}
\Text(-15,0)[]{\ftsz{$\tilde{B}^0$}}
\Text(15,0)[]{\ftsz{$\tilde{B}^0$}}
\Text(0,25)[]{\small{$\tilde{\ell}_{L_\alpha}$}}
\Text(0,-25)[]{\small{$\tilde{f}_R$}}
\end{picture}
\end{center}
\begin{center}
\begin{picture}(110,60)(-55,-30)
\ArrowLine(-25,15)(-45,15)
\ArrowLine(-25,15)(-25,-15)
\ArrowLine(-45,-15)(-25,-15)
\ArrowLine(45,15)(25,15)
\ArrowLine(25,-15)(25,15)
\ArrowLine(25,-15)(45,-15)
\DashArrowLine(-25,15)(25,15){3}
\DashArrowLine(-25,-15)(25,-15){3}
\Text(-42,23)[]{$\tau^c$}
\Text(44,23)[]{$\mu^c$}
\Text(-42,-24)[]{$f$}
\Text(42,-24)[]{$f$}
\Text(-15,0)[]{\ftsz{$\tilde{B}^0$}}
\Text(15,0)[]{\ftsz{$\tilde{B}^0$}}
\Text(0,25)[]{\small{$\tilde{\ell}_{R_\alpha}$}}
\Text(0,-25)[]{\small{$\tilde{f}_L$}}
\end{picture}
\begin{picture}(110,60)(-55,-30)
\ArrowLine(-25,15)(-45,15)
\ArrowLine(-25,15)(-25,0)
\ArrowLine(-25,-15)(-25,0)
\ArrowLine(-25,-15)(-45,-15)
\ArrowLine(45,15)(25,15)
\ArrowLine(25,0)(25,15)
\ArrowLine(25,0)(25,-15)
\ArrowLine(45,-15)(25,-15)
\DashArrowLine(-25,15)(25,15){3}
\DashArrowLine(25,-15)(-25,-15){3}
\Text(-42,23)[]{$\tau^c$}
\Text(44,23)[]{$\mu^c$}
\Text(-42,-24)[]{$f$}
\Text(42,-24)[]{$f$}
\Text(-15,0)[]{\ftsz{$\tilde{B}^0$}}
\Text(15,0)[]{\ftsz{$\tilde{B}^0$}}
\Text(0,25)[]{\small{$\tilde{\ell}_{R_\alpha}$}}
\Text(0,-25)[]{\small{$\tilde{f}_L$}}
\end{picture}
\begin{picture}(110,60)(-55,-30)
\ArrowLine(-25,15)(-45,15)
\ArrowLine(-25,15)(-25,-15)
\ArrowLine(-45,-15)(-25,-15)
\ArrowLine(45,15)(25,15)
\ArrowLine(25,-15)(25,15)
\ArrowLine(25,-15)(45,-15)
\DashArrowLine(-25,15)(25,15){3}
\DashArrowLine(25,-15)(-25,-15){3}
\Text(-42,23)[]{$\tau^c$}
\Text(44,23)[]{$\mu^c$}
\Text(-42,-24)[]{$f^c$}
\Text(42,-24)[]{$f^c$}
\Text(-15,0)[]{\ftsz{$\tilde{B}^0$}}
\Text(15,0)[]{\ftsz{$\tilde{B}^0$}}
\Text(0,25)[]{\small{$\tilde{\ell}_{R_\alpha}$}}
\Text(0,-25)[]{\small{$\tilde{f}_R$}}
\end{picture}
\begin{picture}(110,60)(-55,-30)
\ArrowLine(-25,15)(-45,15)
\ArrowLine(-25,15)(-25,0)
\ArrowLine(-25,-15)(-25,0)
\ArrowLine(-25,-15)(-45,-15)
\ArrowLine(45,15)(25,15)
\ArrowLine(25,0)(25,15)
\ArrowLine(25,0)(25,-15)
\ArrowLine(45,-15)(25,-15)
\DashArrowLine(-25,15)(25,15){3}
\DashArrowLine(-25,-15)(25,-15){3}
\Text(-42,23)[]{$\tau^c$}
\Text(44,23)[]{$\mu^c$}
\Text(-42,-24)[]{$f^c$}
\Text(42,-24)[]{$f^c$}
\Text(-15,0)[]{\ftsz{$\tilde{B}^0$}}
\Text(15,0)[]{\ftsz{$\tilde{B}^0$}}
\Text(0,25)[]{\small{$\tilde{\ell}_{R_\alpha}$}}
\Text(0,-25)[]{\small{$\tilde{f}_R$}}
\end{picture}
\end{center}
\caption{\ftsz Diagrams that contribute to $B^{f_{L,R}}_L$ (first and
second rows) and $B^{f_{L,R}}_R$ (third row). 
In the case of  $B^{\mu_{L,R}}_{L,R}$, {\it i.e.} 
$f=f^d=\mu, f^c=\mu^c$, the following replacements 
are implied: $\ft^u_L \to \tilde{\nu}_\beta, 
\ft_L \to \tilde{\ell}_{L_\beta}, \ft_R \to \tilde{\ell}_{R_\beta}$. }
\label{blrf}
\end{figure}
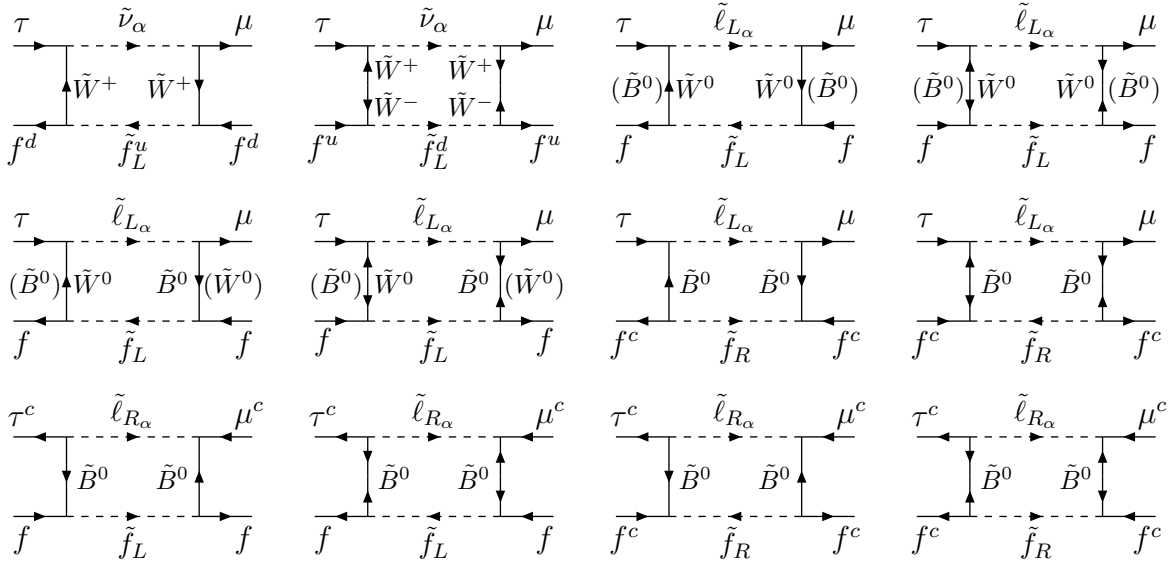

For $f \neq \mu, \tau$:
\bea
{B_L^{f_L} \over s_L c_L} & = &
{g^4 \over 16 \pi^2} \cdot {1 \over 16} \left[
- \left(1 + 4\, \delta_{T^3_{f_L}, -\frac12}\right) 
J_4(M_2^2,M_2^2,\mt^2_{L_2},\mt^2_{\fl}) \right.
\nonumber \\
&  & \left.
- 2 \left(1 + 4\, \delta_{T^3_{f_L}, \frac12}\right)  
M_2^2 I_4(M_2^2,M_2^2,\mt^2_{L_2},\mt^2_{\fl}) \right]
\nonumber \\
& + &  {g^2 g'^2 \over 16 \pi^2} 
\cdot {1 \over 2} (- T^3_{f_L} Y_{f_L}) \left[
J_4(M_1^2,M_2^2,\mt^2_{L_2},\mt^2_{\fl})
+ 2 M_1 M_2 I_4(M_1^2,M_2^2,\mt^2_{L_2},\mt^2_{\fl}) \right]
\nonumber \\
& + &  {g'^4 \over 16 \pi^2}   
\cdot {1 \over 4}  (- Y_{f_L}^2) \left[
J_4(M_1^2,M_1^2,\mt^2_{L_2},\mt^2_{\fl})
+ 2 M_1^2 I_4(M_1^2,M_1^2,\mt^2_{L_2},\mt^2_{\fl}) \right]
\nonumber \\
& - & (L_2 \to L_3)
\\
{B_L^{f_R} \over s_L c_L} & = &
 {g'^4 \over 16 \pi^2}   
\cdot {1 \over 4} \, Y_{f_R}^2  \left[
J_4(M_1^2,M_1^2,\mt^2_{L_2},\mt^2_{\fr})
+ 2 M_1^2 I_4(M_1^2,M_1^2,\mt^2_{L_2},\mt^2_{\fr}) \right]
\nonumber \\
& - &
(L_2 \to L_3)
\\
{B_R^{f_L} \over s_R c_R} & = &
{g'^4 \over 16 \pi^2} Y_{f_L}^2 \left[
J_4(M_1^2,M_1^2,\mt^2_{R_2},\mt^2_{\fl})
+ 2 M_1^2 I_4(M_1^2,M_1^2,\mt^2_{R_2},\mt^2_{\fl}) \right]
\nonumber \\
& - &
(R_2 \to R_3)
\\
{B_R^{f_R} \over s_R c_R} & = &
 {g'^4 \over 16 \pi^2} ( - Y_{f_R}^2 ) \left[
J_4(M_1^2,M_1^2,\mt^2_{R_2},\mt^2_{\fr})
+ 2 M_1^2 I_4(M_1^2,M_1^2,\mt^2_{R_2},\mt^2_{\fr}) \right]
\nonumber \\
& - &
(R_2 \to R_3)
\eea

For $f = \mu$:
\bea
{B_L^{\mu_L} \over s_L c_L} & = &
{g^4 \over 16 \pi^2}  \cdot {1 \over 16} \left[
- c_L^2 \left( 5 J_4(M_2^2,M_2^2,\mt^2_{L_2},\mt^2_{L_2})
+ 2 M_2^2 I_4(M_2^2,M_2^2,\mt^2_{L_2},\mt^2_{L_2}) \right) \right.
\nonumber \\ 
& & \left. + s_L^2 \left( 5 J_4(M_2^2,M_2^2,\mt^2_{L_3},\mt^2_{L_3})
+ 2 M_2^2 I_4(M_2^2,M_2^2,\mt^2_{L_3},\mt^2_{L_3}) \right) \right.
\nonumber \\ 
& & \left. +(c_L^2-s_L^2) \left( 5 J_4(M_2^2,M_2^2,\mt^2_{L_2},\mt^2_{L_3})
+ 2 M_2^2 I_4(M_2^2,M_2^2,\mt^2_{L_2},\mt^2_{L_3}) \right) \right]
\nonumber \\
& + &  {g^2 g'^2 \over 16 \pi^2} \cdot {1 \over 8}  \left[
- c_L^2 \left( J_4(M_1^2,M_2^2,\mt^2_{L_2},\mt^2_{L_2})
+ 2 M_1 M_2 I_4(M_1^2,M_2^2,\mt^2_{L_2},\mt^2_{L_2}) \right) \right.
\nonumber \\ 
& & \left. + s_L^2 \left( J_4(M_1^2,M_2^2,\mt^2_{L_3},\mt^2_{L_3})
+ 2 M_1 M_2 I_4(M_1^2,M_2^2,\mt^2_{L_3},\mt^2_{L_3}) \right) \right.
\nonumber \\ 
& & \left.  +(c_L^2-s_L^2) \left( J_4(M_1^2,M_2^2,\mt^2_{L_2},\mt^2_{L_3})
+ 2 M_1 M_2 I_4(M_1^2,M_2^2,\mt^2_{L_2},\mt^2_{L_3}) \right) \right]
\nonumber \\
& + &  {g'^4 \over 16 \pi^2}  \cdot {1 \over 16}  \left[
- c_L^2 \left( J_4(M_1^2,M_1^2,\mt^2_{L_2},\mt^2_{L_2})
+ 2 M_1^2 I_4(M_1^2,M_1^2,\mt^2_{L_2},\mt^2_{L_2}) \right) \right.
\nonumber \\ 
& & \left. + s_L^2 \left( J_4(M_1^2,M_1^2,\mt^2_{L_3},\mt^2_{L_3})
+ 2 M_1^2 I_4(M_1^2,M_1^2,\mt^2_{L_3},\mt^2_{L_3}) \right) \right.
\nonumber \\ 
& & \left. +(c_L^2-s_L^2) \left( J_4(M_1^2,M_1^2,\mt^2_{L_2},\mt^2_{L_3})
+ 2 M_1^2 I_4(M_1^2,M_1^2,\mt^2_{L_2},\mt^2_{L_3}) \right) \right]
\\
{B_L^{\mu_R} \over s_L c_L} & = &
 {g'^4 \over 16 \pi^2} \cdot {1 \over 4}  \left[
c_R^2 \left( J_4(M_1^2,M_1^2,\mt^2_{L_2},\mt^2_{R_2})
+ 2 M_1^2 I_4(M_1^2,M_1^2,\mt^2_{L_2},\mt^2_{R_2}) \right) \right.
\nonumber \\ 
& & \left. + s_R^2 \left( J_4(M_1^2,M_1^2,\mt^2_{L_2},\mt^2_{R_3})
+ 2 M_1^2 I_4(M_1^2,M_1^2,\mt^2_{L_2},\mt^2_{R_3}) \right) \right]
\nonumber \\ 
& - & (L_2 \to L_3)
\\
{B_R^{\mu_L} \over s_R c_R} & = &
 {g'^4 \over 16 \pi^2} \cdot {1 \over 4}  \left[
c_L^2 \left( J_4(M_1^2,M_1^2,\mt^2_{L_2},\mt^2_{R_2})
+ 2 M_1^2 I_4(M_1^2,M_1^2,\mt^2_{L_2},\mt^2_{R_2}) \right) \right.
\nonumber \\ 
& & \left. + s_L^2 \left( J_4(M_1^2,M_1^2,\mt^2_{L_3},\mt^2_{R_2})
+ 2 M_1^2 I_4(M_1^2,M_1^2,\mt^2_{L_3},\mt^2_{R_2}) \right) \right]
\nonumber \\ 
& - & (R_2 \to R_3)
\\
{B_R^{\mu_R} \over s_R c_R} & = &
 {g'^4 \over 16 \pi^2} \left[
- c_R^2 \left( J_4(M_1^2,M_1^2,\mt^2_{R_2},\mt^2_{R_2})
+ 2 M_1^2 I_4(M_1^2,M_1^2,\mt^2_{R_2},\mt^2_{R_2}) \right) \right.
\nonumber \\ 
& & \left. + s_R^2 \left( J_4(M_1^2,M_1^2,\mt^2_{R_3},\mt^2_{R_3})
+ 2 M_1^2 I_4(M_1^2,M_1^2,\mt^2_{R_3},\mt^2_{R_3}) \right) \right.
\nonumber \\ 
& & \left. + (c_R^2-s_R^2) \left( J_4(M_1^2,M_1^2,\mt^2_{R_2},\mt^2_{R_3})
+ 2 M_1^2 I_4(M_1^2,M_1^2,\mt^2_{R_2},\mt^2_{R_3}) \right) \right]
\eea
\subsection{\large \bf Contributions to $\Delta_{L,R}$ }

The leading contributions to $\Delta_{L,R}$ are of 
 two types:
$\Delta_{L,R} = \Delta^{(b)}_{L,R}  +\Delta^{(c)}_{L,R}$.
The related diagrams are depicted in 
Fig.~\ref{delh} and give: 
\bea
 {\Delta^{(b)}_L \over  s_L c_L} &=& 
\frac{g^2}{16 \pi^2} \cdot \frac32 \mu M_2 
{ I_3}(M^2_2,\mu^2, \tilde{m}^2_{L_2}) 
+ \frac{g'^2}{16\pi^2} \cdot \frac{-1}{2} \mu M_1
{I_3}(M^2_1,\mu^2, \tilde{m}^2_{L_2}) \nonumber\\
& - &    (L_2 \to L_3) \\
 {\Delta^{(b)}_R \over  s_R c_R}& = &
\frac{g'^2}{16 \pi^2} \mu M_1 
{ I_3}(M^2_1,\mu^2, \tilde{m}^2_{R_2}) - 
(R_2\to R_3)
\eea
 
\bea
{\Delta^{(c)}_L \over  s_L c_L} &=&  
\frac{g'^2}{16 \pi^2} \mu M_1 \left[- s^2_R 
{ I_3}(M^2_1, \tilde{m}^2_{R_2}, \tilde{m}^2_{L_2}) 
-c^2_R  { I_3}(M^2_1, \tilde{m}^2_{R_3}, 
\tilde{m}^2_{L_2}) \right]
 -   (L_2 \to L_3) \\
{\Delta^{(c)}_R \over  s_R c_R} & = & 
 \frac{g'^2}{16 \pi^2} \mu M_1 \left[
-s^2_L { I_3}(M^2_1, \tilde{m}^2_{L_2}, \tilde{m}^2_{R_2}) 
- c^2_L {I_3}(M^2_1, \tilde{m}^2_{L_3},\tilde{m}^2_{R_2})\right] 
 -   (R_2 \to R_3)
\eea

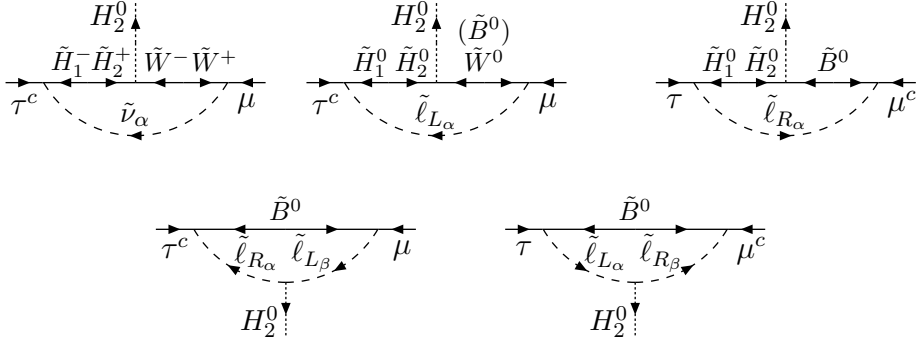
\begin{figure}[htb]
\begin{center}
\begin{picture}(110,60)(-55,-30)
\ArrowLine(-49,0)(-35,0)
\ArrowLine(-18,0)(-35,0)
\ArrowLine(-18,0)(0,0)
\ArrowLine(18,0)(0,0)
\ArrowLine(18,0)(35,0)
\ArrowLine(49,0)(35,0)
\DashLine(0,15)(0,0){1}
\DashArrowLine(0,15)(0,30){1}
\DashArrowArcn(0,20)(40,330,210){3}
\Text(-42,-8)[]{$\tau^c$}
\Text(42,-8)[]{$\mu$}
\Text(-10,25)[]{\small{$H_2^0$}}
\Text(-24,8)[]{\ftsz{$\tilde{H}^-_1$}}
\Text(-9,8)[]{\ftsz{$\tilde{H}^+_2$}}
\Text(12,9)[]{\ftsz{$\tilde{W}^-$}}
\Text(30,9)[]{\ftsz{$\tilde{W}^+$}}
\Text(0,-12)[]{\small{$\tilde{\nu}_{\alpha}$}}
\end{picture}
\begin{picture}(110,60)(-55,-30)
\ArrowLine(-49,0)(-35,0)
\ArrowLine(-18,0)(-35,0)
\ArrowLine(-18,0)(0,0)
\ArrowLine(18,0)(0,0)
\ArrowLine(18,0)(35,0)
\ArrowLine(49,0)(35,0)
\DashLine(0,15)(0,0){1}
\DashArrowLine(0,15)(0,30){1}
\DashArrowArcn(0,20)(40,330,210){3}
\Text(-42,-8)[]{$\tau^c$}
\Text(42,-8)[]{$\mu$}
\Text(-10,25)[]{\small{$H_2^0$}}
\Text(-25,8)[]{\ftsz{$\tilde{H}^0_1$}}
\Text(-9,8)[]{\ftsz{$\tilde{H}^0_2$}}
\Text(18,9)[]{\ftsz{$\tilde{W}^0$}}
\Text(18,20)[]{\ftsz{$(\tilde{B}^0)$}}
\Text(0,-11)[]{\small{$\tilde{\ell}_{L_\alpha}$}}
\end{picture}
\hspace{0.5 cm}
\begin{picture}(110,60)(-55,-30)
\ArrowLine(-49,0)(-35,0)
\ArrowLine(-18,0)(-35,0)
\ArrowLine(-18,0)(0,0)
\ArrowLine(18,0)(0,0)
\ArrowLine(18,0)(35,0)
\ArrowLine(49,0)(35,0)
\DashLine(0,15)(0,0){1}
\DashArrowLine(0,15)(0,30){1}
\DashArrowArc(0,20)(40,210,330){3}
\Text(-42,-7)[]{$\tau$}
\Text(44,-8)[]{$\mu^c$}
\Text(-10,25)[]{\small{$H_2^0$}}
\Text(-25,8)[]{\ftsz{$\tilde{H}^0_1$}}
\Text(-9,8)[]{\ftsz{$\tilde{H}^0_2$}}
\Text(18,9)[]{\ftsz{$\tilde{B}^0$}}
\Text(0,-11)[]{\small{$\tilde{\ell}_{R_\alpha}$}}
\end{picture}
\end{center}
\begin{center}
\vspace{-1.cm}
\begin{picture}(110,80)(-55,-40)
\ArrowLine(-49,0)(-35,0)
\ArrowLine(0,0)(-35,0)
\ArrowLine(0,0)(35,0)
\ArrowLine(49,0)(35,0)
\DashArrowLine(0,-20)(0,-40){1}
\DashArrowArcn(0,20)(40,330,270){3}
\DashArrowArcn(0,20)(40,270,210){3}
\Text(-42,-7)[]{$\tau^c$}
\Text(44,-8)[]{$\mu$}
\Text(0,8)[]{\ftsz{$\tilde{B}^0$}}
\Text(-10,-35)[]{\small{$H_2^0$}}
\Text(-11,-9)[]{\small{$\tilde{\ell}_{R_\alpha}$}}
\Text(10,-9)[]{\small{$\tilde{\ell}_{L_\beta}$}}
\end{picture}
\hspace{0.5 cm}
\begin{picture}(110,80)(-55,-40)
\ArrowLine(-49,0)(-35,0)
\ArrowLine(0,0)(-35,0)
\ArrowLine(0,0)(35,0)
\ArrowLine(49,0)(35,0)
\DashArrowLine(0,-20)(0,-40){1}
\DashArrowArc(0,20)(40,210,270){3}
\DashArrowArc(0,20)(40,270,330){3}
\Text(-42,-7)[]{$\tau$}
\Text(44,-8)[]{$\mu^c$}
\Text(0,8)[]{\ftsz{$\tilde{B}^0$}}
\Text(-10,-35)[]{\small{$H_2^0$}}
\Text(-11,-9)[]{\small{$\tilde{\ell}_{L_\alpha}$}}
\Text(10,-9)[]{\small{$\tilde{\ell}_{R_\beta}$}}
\end{picture}
\end{center}
\caption{\ftsz Diagrams that contribute to $\Delta^{(b)}_{L,R}$ (first row)
and  $\Delta^{(c)}_{L,R}$ (second row).}
\label{delh}
\end{figure}

\subsection{\large \bf Contributions to muon $g-2$ \label{ag2}}
The MSSM contributions to the muon anomalous magnetic moment are 
described by diagrams analogous to those for $D^\ga$ 
(Figs.~\ref{dgalra},~\ref{dgalrb},~\ref{dgalrc}), 
with the replacements $\tau (\tau^c) \to \mu (\mu^c)$.
The result can be decomposed into  five terms, {\it i.e.}
$a^{\rm MSSM}_\mu = a^{(a)}_{\mu L} +a^{(a)}_{\mu R} +
a^{(b)}_{\mu L}+a^{(b)}_{\mu R}+a^{(c)}_{\mu L R}$,  given as follows:
\bea
{ a^{ (a)}_{\mu L} \over 2 m^2_\mu} & = & 
{g^2 \over 16 \pi^2} \cdot {c^2_L \over 4}  \left[ 
\mt^2_{L_2} J_5(M_2^2,\mt^2_{L_2},\mt^2_{L_2},\mt^2_{L_2},\mt^2_{L_2})
-2 M_2^2 J_5(M_2^2,M_2^2,M_2^2,M_2^2,\mt^2_{L_2}) \right]
\nonumber \\
& + & 
{g'^2 \over 16 \pi^2} \cdot {c^2_L \over 4}  
\mt^2_{L_2} J_5(M_1^2,\mt^2_{L_2},\mt^2_{L_2},\mt^2_{L_2},\mt^2_{L_2})
+ (c^2_L\to s^2_L, L_2 \to L_3)
\\
{ a^{ (a)}_{\mu R} \over 2 m^2_\mu   } & = & 
{g'^2 \over 16 \pi^2}
c^2_R \mt^2_{R_2} J_5(M_1^2,\mt^2_{R_2},\mt^2_{R_2},\mt^2_{R_2},\mt^2_{R_2})
+ (c^2_R\to s^2_R, R_2 \to R_3) \\
{  a^{ (b)}_{\mu L} \over 2 m^2_\mu  } & = & 
{g^2 \over 16 \pi^2}  \cdot {c^2_L \over 2} \left[ 
- \mt^2_{L_2} 
J_5(M_2^2,\mu^2,\mt^2_{L_2},\mt^2_{L_2},\mt^2_{L_2})
- 2 \mt^4_{L_2} I_5(M_2^2,\mu^2,\mt^2_{L_2},\mt^2_{L_2},\mt^2_{L_2})
\right.
\nonumber \\
& & \left.
+ M_2 \mu \tan\beta \left(   \mt^2_{L_2} 
I_5(M_2^2,\mu^2,\mt^2_{L_2},\mt^2_{L_2},\mt^2_{L_2})
+ 2 J_5(M_2^2,M_2^2,\mu^2,\mu^2,\mt^2_{L_2}) \right. \right.
\nonumber \\
& & \left. \left.
+ 2 J_5(M_2^2,M_2^2,M_2^2,\mu^2,\mt^2_{L_2})
+ 2 J_5(M_2^2,\mu^2,\mu^2,\mu^2,\mt^2_{L_2})
\right) \right]
\nonumber \\
& + & 
{g'^2 \over  16 \pi^2} \, { c^2_L \over 2} \mt^2_{L_2}  
\left[ J_5(M_1^2,\mu^2,\mt^2_{L_2},\mt^2_{L_2},\mt^2_{L_2}) 
- M_1 \mu \tan\beta \,
I_5(M_1^2,\mu^2,\mt^2_{L_2},\mt^2_{L_2},\mt^2_{L_2}) \right]
\nonumber \\
& + & (c^2_L\to s^2_L, L_2 \to L_3)
\\
{  a^{ (b)}_{\mu R} \over 2 m^2_\mu   } & = & 
{g'^2 \over  16 \pi^2} c^2_R \mt^2_{R_2} 
\left[ - J_5(M_1^2,\mu^2,\mt^2_{R_2},\mt^2_{R_2},\mt^2_{R_2}) 
+ M_1 \mu \tan\beta \,
I_5(M_1^2,\mu^2,\mt^2_{R_2},\mt^2_{R_2},\mt^2_{R_2}) \right]
\nonumber \\
& + & (c^2_R \to s^2_R, R_2 \to R_3) \\
{  a^{ (c)}_{\mu L R} \over 2 m^2_\mu   }  & = & 
- {g'^2 \over 16 \pi^2} M_1^3 \, \times
\nonumber \\
& & \left\{
\left[ \frac{m_\tau}{m_\mu}\left(c^2_R s_L c_L A^R_{\mu \tau}+
s_R c_R c^2_L A^L_{\mu \tau} + s_R c_R s_L c_L
 [A_{\tau} + \mu \tan\beta ]\right) \right. \right. \nonumber\\
& & \left. \left.  
+ c^2_R  c^2_L (A_{\mu} + \mu \tan\beta ) \right]
I_5(M_1^2,M_1^2,M_1^2,\mt^2_{L_2},\mt^2_{R_2}) \right.
\nonumber \\
& & + \left. 
\left[ \frac{m_\tau}{m_\mu}\left(- c^2_R s_L c_L A^R_{\mu \tau}+
s_R c_R s^2_L A^L_{\mu \tau} - s_R c_R s_L c_L
 [A_{\tau} + \mu \tan\beta ]\right) \right. \right. \nonumber\\
& & \left. \left.  
+ c^2_R  s^2_L (A_{\mu} + \mu \tan\beta ) \right]
I_5(M_1^2,M_1^2,M_1^2,\mt^2_{L_3},\mt^2_{R_2}) \right.
\nonumber \\
& &  \left.
+\left[ \frac{m_\tau}{m_\mu}\left(s^2_R s_L c_L A^R_{\mu \tau} -
s_R c_R c^2_L A^L_{\mu \tau} - s_R c_R s_L c_L
 [A_{\tau} + \mu \tan\beta ]\right) \right. \right. \nonumber\\
& & \left. \left.  
+s^2_R  c^2_L (A_{\mu} + \mu \tan\beta ) \right]
I_5(M_1^2,M_1^2,M_1^2,\mt^2_{L_2},\mt^2_{R_3}) \right.
\nonumber \\
& &\left.
+\left[ \frac{m_\tau}{m_\mu}\left(- s^2_R s_L c_L A^R_{\mu \tau} -
s_R c_R s^2_L A^L_{\mu \tau} + s_R c_R s_L c_L
 [A_{\tau} + \mu \tan\beta ]\right) \right. \right. \nonumber\\
& & \left. \left.  
+ s^2_R  s^2_L (A_{\mu} + \mu \tan\beta ) \right]
I_5(M_1^2,M_1^2,M_1^2,\mt^2_{L_3},\mt^2_{R_3}) \right\} 
\nonumber \\
& & 
\eea

\end{document}